# I/O EFFICIENT ALGORITHMS FOR MATRIX COMPUTATIONS

A Thesis Submitted

in Partial Fulfilment of the Requirements

for the Degree of

## DOCTOR OF PHILOSOPHY

*by*

**Sraban Kumar Mohanty**

**(Roll Number: 03610107)**

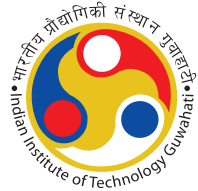

*to the*

**DEPARTMENT OF COMPUTER SCIENCE AND ENGINEERING**

**INDIAN INSTITUTE OF TECHNOLOGY GUWAHATI**

**GUWAHATI - 781039, INDIA**

December, 2009

# CERTIFICATE

It is certified that the work contained in the thesis entitled "**I/O EFFICIENT ALGORITHMS FOR MATRIX COMPUTATIONS**" by **Sraban  Kumar Mohanty** has been carried out under my supervision and that this work has not been submitted elsewhere for a degree.

December 2009

**Dr. Sajith Gopalan**
Associate Professor
DEPARTMENT OF COMPUTER SCIENCE AND ENGINEERING
IIT Guwahati

# Acknowledgements

The thesis submission marks the end of a long and eventful journey with awful turns and little triumphs over them, for which I would like to acknowledge many people for their support along the way.

From my first step into IIT Guwahati to the thesis submission, who has been perpetually supporting me with his proper guidance balanced by the freedom to express myself, trenchant analysis to probing questions, remarkable patience to placid composure, tireless editing of thesis to imparting intellection, he is my supervisor, Dr. Sajith Gopalan. Thank you Sir.

I would like to thank my doctoral committee members, Dr. Rafikul Alam, Dr. S. V. Rao, and Dr. D. Goswami for their support. I am obliged to Prof. Sukumar Nandi for his help and advises. I am indebted to Prof. Robert A. van de Geijn and Prof. Jack J. Dongarra for their support in answering my queries with profound clarity through e-mails during initial days of my research.

I thank the Council of Scientific & Industrial Research, Govt. of India for providing me with financial support through Senior Research Fellowship. Thanks are also due to the Department of Computer Science and Engineering, IIT Guwahati for all facilities provided to me during my research work. Mr. N. Malakar of CSE department deserve special thanks for his assistance in all official matters.

My gratitude to my idol, my philosopher and my ideal teacher, Prof. Swadhinananda Pattanayak of Institute of Mathematics and Applications, Bhubaneswar, for his motivation to pursue research. I am obliged to Prof. R. Balasubramanian of Institute of Mathematical Sciences Chennai for his inspirational teachings and his encouragement to pursue research during my M.Tech dissertation and there after.

My earnest respect to all my teachers, starting from my remote village primary school to University, in this long journey they have guided, showed the path towards success and enriched my personality by imparting ethical and academic knowledge. I would never forget your love and care, the discipline, advice and suggestions whenever I need them and you all deserve to be the true role models.

I am indebted to Dr Sujit Dash and Mana bhai for their elderly advices. Special regards to Sujata didi of Vivekanda Kendra for teaching me Yoga and she has been an important and indispensable source of spiritual support in my life out side research. At a thousand miles distance from home, Utkalika, the Odia family of IIT Guwahati has been most significant in making me feel at home during my stay.

I owe collective and individual acknowledgments to my colleagues: cool Alka, brilliant Arup sir, discussant Bidyut, sober Godfrey, conscious Lipika, happy Mausumi ma'am, homey Neminath, definitive Nitya sir, pleasing Pallav, jubilant Pravati, childish Rajendra, shy Rohit, hardworking Satya, ready Suresh, smiling Suddhasil, and respectful Tezmani sir of CSE Department whose presence somehow perpetually refreshed, helped, and made my stay memorable. I would miss you all especially during the evening tea.

My gratitude to Shree, Piu and Sridhar Samal for their kind hospitality. I would like to remember Safiq bhai, Martha, Jugal, Subash, Sumant, and Smruti for their goodness. Sweet memories of friendship, who were always ready to share hard realities of life with a helping hand are also due to philosophical Tara, youngish Biswa, and gushing Raju.

The lazy evenings spent on the mythical rocks present on the shore of grand Brahmaputra and soothed by the meditative rhythmic sound, the sunset in the dusky horizon and the illuminated full moon nights over the hilltop with serene calmness, the chirping birds in the tranquil lake, the omnipresent cheerful trees, mild earthquakes in mid nights, silverish wet fog entering the hostel rooms in quiet winter nights, cruising in mighty Brahmaputra; each is a memorable addendum to my IIT Guwahati experience. These memories are heartily cherished.

I wish to thank my childhood friends Dusa, Sanjaya, Bijaya, Fani, Kamal for helping me get through the difficult times, and for all the emotional support, camaraderie, fun, and caring they provided. Thanks for being there for me whenever I needed you.

My deepest gratitude goes to my parents for their unflagging love and support throughout my life. In spite of many hardships my parents spare no effort to provide the best possible environment for me to grow up and attaining education. I have no suitable words that can fully describe their everlasting love to me. The constant love and affection of my sisters Manu and Binu is sincerely acknowledged.

Lastly, but most importantly my appreciation to my wife Mami whose love, selfless sacrifice, ever compromising attitude, and care helped me to concentrate on completing this work.

With a limited sentiency, my heartiest gratitude to all of them, who have been a part of my life but I failed to mention them here.

December 2009                                                  **Sraban Kumar Mohanty**

# Abstract

Algorithms for large data sets, unlike in-core algorithms, have to keep the bulk of their data in the secondary memory, which typically is much slower than the main memory. In designing these out-of-core algorithms, the goal is therefore to minimise the number of I/Os executed. The literature is rich in efficient out-of-core algorithms for matrix computation. But very few of them are designed on the external memory model of Aggarwal and Vitter, and as such attempt to quantify their performances in terms of the number of I/Os performed. This thesis makes some contributions in that direction.

We analyse some QR decomposition algorithms, and show that the I/O complexity of the tile based algorithm is asymptotically the same as that of matrix multiplication. This algorithm, we show, performs the best when the tile size is chosen so that exactly one tile fits in the main memory. We propose a constant factor improvement, as well as a new recursive cache oblivious algorithm with the same asymptotic I/O complexity.

The traditional unblocked and blocked Hessenberg, tridiagonal, and bidiagonal reductions are not I/O efficient because vector-matrix operations dominate their performances. We design Hessenberg, tridiagonal, and bidiagonal reductions that use banded intermediate forms, and perform only asymptotically optimal numbers of I/Os; these are the first I/O optimal algorithms for these problems.

In particular, we show that known slab based algorithms for two sided reductions all have suboptimal asymptotic I/O performances, even though they have been reported to do better than the traditional algorithms on the basis of empirical evidence.

We propose new tile based variants of multishift QR and QZ algorithms that under certain conditions on the number of shifts, have better seek and I/O complexities than all known variants.

We show that techniques like rescheduling of computational steps, appropriate choosing of the blocking parameters and incorporating of more matrix-matrix operations, can be used to improve the I/O and seek complexities of matrix computations.

# Contents

# List of Figures

# List of Tables

# Chapter 1

# Introduction

Computing applications that deal with massive data sets abound today. Examples can be found in databases, geographic information systems, VLSI verification, computerised medical treatment, astrophysics, geophysics, multimedia, computer graphics, virtual reality, 3D simulation and modelling, sensors, web applications, network traffic analysis, constraint logic programming, computational biology, to name a few.

Traditional "in-core" algorithms assume that the main memory is infinite in size and allows random uniform access to all its locations, and therefore, that all their data fit in the main memory. Algorithms for large data sets (often called out-of-core (OOC) or external memory algorithms) cannot assume this. In reality, the main memory is limited; so, these algorithms have to keep the bulk of their data in the secondary memory, which typically is much slower than the main memory.

The usual performance metric for in-core algorithms is the number of instructions executed. However, for OOC algorithms, the number of inputs/outputs (I/Os) executed would be a more appropriate performance metric. So, in OOC algorithm design, the goal is to minimise the number of I/Os. OOC algorithms are therefore sometimes called I/O efficient algorithms.

In the last few decades, computers have become a lot faster, and the amount of main memory they have has grown. But the issue of the main memory being limited has only become more relevant, because applications have grown even faster in size. Also, small computing devices (e.g., sensors, smart phones) with limited memories have found several uses.

## 1.1   The External Memory Model

Designing of I/O efficient algorithms has, therefore, been an actively researched area in the last few years. For many a problem, it has been shown that the traditionally efficient algorithm is not very I/O efficient, and that novel and very different design techniques can be used to produce a much more I/O efficient algorithm. The external memory model of Aggarwal and Vitter [4,107] has been used to design many of these algorithms. This model has a single processor and a two level memory. It is assumed that the bulk of the data is kept in the secondary memory (disk) which is a permanent storage. The secondary memory is divided into blocks. An I/O is defined as the transfer of a block of data between the secondary memory and a volatile main memory, which is limited in size. The processor's clock period and main memory access time are negligible when compared to secondary memory access time. The measure of performance of an algorithm is the number of I/Os it performs. The model defines the following parameters: the size of the main memory ($M$), and the size of a disk block ($B < M$). (The two levels might as well be the cache and the main memory.)

We add another performance metric, namely, the number of seeks, to the model. A seek is the task of positioning the read/write head of the disk to the required data block. In reality, the cost of seeking a block will depend on the location of the block on the disk. We make the simplistic assumption that every block can be sought at the same cost. We also assume that, irrespective of the size of the data block, all contiguous data can be accessed using a single seek. We define the seek complexity of an algorithm as the number of times the algorithm leaves one data locality for another, (or, in other words, chooses to read or write a block that is not physically contiguous with the last block read or written) thereby necessitating a seek.

Suppose that an algorithm performs $k$ seeks, and that after the $i$-th seek it moves $s_i$ data items between the disk and the main memory. Let $n_i = \lceil \frac{s_i}{B} \rceil B$, and $s = \sum_{i=1}^{k} s_i$, and $n = \sum_{i=1}^{k} n_i$. Then $\frac{n}{B} = \sum_{i=1}^{k} \frac{n_i}{B} \leq \sum_{i=1}^{k} \lceil \frac{s_i}{B} \rceil \leq \sum_{i=1}^{k} (\frac{s_i}{B} + 1) = \frac{s}{B} + k$. That is, the number of I/Os performed by the algorithm is at most the number of blocks of data items moved plus the number of seeks.

Throughout this thesis, unless stated otherwise, we assume that $M \geq B^2$. This is a realistic assumption to make, and is called the "tall cache assumption" [54].

Algorithms that are oblivious of the memory hierarchy, and nevertheless use the hierarchy efficiently have received particular attention because they are portable [54]. A hierarchy oblivious algorithm leaves the page replacement decisions to the operating system (and still executes efficiently), whereas a hierarchy aware algorithm is assumed to have complete control over the secondary memory. Hierarchy oblivious algorithms are often called "cache oblivious" in literature.

## 1.2   OOC Algorithms for Matrix Computations

Early work on external memory algorithms were largely concentrated on fundamental problems like sorting, permutating, graph problems, computational geometry and string matching problems [4, 9, 29, 30, 40, 59, 94, 106, 108]. External memory algorithms for fundamental matrix operations like matrix multiplication and matrix transposition were proposed in [4, 108]. Not many linear algebra algorithms have been designed or analysed explicitly on the external memory model.

Software libraries that implement various linear algebra operations have been in development over many years. BLAS [15, 41, 42] provides kernel routines to implement a variety of fundamental vector and matrix operations. The software library LAPACK [7], based on BLAS routines and designed for sequential computers, implements a variety of advanced linear algebra routines. LAPACK does not provide explicit OOC capabilities, and has to rely on virtual memory systems to perform out-of-core computations. For larger problems, an extension of LAPACK called ScaLAPACK [31] is available. It is designed for distributed memory parallel architectures, and includes prototype OOC implementations of some liner system solvers. SOLAR [101] further extends LAPACK and ScaLAPACK, by adding OOC capabilities to them in a portable fashion. SOLAR routines use ScaLAPACK routines for in-core computations and manage matrix inputs and outputs using an I/O layer. PLA-PACK [5, 103], an extension of LAPACK for parallel computations, provides for coding parallel linear algebra algorithms at a high level of abstraction. POOCLAPACK [93] is an OOC extension of PLAPACK. It provides parallel implementations of OOC linear algebra operations. It uses PLAPACK routines for in-core parallel computations.

Even apart from the above, the literature is rich in efficient out-of-core algorithms. For example, dense-matrix factorizations like LU-factorization, Cholesky factorization, QR-factorization, solving of systems of linear equations, standard and generalized eigenvalue problems have been investigated from the out-of-core perspective. But very few of them are designed on the external memory model, and as such attempt to quantify the performance in terms of the number of I/Os performed. Without such analyses, it is often difficult to say for sure which algorithm will do better under a given circumstance. This thesis makes some contributions in that direction.

See [100] for a survey on out-of-core algorithms in linear algebra. Also see [50] for a survey on recursive blocked algorithms and hybrid data structures for problems like Cholesky and QR decompositions, and solving of general linear systems.

Most of the computations involving vector-vector (V-V) or vector-matrix (V-M) operations, perform $O(N^2/B)$ I/Os while executing $O(N^2)$ flops thus giving the so called *surface-to-surface* effect to the ratio of flops to I/Os [58, 115]. It has been observed that casting computations in terms of matrix-matrix (M-M) operations improves performance by reducing I/Os [23,44,58,115]. The reason is that M-M operations perform only $O(N^2/B)$ I/Os while executing $O(N^3)$ flops giving the so called *surface-to-volume* effect to the ratio of flops to I/Os [44].

Algorithms can be recast to be rich in M-M operations using either the *slab* or the *tile* approach. In the former, the matrix is partitioned into vertical slabs, which are processed one after the other; between two slabs the whole of the matrix may have to be updated using an aggregate of transformations generated from the processing of the slab; this updation could be done using M-M operations. Slab based algorithms for Cholesky, LU and QR decompositions that use aggregations of Gauss elimination or Householder transformations have been reported [41, 48, 62]. In the tile approach, the matrix is typically partitioned into square tiles. It has been observed that the tile approach results in more scalable out-of-core algorithms for Cholesky [63, 93, 101] and QR decompositions [64], and in more granular and asynchronous parallel algorithms on multicore architectures for LU, Cholesky, and QR decompositions [28].

## 1.3   Contributions of this Thesis

In this thesis, we present I/O efficient algorithms for some matrix problems. We also analyse some known algorithms on the external memory model.

## QR Decomposition

For every matrix $A \in \mathbb{R}^{N \times P}$, there exist an orthogonal matrix $Q \in \mathbb{R}^{N \times N}$ and upper triangular matrix $R \in \mathbb{R}^{N \times P}$ such that $A = QR$. The decomposition of $A$ into $Q$ and $R$ is called a QR-decomposition of $A$.

QR-decomposition is useful in solving eigenvalue problems, systems of linear equations and linear least-square problems. We investigate the various out-of-core QR-decomposition algorithms in Chapter 2.

A QR decomposition of a matrix can be obtained using Givens rotations, orthogonalisation via Gram-Schmidt and Modified Gram-Schmidt methods, or Householder transformations [58, 115]. Stability of the system, dimensions of the matrix, architecture of the machine and how the factorization is to be subsequently used are some of the factors that influence the choice [64].

We focus on Householder based methods for QR-decomposition, because these exploit locality of reference better than rotation based algorithms [21, 78, 79] and therefore are more amenable to OOC computations.

The traditional QR decomposition based on Householder transformations is rich in matrix-vector, and not matrix-matrix operations. Efficient parallel and OOC computations require algorithms rich in the latter. Exploiting data locality to recast algorithms in terms of matrix-matrix multiplication has received much attention [7, 23, 39, 41, 43, 96, 101]. For QR decomposition using Householder transformations, slab based [7, 23, 91, 96] and tile based [28, 64] algorithms have been suggested. A slab based algorithm reads a slab of columns into the main memory at a time. A tile based algorithm partitions the matrix into a number of square tiles, and processes one tile at a time.

We analyse some QR decomposition algorithms for their I/O complexity. We show that the traditional algorithm based on Householder transformations [115] has an I/O complexity of $O(NP\min\{N, P\}/B)$.

The slab based algorithm of [96] is analysed. We choose this algorithm because of its simplicity. (There have been slab based algorithms that build on this. Recursive in-core algorithms that partition the matrix vertically are given in [48, 49]. An OOC implementation of these is given in [91].) We consider an implementation of the algorithm where matrix multiplication is done using the blocked algorithm of [58]. We find that the I/O cost of this implementation is of the same order as that of matrix multiplication for most choices of the problem parameters, and that the optimal choice for slab width is not necessarily the number of columns that can be fit in the main memory.

We also analyse the tile based algorithm of [64], and show that its I/O complexity is of the same order as that of matrix multiplication. We also suggest a new tile based algorithm that improves the I/O complexity by an $O(1)$ factor.

We also present a new cache-oblivious QR-decomposition algorithm, and its analysis. This algorithm is useful when $Q$ needs to be reported explicitly. Its I/O complexity is of the same order as that of the tile based algorithm.

## Banded Hessenberg reduction and related problems

Computing of eigenvalues and singular values of matrices have wide-ranging applications in engineering and computational sciences.

The eigenvalues of a square matrix are typically computed in a two stage process. In the first stage the matrix is reduced, through a sequence of orthogonal similarity transformations (OSTs), to a condensed form (Hessenberg form, if the matrix is nonsymmetric, and tridiagonal form, otherwise), and in the second stage an iterative method called the $QR$ algorithm is applied to the condensed form [58, 115]. The reduction of a nonsymmetric $N \times N$ matrix to upper Hessenberg form as well as a symmetric $N \times N$ matrix to tridiag-onal form using OSTs takes $O(N^3)$ flops. Each iteration of the QR algorithm takes $O(N^2)$ and $O(N)$ flops respectively, when applied on Hessenberg and tridiagonal matrices.

The singular values of a matrix are also typically computed in a similar two stage process. First the matrix is reduced to bidiagonal form using orthogonal equivalence transformations (OETs); this takes $O(N^3)$ flops. Then the modified QR algorithm [57] is applied to the bidiagonal matrix. Each iteration of the QR algorithm takes $O(N)$ flops.

(An OST on a square matrix $A \in \mathbb{R}^{N \times N}$ transforms $A$ into $Q^T A Q$ using an orthogonal matrix $Q$. An OET on a matrix $A \in \mathbb{R}^{N \times P}$ transforms $A$ into $Q^T A P$ using orthogonal matrices $Q$ and $P$. OSTs and OETs preserve eigenvalues.)

Thus, in each case, the first stage of the computation, which costs $O(N^3)$ flops, forms a bottleneck. Each of these algorithms also needs $O(N^3/B)$ I/Os to perform its $O(N^3)$ operations.

The traditional Hessenberg, tridiagonal, and bidiagonal reduction algorithms based on Householder transformations (Householders) or rotators are rich in V-V and V-M operations, and not in M-M operations [58, 115]. Householder based sequential and parallel algorithms for Hessenberg, tridiagonal and bidiagonal reductions using the slab approach have been proposed [45, 46, 89]. But even in these, V-M operations dominate the performance: after the reduction of each column of a slab, the rest of the matrix needs to be read in to update the next column before it will be ready to be reduced. This is because of the two sidedness of the transformations involved. This also makes it difficult to design tile based versions of these algorithms. To the best of our knowledge, no tile based algorithm has been proposed for the above direct reductions.

Due to the above reasons, it has been proposed that the reduction in the first stage be split into two steps [17, 19, 21, 77, 80], the first reducing the matrix to a block condensed form (banded Hessenberg, symmetric banded or banded, as is relevant) [11, 61, 81], and the second further reducing it to Hessenberg, tridiagonal or bidiagonal form [21, 77, 78, 80]. Almost all operations of the first step are performed using M-M operations. Though the modified first stage takes more flops, its richness in M-M operations makes it more efficient on machines with multiple levels of memory.

The first steps of these reductions, namely the banded Hessenberg and related reductions, are discussed in Chapter 3.

**Reduction of a nonsymmetric matrix to banded Hessenberg form**   We study the slab based algorithm of [58], and analyse it for its I/O complexity for varying values of $M$ and $N$, for a given value of $t$, the desired bandwidth. This algorithm assumes that the slab size $k = t$. We generalise this into an algorithm with $k$ not necessarily the same as $t$. We find that the algorithm performs the best when $k = \min\{t, \sqrt{M}\}$.

We also investigate the I/O complexity of some tile based algorithms [27, 64, 90].

These algorithms, typically choose $t$ as the tile size, and also at any time, maintain only a small number of tiles in the main memory. If $t \ll \sqrt{M}$, then most of the main memory stays unused. We propose an algorithm that allows the number of tiles kept in the main memory to depend on the tile size. We also propose an algorithm for $t > \sqrt{M}$, which partitions the matrix into tiles of size $\sqrt{M/3}$; this gives a constant factor improvement over the choice of $t$ as the tile size. We conclude that banded Hessenberg reduction can be achieved in $O(N(N-t)^2/B\min\{t, \sqrt{M}\})$ I/Os using the tile approach, when $(N-t) = \Theta(N)$.

**Reduction of a symmetric matrix to symmetric banded form**   We recall the standard slab based algorithm for the problem, and draw conclusions similar to the above. Our tile based algorithms for banded Hessenberg reduction work for symmetric band reduction too, and perform about half as many I/Os.

**Reduction to banded form**    We describe some known slab and tile based algorithms for the problem and analyse them for their I/O complexities, as in the cases of banded Hessenberg and symmetric band reductions.

## Reduction from banded Hessenberg form to Hessenberg form, and related problems

These are studied in Chapter 4.

We recall an unblocked and a blocked direct Hessenberg reduction algorithm, and show that they both have an I/O complexity of $O(N^3/B)$. For the banded Hessenberg to Hessenberg reduction we propose two algorithms. The first one requires $O(N^3/B)$ I/Os, but the second one, an extension of the first using tight packing of bulges requires only $O\left(\frac{t^2 N^2}{B} + \frac{N^3}{tB}\right)$ I/Os when $t$, the bandwidth, is at most $\sqrt{M}$. Combining this with the results of Chapter 3, we show that the Hessenberg reduction can be performed in $O(N^3/\sqrt{M}B)$ I/Os, when $N$ is sufficiently large. This matches the known lower bound on the data movement [13].

We recall unblocked and blocked tridiagonal reductions, and show that they have an I/O complexity of $O(N^3/B)$. For the symmetric banded form to tridiagonal form reduction, we show, the known algorithms take $O(N^2 t/B)$ I/Os. Combining this with the results of Chapter 3, we show that the tridiagonal reduction can be performed in $O(N^3/\sqrt{M}B)$ I/Os, when $N$ is sufficiently large. This matches the known lower bound on the data movement [13].

We recall unblocked and a blocked bidiagonal reductions, and show that they have an I/O complexity of $O(NP\min\{N,P\}/B)$, when applied on an $N \times P$ matrix. For the banded to bidiagonal reduction, we show, the known algorithms take $O(\min\{N,P\}^2 t)/B)$ I/Os, when $t$ is the bandwidth. Combining this with the results of Chapter 3, we show that the bidiagonal reduction can be performed in $O\left(\frac{\min\{N,P\}.NP}{\sqrt{M}B}\right)$ I/Os when the matrix is sufficiently large. This matches the known lower bound on data movement [13].

## The generalised eigenvalue problem

Chapter 5 deals with the generalized eigenvalue problem. We analyze both the unblocked and blocked algorithms for Hessenberg-Triangular reduction using Givens rotations [70,86] and show that their I/O complexity is $O(N^3/B)$. We then analyse the various known tile based algorithms [37,70] for reducing a matrix pair to banded Hessenberg-Triangular pair with bandwidth $t$. Then we study the reduction of a banded Hessenberg-Triangular pair to Hessenberg-Triangular pair [37,70], and obtain new algorithms analogous to those in Chapter 4. Then we put the two steps together, and find, for given values of $M$ and $N$, the band width $t$ that will minimise the I/O cost of the two step reduction. For sufficiently large matrices, our reduction matches the known lower bound on the data movement [13].

## The QR and QZ algorithms

The QR algorithm solves the standard eigenvalue problem. Analogously, the QZ-algorithm solves the generalised eigenvalue problem. Both are iterative algorithms. Chapter 6 deals with both of them. We analyse the multishift QR algorithm of [12]; this algorithm chases $m \times m$ bulges, where $m$ is the number of shifts, using both matrix-vector and matrix-matrix operations. We also investigate the small-bulge multishift QR algorithm [25] which was proposed to avoid the phenomenon called shift blurring. We propose a tile based algorithm that under certain conditions of the number of shifts has better seek and I/O complexities. We prove analogous results for the QZ algorithm [69] too.

## 1.4    Preliminaries

Most of the algorithms in this thesis, invoke matrix transposition and multiplication. Therefore, we analyse the well-known cache-aware algorithms for these problems here [68, 107, 108].

A matrix is said to be in row (column) major order if its rows (columns) appear contiguously on the disk.

### 1.4.1    Matrix Transposition

The standard memory-hierarchy aware matrix transposition algorithm divides the input and output matrices into tiles so that two tiles fit in the main memory together, and transposes the input tiles one at a time. Formally, on a $P \times Q$ matrix $E$:

---

**Algorithm 1.1.** *Matrix Transposition*

---

*Input: Matrices $E \in \mathbb{R}^{P \times Q}$ and $F \in \mathbb{R}^{Q \times P}$*
*Output: $F = E^T$*

---

*Let $s = \sqrt{M/2}$;*
**for** $i = 1$ *to* $\lceil P/s \rceil$ **do**
    **for** $j = 1$ *to* $\lceil Q/s \rceil$ **do**
        $F_{ji} = E_{ij}^T$;
    **endfor**
**endfor**

---

Assume that $B < P, Q, \sqrt{M/2}$. Also assume that $E$ and $F$ are in row-major order. The data movement amounts to $2PQ$. Therefore, the I/O complexity $T \leq \frac{2PQ}{B} + S$ where $S$ is the number of seeks.

We now estimate the number of seeks.

- $P, Q \leq \sqrt{M/2}$. The two matrices fit in the main memory together. Read $E$ in, compute $F$ and write $F$. If a seek is needed between two rows, then $S = P + Q$. If rows follow one another contiguously, then $S = 2$.

- $P \leq \sqrt{M/2}$ and $Q > \sqrt{M/2}$. Divide $E$ and $F$ into slabs of width $s$ each, $E$ vertically and $F$ horizontally, so that a slab each of $E$ and $F$ will fit in the main memory together. Let $q = \lfloor Q/s \rfloor s$ and $x_q = Q \bmod s$. If a seek is needed between two rows, then $S = \frac{q}{s}(P + s) + P + x_q = \frac{Pq}{s} + P + Q$.

  If rows follow one another contiguously, then $S = \frac{q}{s}(P+1) + P + 1 = \frac{Pq}{s} + P + q/s + 1$

- $Q \leq \sqrt{M/2}$ and $P > \sqrt{M/2}$. Similar to the case above. Interchange the roles of $P$ and $Q$.

- $P, Q > \sqrt{M/2}$. Divide $E$ and $F$ into tiles of width $s$ each, so that a tile each of $E$ and $F$ will fit in the main memory together. Let $q = \lfloor Q/s \rfloor s$, $x_q = Q \bmod s$, $p = \lfloor P/s \rfloor s$ and $x_p = P \bmod s$. In this case, it does not matter whether the rows follow one another contiguously. $S = \frac{p}{s}\frac{q}{s}(2s) + \frac{q}{s}(s + x_p) + \frac{p}{s}(s + x_q) + x_q + x_p = \frac{Pq + Qp}{s} + P + Q$.

If $B$ divides $s$, and $s$ divides both $P$ and $Q$ then, the seeks will not cause additional I/Os, and the I/O complexity would be exactly $\frac{2PQ}{B}$.

When $P = Q = N$, $T \leq \frac{2N^2}{B} + 2N$.

Table 1.1: The seek complexity of transposing a $P \times Q$ matrix $E$ into a $Q \times P$ matrix $F$, when both $E$ and $F$ are in row major order. If a seek is needed between two consecutive rows the seeks are as in column #1. Otherwise, the seeks are as in column #2.

| Seeks for $E$, $F$ in RMO | # 1 | # 2 |
|:---:|:---:|:---:|
| $P, Q \leq \sqrt{M/2}$ | $P + Q$ | 2 |
| $P \leq \sqrt{M/2}$ and $Q > \sqrt{M/2}$ | $\frac{Pq}{s} + P + Q$ | $\frac{Pq}{s} + P + \frac{q}{s} + 1$ |
| $Q \leq \sqrt{M/2}$ and $P > \sqrt{M/2}$ | $\frac{Qp}{s} + Q + P$ | $\frac{Qp}{s} + Q + \frac{p}{s} + 1$ |
| $P, Q > \sqrt{M/2}$ | $\frac{Pq+Qp}{s} + P + Q$ | $\frac{Pq+Qp}{s} + P + Q$ |

## 1.4.2 Matrix Multiplication

Here the input and output matrices are divided into tiles so that three tiles fit in the main memory together [100, 107, 108]. Formally, on a $P \times Q$ matrix $E$, and $Q \times R$ matrix $F$:

---

**Algorithm 1.2.** *Matrix Multiplication*

---

*Input: Matrices $E \in \mathbb{R}^{P \times Q}$, $F \in \mathbb{R}^{Q \times R}$ and $G \in \mathbb{R}^{P \times R}$*
*Output: $G = E * F$*

---

Let $s = \sqrt{M/3}$;
**for** $i = 1$ *to* $\lceil P/s \rceil$ **do**
    **for** $j = 1$ *to* $\lceil R/s \rceil$ **do**
        $G_{ij} = 0$;
        **for** $k = 1$ *to* $\lceil Q/s \rceil$ **do**
            $G_{ij} += E_{ik} * F_{kj}$;
        **endfor**
    **endfor**
**endfor**

---

Assume that $B < P, Q, R, \sqrt{M/3}$. (Also, assume that $E$, $F$ and $G$ are in row-major order; while the elements of a row are contiguous, a seek may be necessary between rows.) Let $q = (\lceil Q/s \rceil - 1)s$, $x_q = Q - sq$, $r = (\lceil R/s \rceil - 1)s$, $x_r = R - sr$, $p = (\lceil P/s \rceil - 1)s$ and $x_p = P - sp$. Let $D$ denote the amount of data moved, $S$ the number of seeks, and $T = D + S$ the I/O complexity.

- $P, Q, R \leq \sqrt{M/3}$. All three matrices fit in the main memory together. Read $E$ and $F$ in, compute $G$ and write $G$. The data moved is $D = PQ + QR + PR$. The number of seeks $S = 2P + Q$ or 3, depending on whether a seek is needed between two consecutive rows or not. $T \leq \frac{PQ + QR + PR}{B} + 2P + Q$.

- $P, Q \leq \sqrt{M/3}$ and $R > \sqrt{M/3}$. $x_p = P$. Divide $F$ and $G$ into tiles of width $s$ each, so that a tile each of $F$ and $G$ will fit in the main memory along with $E$. $F$ is read in and $G$ is written out tile by tile. We have $D = PQ + QR + PR$. $S = P + (\frac{r}{s} + 1)(P + Q) = 2P + Q + \frac{Pr + Qr}{s}$ or $S = 1 + (\frac{r}{s} + 1)(P + Q) = 2P + Q + \frac{Pr + Qr}{s}$, depending on whether a seek is needed between two consecutive rows or not. Thus, $T \leq \frac{PQ+QR+PR}{B} + 2P + Q + \frac{Pr+Qr}{s}$.

- $P, R \leq \sqrt{M/3}$ and $Q > \sqrt{M/3}$. Divide $E$ and $F$ into tiles of width and height respectively of $s$ each, so that a tile each of $E$ and $F$ will fit in the main memory along with $G$. Read $E$ and $F$ tile by tile, and write $G$. We have $D = PQ+QR+PR$. $S = P+(\frac{q}{s}+1)(P+s) = 2P+q+s+\frac{Pq}{s}$. $S = 1+(\frac{q}{s}+1)(P+s) = 2P+q+s+\frac{Pq}{s}$. Thus, $T \leq \frac{PQ+QR+RP}{B} + \frac{Pq}{s} + 2P + Q + s$

- $Q, R \leq \sqrt{M/3}$ and $P > \sqrt{M/3}$. Divide $E$ and $G$ into tiles of height $s$ each, so that a tile each of $E$ and $G$ will fit in the main memory along with $F$. Read $F$. Read $E$ and write $G$ tile by tile. We have $D = PQ+QR+PR$. $S = Q+(\frac{p}{s}+1)(s+s) = 2p+2s+Q$ or $S = 2p+2s+1$, depending on whether a seek is needed between two consecutive rows or not. We have $T \leq \frac{PQ+QR+RP}{B} + 2p + 2s + Q$.

- $P, Q > \sqrt{M/3}$ and $R \leq \sqrt{M/3}$. Divide $E$ into tiles of size $s \times s$, $F$ and $G$ into tiles of height $s$ each, so that a tile each of $E$, $F$ and $G$ will fit in the main memory. We have $D = \frac{p}{s}(sQ+QR) + (x_pQ+QR) + PR = \frac{pQR}{s} + PQ + QR + PR$. $S = \frac{p}{s}(s+(\frac{q}{s}+1)s+Q) + x_p + (\frac{q}{s}+1)x_p + Q = 2P+Q+\frac{pQ+Pq}{s}$. If the rows follow each other contiguously, then $Q$ in the above becomes $(\frac{q}{s}+1)$. Thus, $T \leq \frac{pQR}{sB} + \frac{PR+PQ+QR}{B} + 2P + Q + \frac{pQ+Pq}{s}$.

- $Q, R > \sqrt{M/3}$ and $P \leq \sqrt{M/3}$. Divide $F$ into tiles of size $s \times s$, $E$ and $G$ into tiles of width $s$ each, so that a tile each of $E$, $F$ and $G$ will fit in the main memory. We have $D = \frac{r}{s}(PQ+Qs) + (PQ+Qx_r) + PR = \frac{PQr}{s} + PQ + QR + PR$. $S = (P+P(\frac{q}{s}+1)+Q)(\frac{r}{s}+1) = \frac{Qr+2Pr+Pq}{s} + \frac{Pqr}{s^2} + Q + 2P$. $S$ remains the same even if row seeks are not needed. Thus, $T \leq \frac{PQr}{sB} + \frac{PQ+QR+PR}{B} + \frac{Qr+2Pr+Pq}{s} + \frac{Pqr}{s^2} + Q + 2P$

- $P, R > \sqrt{M/3}$ and $Q \leq \sqrt{M/3}$. Divide $G$ into tiles of size $s \times s$, $E$ and $F$ into tiles of height and width respectively of $s$ each, so that a tile each of $E$, $F$ and $G$ will fit in the main memory. We have $D = \frac{p}{s}\frac{r}{s}2Qs + \frac{r}{s}(x_pQ + Qs) + \frac{p}{s}(x_rQ + Qs) + (x_pQ + x_rQ) + PR = \frac{PQr + pQR}{s} + QR + PQ + PR$. $S = \frac{p}{s}\frac{r}{s}(2s + Q) + \frac{r}{s}(2x_p + Q) + \frac{p}{s}(2s + Q) + 2x_p + Q = \frac{pQr}{s^2} + \frac{2Pr + Qr + pQ}{s} + 2P + Q$. If rows follow each other contiguously, then $S$ is $\frac{pQr + pr}{s^2} + \frac{Pr + Qr + pQ + r + p}{s} + P + Q + 1$ Thus, $T \leq \frac{PQr + pQR}{sB} + \frac{pQr}{s^2} + \frac{2Pr + Qr + pQ}{s} + 2P + Q$.

- $P, Q, R > \sqrt{M/3}$. Divide $E$, $F$ and $G$ into tiles of size $s \times s$, so that a tile each of $E$, $F$ and $G$ will fit in the main memory. We have $D = \frac{p}{s}\frac{r}{s}2Qs + \frac{r}{s}(x_pQ + Qs) + \frac{p}{s}(x_rQ + Qs) + (x_pQ + x_rQ) + PR = \frac{PQr + pQR}{s} + QR + PQ + PR$. $S = \frac{p}{s}(\frac{r}{s} + 1)((\frac{q}{s} + 1)s + Q) + (\frac{r}{s} + 1)((\frac{q}{s} + 1)x_p + Q) + P(\frac{r}{s} + 1) = \frac{pQr + Pqr}{s^2} + \frac{2Pr + Qr + pQ + Pq}{s} + 2P + Q$. $S$ remains the same even if row seeks are not needed. Thus, $T \leq \frac{PQr + pQR}{sB} + \frac{PQ + QR + PR}{B} + \frac{pQr + Pqr}{s^2} + \frac{2Pr + Qr + pQ + Pq}{s} + 2P + Q$.

When $P = Q = R = N$, and $s$ divides $N$, there are only two cases:

- $N \leq \sqrt{M/3}$. $T \leq \frac{3N^2}{B} + 3N$.

- $N > \sqrt{M/3}$. We have $D = \frac{2N^3}{s} + N^2$. $S = \frac{2N^3}{s^2} + \frac{N^2}{s} - 2N$. Thus, $T \leq \frac{2N^3}{sB} + \frac{N^2}{B} + \frac{2N^3}{s^2} + \frac{N^2}{s} - 2N$.

In the above we assume that the matrices are in row major order. The case of column major order can be analysed similarly. The following Table 1.2 summarises the asymptotic seek complexities for the various cases.

In general, matrix multiplication takes $\Theta(\frac{PQR}{s^2} + \frac{PQ + QR + RP}{s} + Q + R)$ seeks if the matrices are in CMO.

## 1.5    Organisation of this Thesis

In Chapter 2 we study QR decomposition. Chapter 3 is devoted to banded Hessenberg reduction and related problems. Chapter 4 deals with the reduction of banded Hessenberg form to Hessenberg form. In Chapter 5, the generalised eigenvalue problem is studied. In Chapter 6, we study the QR and QZ algorithms.

Table 1.2: The asymptotic seek complexity of multiplying a $P \times Q$ matrix with a $Q \times R$ matrix, when all matrices are kept in row major order (RMO) or all are kept in column major order (CMO)

|  | RMO | CMO |
|---|---|---|
| $P, Q, R \leq s$ | $P + Q$ or 1 | $Q + R$ or 1 |
| $P, Q \leq s$ and $R > s$ | $\frac{(P+Q)R}{s}$ | $R$ |
| $P, R \leq s$ and $Q > s$ | $Q$ | $Q$ |
| $Q, R \leq s$ and $P > s$ | $P$ | $\frac{P(Q+R)}{s}$ |
| $P, Q > s$ and $R \leq s$ | $\frac{PQ}{s}$ | $\frac{PQ}{s}$ |
| $P, R > s$ and $Q \leq s$ | $\frac{PR}{s}$ | $\frac{PR}{s}$ |
| $Q, R > s$ and $P \leq s$ | $\frac{QR}{s}$ | $\frac{QR}{s}$ |
| $P, Q, R > s$ | $\frac{PQR}{s^2}$ | $\frac{PQR}{s^2}$ |

# Chapter 2

# QR Decomposition

## 2.1   Introduction

In this chapter, we study the QR-decomposition problem. For $A \in \mathbb{R}^{N \times P}$, the problem seeks to find an orthogonal matrix $Q \in \mathbb{R}^{N \times N}$, and an upper triangular matrix $R \in \mathbb{R}^{N \times P}$ so that $A = QR$. QR-decomposition is a classical factorization method like LU and Cholesky decompositions. It is useful in solving the least squares problem, the Eigenvalue problem and systems of linear equations. A QR decomposition can be obtained using Givens rotations, orthogonalisation via Gram-Schmidt and Modified Gram-Schmidt methods, or Householder transformations [58, 115]. Stability of the system, dimensions of the matrix, architecture of the machine and how the factorization is to be subsequently used are some of the factors that influence the choice [64].

We focus on Householder based methods for QR-decomposition, because these exploit locality of reference better than rotation based algorithms [21, 78, 79], and therefore are more amenable to OOC computations.

The traditional QR decomposition based on Householder transformations is rich in matrix-vector, and not matrix-matrix operations. Efficient parallel and OOC computations require algorithms rich in the latter. Exploiting data locality to recast algorithms in terms of matrix-matrix multiplication has received much attention [7, 23, 39, 41, 43, 96, 101]. For QR decomposition using Householder transformations, slab based [7, 23, 91, 96] and tile based [28, 64] algorithms have been suggested. A slab based algorithm reads a slab of columns into the main memory at a time. A tile based algorithm partitions the matrix into a number of square tiles, and processes one tile at a time.

We analyse some QR decomposition algorithms for their I/O complexity. We show that the traditional algorithm based on Householder transformations [115] has an I/O complexity of $O(NP \min\{N, P\}/B)$.

The slab based algorithm of [96] is analysed. We choose this algorithm because of its simplicity. (There have been slab based algorithms that build on this. Recursive in-core algorithms that partition the matrix vertically are given in [48, 49]. An OOC implementation of these is given in [91].) We consider an implementation of the algorithm where matrix multiplication is done using the blocked algorithm of [58]. We find that the I/O cost of this implementation is of the same order as that of matrix multiplication for most choices of the problem parameters, and that the optimal choice for slab width is not necessarily the number of columns that can be fit in the memory.

We also analyse the tile based algorithm of [64], and show that its I/O complexity is of the same order as that of matrix multiplication. (A tile based algorithm that uses techniques similar to those of [64] has been designed for multicore architectures [28].)

We also present a new cache-oblivious QR-decomposition algorithm, and its analysis. This algorithm is useful when $Q$ needs to be reported explicitly. Its I/O complexity is of the same order as that of the tile based algorithm.

### 2.1.1  Organisation of this Chapter

The rest of the chapter is organised as follows: Section 2 presents some preliminaries. Section 3 analyses the traditional QR-decomposition that uses Householder transformations. Sections 4 and 5 analyse a slab based and tile based algorithm respectively. In Section 6, we present a new cache-oblivious QR-decomposition algorithm and analyse it.

## 2.2  Preliminaries

If $u$ is a unit vector in $\mathbb{R}^N$, then there exists a symmetric orthogonal matrix $Q_u \in \mathbb{R}^{N \times N}$, called the Householder of $u$, such that $Q_u u = -u$ and, for any vector $v$ orthogonal to $u$, $Q_u v = v$.

Here $u$ can be thought of as characterizing the $(N-1)$-dimensional hyperplane that passes through the origin and is orthogonal to $u$. Then, for any $x$, $Q_u x$ would be the reflection of $x$ in this hyperplane. It is easy to see that $Q_u = (I - 2uu^T)$ satisfies the requirement. If $||u||_2 \neq 1$, then let $\hat{u} = \frac{u}{||u||_2}$, the unit vector in the direction of $u$, and $Q_u = Q_{\hat{u}} = \left(I - \frac{2}{||u||_2^2} uu^T\right) = (I - \beta uu^T)$.

For any two vectors $x$ and $y$ with equal norms, there exists a $u$ such that $Q_u x = y$. This can be shown as follows: Let $u = x - y$ and $v = x + y$. As $(x-y)(x+y) = ||x||_2^2 - ||y||_2^2 = 0$, $u$ and $v$ are orthogonal. So, $Q_u u = -u$ and $Q_u v = v$. But $x = (x+y)/2 + (x-y)/2 = (u+v)/2$. Therefore, $Q_u x = Q_u(v+u)/2 = (v-u)/2 = y$. In particular, for a given $x = [x_1 \ \cdots \ x_N]^T$, let $\sigma = \text{sign}(x_1).||x||_2$, and $y = [-\sigma \ 0 \ \cdots \ 0]^T$. Clearly, $x$ and $y$ have equal norms. If $u = (x - y) = [x_1 + \sigma \ x_2 \ \cdots \ x_N]^T$, then $Q_u x = y$. Also, $||u||_2^2 = (x_1 + \sigma)^2 + x_2^2 + \ldots + x_N^2 = ||x||_2^2 + 2x_1\sigma + \sigma^2 = 2\sigma^2 + 2x_1\sigma = 2\sigma u_1$. Therefore, $Q_u = (I - (1/\sigma u_1)uu^T)$.

Thus, we have the following algorithmic result.

**Algorithm 2.1.** *Given vectors $u$ and $v$, $Q_u v$ can be found in $O(N)$ operations and $O(N/B)$ I/Os.*

*Proof.* $Q_u v = (v - \frac{2}{||u||_2} u u^T v)$. $u^T v$ can be found by making a pass each through $v$ and $u$; $||u||_2$ can be found on the fly; Let $a = \frac{2}{||u||_2} u^T v$. Now, $v - au$ can be found by scanning $u$ and $v$ a second time. That totals to $4N/B$ I/Os, without counting the $N/B$ I/Os needed to write the result. Also, the number of seeks needed is $O(N/M)$. □

**Algorithm 2.2.** *Given a vector $x$, a vector $u$ such that $Q_u x$ is along dimension 1 and has a sign opposite to that of $x_1$ can be found in $O(N)$ operations and $O(N/B)$ I/Os.*

*Proof.* $\max_i\{x_i\}$ can be found from a single scan of $x$. Dividing of every $x_i$ with this, calculating of $||x||_2$, $\sigma = \text{sign}(x_1) . ||x||_2$, and $u = [x_1 + \sigma \ x_2 \ \cdots \ x_N]^T$ can all be done with a further scan. That totals to $2N/B$ I/Os, without counting the $N/B$ I/Os needed to write $u$ normalised to have a first component of 1. Also, the number of seeks needed is $O(1)$. □

## 2.3 The Traditional Algorithm for $QR$ Decomposition using Householder Transformations, and its I/O Complexity

If $A \in \mathbb{R}^{N \times P}$, then there exists an orthogonal matrix $Q \in \mathbb{R}^{N \times N}$ and an upper triangular matrix $R \in \mathbb{R}^{N \times P}$ such that $A = QR$ [115]. This can be proved by induction on $N$ and $P$. If $N = 1$, then $A$ is already upper triangular; $A = [1]A$. If $P = 1$, then $A$ is a vector. By the discussion above, there exists a Householder transformation $Q$ such that $Q^T A = R$ where vector $R$ is upper triangular; all its components, except the first, are zero. Thus, $A = QR$. These form the basis. Suppose the statement is true for all matrices with fewer rows or columns than $A$. Choose $k < P$. If $k \geq N$, then let $A = (A_1, A_2)$, where $A_1$ has $k$ columns; $1 \leq k < P$. Then, by the hypothesis, there exists an orthogonal matrix $Q_1 \in \mathbb{R}^{N \times N}$ and an upper triangular matrix $R_1 \in \mathbb{R}^{N \times k}$ such that $A_1 = Q_1 R_1$. Then say, $R = Q_1^T A = (Q_1^T A_1 \quad Q_1^T A_2) = (R_1 \quad Q_1^T A_2)$. $R$ is upper triangular, and therefore $A = Q_1 R$. If $k < N$, let $A = \begin{pmatrix} A_{11} & A_{12} \\ A_{21} & A_{22} \end{pmatrix}$, where $A_{11} \in \mathbb{R}^{k \times k}$. Then, by the hypothesis, there exists an orthogonal matrix $Q_1 \in \mathbb{R}^{N \times N}$ and an upper triangular matrix $R_1 = \begin{pmatrix} \hat{R}_1 \\ 0 \end{pmatrix}$ with $R_1 \in \mathbb{R}^{N \times k}$ and $\hat{R}_1 \in \mathbb{R}^{k \times k}$ such that $\begin{pmatrix} A_{11} \\ A_{21} \end{pmatrix} = Q_1 \begin{pmatrix} \hat{R}_1 \\ 0 \end{pmatrix}$. Then say, $B = Q_1^T A = \begin{pmatrix} \hat{R}_1 & B_{12} \\ 0 & B_{22} \end{pmatrix}$. Let $B_{22} = \hat{Q}_2 \hat{R}_2$. Then, with $Q_2 = \begin{pmatrix} I_k & 0 \\ 0 & \hat{Q}_2 \end{pmatrix}$, $Q_2^T \begin{pmatrix} \hat{R}_1 & B_{12} \\ 0 & B_{22} \end{pmatrix} = \begin{pmatrix} \hat{R}_1 & B_{12} \\ 0 & \hat{R}_2 \end{pmatrix} = R$ is upper triangular. That is $Q_2^T Q_1^T A = R$. Hence, $A = QR$, where $Q = Q_1 Q_2$ is an orthogonal matrix.

The traditional algorithm for QR-decomposition stems directly from this result.

**Algorithm 2.3.** *For any $A \in \mathbb{R}^{N \times P}$, the QR-decomposition of $A$ can be achieved in $O(NPX)$ operations and $O(NPX/B)$ I/Os and $O(NPX/M)$ seeks, where $X = \min\{N, P\}$.*

*Proof.* If $N = 1$ there is nothing to do. If $N > 1$, Let $k = 1$ in the proof of the above lemma. The Householder $Q_1$ can be found in $O(N)$ operations and $3N/B$ I/Os. We assume that $u$ that characterises the Householder overwrites the first column of $A$ except for $A_{11}$; since $u$ is normalised, its first component need not be stored. The matrix $B$ can be calculated by applying $Q_1$ to each column of $A$ in a total of $O(NP)$ operations and $5N(P-1)/B$ I/Os. Now the algorithm recurses on the $(N-1) \times (P-1)$ matrix $A_{22}$. Clearly, the algorithm hits the basis after $X$ levels of recursion.

$T(N, P) \leq \frac{5}{B} \sum_{i=0}^{X/B-1} \sum_{j=0}^{B-1} (N - iB)(P - iB - j) = O(\frac{NPX}{B})$

When $N > P = \omega(B)$, the number of I/Os is $\frac{5P^2}{2B}(N - \frac{P}{3} + o(N))$

When $P > N = \omega(B)$, the number of I/Os is $\frac{5N^2}{2B}(P - \frac{N}{3} + o(P))$ $\qquad \square$

## 2.4   The Slab Approach

Here we analyse an implementation of the algorithm of [96], which is based on the following "Compact $WY$ representation": A product $P_k = Q_1 \ldots Q_k$ of $k \leq N$ Householders, where each $Q_i = (I - \beta_i u_i u_i^T) \in \mathbb{R}^{N \times N}$, can be written in the form $(I + Y_k T_k Y_k^T)$ for a $Y_k \in \mathbb{R}^{N \times k}$ and an upper triangular $T_k \in \mathbb{R}^{k \times k}$.

This can be proved by induction on $k$. For the basis, let $k = 1$. $P_1 = Q_1 = (I - \beta_1 u_1 u_1^T) = (I + u_1[-\beta_1]u_1^T)$. Therefore, choosing $Y_1 = u_1$ and $T_1 = [-\beta_1]$ will do. Say, we have $Y_{k-1}$ and $T_{k-1}$ so that $P_{k-1} = Q_1 \ldots Q_{k-1} = (I + Y_{k-1} T_{k-1} Y_{k-1}^T)$. Let $Y_k = (Y_{k-1} \; u_k)$, and $T_k = \begin{pmatrix} T_{k-1} & z \\ 0 & \rho \end{pmatrix}$, where $\rho = -\beta_k$ and $z = -\beta_k T_{k-1} Y_{k-1}^T u_k$. Then, $I + Y_k T_k Y_k^T = P_{k-1} Q_k = P_k$.

An implementation of the algorithm of [96] is described now: Let $A$ be the input $N \times P$ matrix. Divide it into vertical slabs each consisting of $k$ columns, where $k < P$ is a parameter to be chosen. Let $X = \min\{N, P\}$ and $y = \min\{N, k\}$. Let us call the leftmost slab $A_1$ and the rest of the matrix $A_2$. That is, $A = (A_1 \; A_2)$. Perform a QR decomposition of $A_1$ using the traditional algorithm. Also, on the fly, aggregate the Householders that the algorithm produces. While processing the $i$-th column of the slab, $T_i$ can be computed from $Y_{i-1}$, $T_{i-1}$ and $Q_i$ in $O(iN)$ operations, $O(iN/B)$ I/Os and $O(iN/M)$ seeks; $Y_i$ is obtained from $Y_{i-1}$ by merely adding $u_i$ as an extra column. The cost of aggregating $y$ Householders would therefore be $O(y^2 N)$ operations, $O(y^2 N/B)$ I/Os and $O(y^2 N/M)$ seeks. Pre-multiply $A_2$ with this aggregate, and if $N \leq k$, then halt. Otherwise, recurse with the lower $N - k$ rows of $A_2$. Assume that a matrix multiplication is done using the blocked algorithm of [58], when neither matrix fits in the main memory,

The algorithm chooses a width $k$ for the slab. The I/O complexity of the algorithm will depend on the choice. We analyse the I/O complexity of the algorithm for the various possible choices of $k$:

In the tables below, $\alpha$, $\beta$, and $\gamma$ stand for the following operations in that order: QR-

Table 2.1: Blocks of data moved in scenarios $A$, $B$, $C$ and $D$.

| Scenario | $\alpha$ | $\beta$ | $\gamma$ | $\nu$ | Total I/Os |
|---|---|---|---|---|---|
| A | $Nk^2/B$ | $Nk^2/B$ | $\frac{kNP}{B\sqrt{M}} + \frac{NP}{B}$ | $X/k$ | $\frac{NPX}{B\sqrt{M}} + \frac{NX}{B}(k + \frac{P}{k})$ |
| B | $N^2k/B$ | $N^3/B$ | $\frac{N^2P}{B\sqrt{M}} + \frac{NP}{B}$ | $1$ | $\frac{N^2P}{B\sqrt{M}} + \frac{N}{B}(Nk + P)$ |
| C | $Nk/B$ | $Nk/B$ | $\frac{NP}{B}$ | $X/k$ | $\frac{NPX}{kB}$ |
| D | $Nk/B$ | $N^2/B$ | $\frac{NP}{B}$ | $1$ | $\frac{NP}{B}$ |

Table 2.2: Seeks performed in scenarios $A$, $B$, $C$ and $D$.

| Scenario | $\alpha$ | $\beta$ | $\gamma$ | $\nu$ | Total seeks |
|---|---|---|---|---|---|
| A | $Nk^2/M$ | $Nk^2/M$ | $\frac{kNP}{M} + \frac{NP}{\sqrt{M}} + P$ | $X/k$ | $\frac{NPX}{M} + \frac{NPX}{k\sqrt{M}} + \frac{PX}{k}$ |
| B | $N^2k/M$ | $N^3/M$ | $\frac{N^2P}{M} + \frac{NP}{\sqrt{M}} + P$ | $1$ | $\frac{N^2P}{M} + \frac{NP}{\sqrt{M}} + P$ |
| C | $k$ | $k$ | $P$ | $X/k$ | $\frac{PX}{k}$ |
| D | $k$ | $N$ | $P$ | $1$ | $P$ |

decomposing a slab, aggregating the Householders generated during the QR-decomposition of a slab, and multiplying the aggregate of a slab with the rest of the matrix. $\nu$ stands for the number of slabs to be processed. Table 2.1 and 2.2 summarise, respectively, the number of blocks of data moved and the number of seeks performed by the operations under the scenarios A, B, C and D listed below. The total is calculated as $\nu(\alpha + \beta + \gamma)$ in each case. For brevity, we drop the order notation from the table.

Scenario A: $M < Nk$ (the slab does not fit in the main memory) and $k \leq N$ (the slab is tall). Here $k \leq N, P$; so, $y = k$. $\gamma$ is dominated by the cost of calculating the product $YTY^TA$, which is a multiplication chain of $N \times y$, $y \times y$, $y \times N$ and $N \times P$ matrices. It is easy to see that the cost of the multiplication is as given, once we note that $Y^T$ is available in row-major order.

Scenario B: $M < Nk$ (the slab does not fit in the main memory) and $N \leq k \leq P$ (the slab is fat). Here, $X = y = N$. $Y$ and $T$ are both $N \times N$ matrices.

Scenario C: $Nk \leq M$ (the slab fits in the main memory), and $k \leq N$ (the slab is tall). Here $k \leq N, P$; so $y = k$. The QR decomposition and the aggregation of the $Y$ matrix can be done in internal memory. $YTY^{T}A$ can be calculated one column at a time.

Scenario D: $Nk \leq M$ (the slab fits in the main memory), and $N \leq k$ (the slab is fat); $N \leq k \leq P$; so $X = y = N$, and $N^2 < M$. $Y$ and $T$ are both $N \times N$ matrices. So is $YTY^{T}$, and thus it can stay in the main memory. Read $A$ column by column, update and write back.

Now we find the choices of $k$ that, for the given values of $N$, $P$ and $M$, minimise the I/O complexity under various conditions. In all but two of the cases below, with "the tall cache assumption", the number of blocks of data moved dominates the number of seeks performed. In addition to the optimal I/O bounds and the associated seek bounds, we also calculate the I/O and seek bounds for $k = \sqrt{M}$.

If $N > M$, there is no choice of $k$ for which the slab will fit in the main memory. If, in addition, $N \geq P$, then $N \geq k$ and $X = P$; the slabs are tall; we have Scenario A; the I/O cost is $O(\frac{NP}{B}(k + \frac{P}{k} + \frac{P}{\sqrt{M}}))$. Clearly, the choice that minimises this is $k = \sqrt{P}$, which costs $O(\frac{NP^2}{B}(\frac{1}{\sqrt{P}} + \frac{1}{\sqrt{M}}))$ I/Os, and $O(\frac{NP^2}{\sqrt{M}}(\frac{1}{\sqrt{P}} + \frac{1}{\sqrt{M}}))$ seeks.

If we were to choose $k = \sqrt{M}$, the number of I/Os would be $O(\frac{NP}{B}(\frac{P+M}{\sqrt{M}}))$, and the number of seeks $O(\frac{NP^2}{M})$.

If $N > M$, but $N < P$, a choice of $k \geq N$ is also possible. If $k \in [N, P]$, then we have Scenario B, and the I/O cost is $O(\frac{N^2P}{B\sqrt{M}} + \frac{N}{B}(Nk + P))$; $k = N$ is the best choice. If $k \leq N$ then we have Scenario A, and the cost is $O(\frac{N^2}{B}(k + \frac{P}{k} + \frac{P}{\sqrt{M}}))$. The best choice of $k$ from $[1, N]$ is $\sqrt{P}$ or $N$ depending on whether $\sqrt{P} < N$ or not. That is, the best $k$ is $\sqrt{P}$, if $\sqrt{P} < N$ [I/O: $O(\frac{N^2P}{B}(\frac{1}{\sqrt{P}} + \frac{1}{\sqrt{M}})) = O(\frac{N^2P}{B\sqrt{M}})$; Seek: $O(\frac{N^2P}{M})$] and $N$ otherwise [I/O:

$O(\frac{N^2}{B}(N + \frac{P}{N} + \frac{P}{\sqrt{M}})) = O(\frac{N^2 P}{B\sqrt{M}})$; Seek: $O(\frac{N^2 P}{M} + \frac{N^2 P}{N\sqrt{M}}) = O(\frac{N^2 P}{M})]$.

With $k = \sqrt{M}$, the number of I/Os is $O(\frac{N^2}{B}(\frac{P+M}{\sqrt{M}})) = O(\frac{N^2 P}{B\sqrt{M}})$, and the number of seeks is $O(\frac{N^2 P}{M})$.

If $N \leq M$ and $N \leq M/N$, then a $k \in [N, M/N]$ can be chosen; Scenario D; blocks of data moved: $O(NP/B)$; seeks: $O(P)$. In this case, seeks may dominate the I/O complexity.

If $N \leq M$ and $M/N < N$, $(M/N \leq \sqrt{M} < N \leq M)$, then the choice of $k$ can satisfy (i) $k \leq M/N$, (ii) $k \in (M/N, N]$, or (iii) $k > N$.

If $P \leq N$, then (iii) is ruled out.

If $M/N \geq P$, the matrix fits in the main memory; Scenario C; $O(NP/B)$ blocks of data are moved; seeks: $O(P)$. Here too, seeks may dominate the I/O complexity.

Now assume that $M/N < P$. That is, $M/N < P \leq N \leq M$. If $k \leq M/N$, we have Scenario C, and so the cost is $O(\frac{NP^2}{kB})$. The best choice for $k$ in $[1, M/N]$ thus is $M/N$; I/O cost: $O(\frac{N^2 P^2}{BM})$; seeks: $O(\frac{NP^2}{M})$. For $k \geq M/N$, we have Scenario A; the I/O cost is $O(\frac{NP}{B}(\frac{P}{\sqrt{M}} + k + \frac{P}{k}))$, which minimises at $k = \sqrt{P}$, and grows monotonically thereafter.

If $\sqrt{P} > M/N$, then the minimum I/O in $[M/N, P]$ is given by $k = \sqrt{P}$ and is $O(\frac{NP^2}{B\sqrt{P}}) = O(\frac{N^2 P^2}{BM})$. Thus $k = \sqrt{P}$ is the best choice in $[1, P]$. Interestingly, here a slab size that does not fit in the main memory turns out to be the best choice. The number of seeks in this case is $O(\frac{NP^2}{\sqrt{P}\sqrt{M}})$.

If $\sqrt{P} \leq M/N$, then the smallest value in $[M/N, P]$ is provided by $k = M/N$, which gives a cost that is worse than $O(\frac{N^2 P^2}{BM})$.

A choice of $k = \sqrt{M}$ makes sense only if $P \geq \sqrt{M} \geq M/N$; even then the cost will be $O(\frac{NPM}{B\sqrt{M}})$. Seeks: $O(\frac{NP^2}{M})$.

Let $N \leq M$ and $M/N \leq N < P$. First assume that $P < M$. That is, $M/N \leq N <$

$P < M$. If $k \leq M/N$ (Scenario C), then the I/O cost is $O(\frac{N^2 P}{kB})$. The best choice for $k$ in $[1, M/N]$ thus is $M/N$; cost: $O(\frac{N^3 P}{BM})$. For $k \in [M/N, N]$ (Scenario A), the I/O cost is $O(\frac{N^2}{B}(\frac{P}{\sqrt{M}} + k + \frac{P}{k}))$, which minimises at $k = \sqrt{P}$, and grows monotonically thereafter.

If $\sqrt{P} \in [M/N, N]$, then the least I/O in $[M/N, N]$ is given by $k = \sqrt{P}$ and is $O(\frac{N^2 P}{B}(\frac{1}{\sqrt{M}} + \frac{1}{\sqrt{P}})) = O(\frac{N^2 P}{B}\frac{1}{\sqrt{P}})$; this is $O(\frac{N^3 P}{BM})$. Thus $k = \sqrt{P}$ is the best choice in $[1, N]$. Seeks: $O(\frac{N^2 P}{\sqrt{P}\sqrt{M}})$.

If $\sqrt{P} < M/N$, then the least I/O for $k \in [M/N, N]$ is provided by $k = M/N$, which is worse than $O(\frac{N^3 P}{BM})$. Seeks: $O(\frac{N^2 P}{M})$.

If $\sqrt{P} > N$, then the best I/O bound is $O(\frac{N^2 P}{B\sqrt{M}})$; the choice $k = N$ gives this. Seeks: $O(\frac{N^2 P}{M})$.

Now consider $k \in [N, P]$. Scenario B. The cost is $O(\frac{N^2 P}{B\sqrt{M}} + \frac{N}{B}(P + Nk))$. The best $k$ is $N$; cost: $O(\frac{N^2 P}{B\sqrt{M}} + \frac{NP + N^3}{B})$. This does not improve on the minimum of $[1, N]$, in any of the cases.

A choice of $k = \sqrt{M}$ causes the cost to be $O(\frac{NPM}{B\sqrt{M}})$. Seeks: $O(\frac{N^2 P}{M})$.

It now remains to consider the case of $M/N \leq \sqrt{M} \leq N \leq M \leq P$. If $k \leq M/N$, then the cost is $O(\frac{N^2 P}{kB})$. The best choice for $k$ in $[1, M/N]$ thus is $M/N$; cost: $O(\frac{N^3 P}{BM})$. For $k = \sqrt{M} \in [M/N, P]$, the I/O cost is $O(\frac{N^2 P}{B\sqrt{M}}) = O(\frac{N^3 P}{BM})$. (Seek: $O(\frac{N^2 P}{M})$.) A detailed analysis shows that the best choices for $k$ are $\sqrt{P}$ and $N$ when $\sqrt{P} \in [M/N, N]$ and $\sqrt{P} > N$ respectively; but the I/O and seek bounds for these choices are of the same order as those of $k = \sqrt{M}$ asymptotically.

The cost for a $k$ from $[N, P]$ is $O(\frac{N^2 P}{B\sqrt{M}} + \frac{N}{B}(P + Nk))$. The best choice for $k$ is $N$; cost: $O(\frac{N^2 P}{B\sqrt{M}} + \frac{NP + N^3}{B})$. This does not improve on the minimum of $[1, N]$.

The above findings are summarised in Table 2.3. In particular, for a square matrix with $N = P$, the I/O and seek complexities are as given in Table 2.4.

Table 2.3: Summary of the analysis of the slab based algorithm

| | $k$ | I/Os | Seeks |
|---|---|---|---|
| $N > M,\ N \geq P$ | $\sqrt{P}$ | $\frac{NP^2}{B}(\frac{1}{\sqrt{P}} + \frac{1}{\sqrt{M}})$ | $\frac{NP^2}{\sqrt{M}}(\frac{1}{\sqrt{P}} + \frac{1}{\sqrt{M}})$ |
| $N > M,\ N < P,\ \sqrt{P} \leq N$ | $\sqrt{P}$ | $\frac{N^2 P}{B\sqrt{M}}$ | $\frac{N^2 P}{M}$ |
| $N > M,\ N < P,\ \sqrt{P} > N$ | $N$ | $\frac{N^2 P}{B\sqrt{M}}$ | $\frac{N^2 P}{M}$ |
| $N \leq M,\ \frac{M}{N} \geq N$ | $\frac{M}{N}$ | $\frac{NP}{B} + P$ | $P$ |
| $N \leq M,\ \frac{M}{N} < N,\ P \leq N,\ \frac{M}{N} \geq P$ | $P$ | $\frac{NP}{B}$ | $P$ |
| $N \leq M,\ \frac{M}{N} < N,\ P \leq N,\ \frac{M}{N} < P,\ \frac{M}{N} < \sqrt{P}$ | $\sqrt{P}$ | $\frac{NP^2}{B\sqrt{P}}$ | $\frac{NP^2}{\sqrt{P}\sqrt{M}}$ |
| $N \leq M,\ \frac{M}{N} < N,\ P \leq N,\ \frac{M}{N} < P,\ \frac{M}{N} \geq \sqrt{P}$ | $\frac{M}{N}$ | $\frac{N^2 P^2}{BM}$ | $\frac{NP^2}{M}$ |
| $N \leq M,\ \frac{M}{N} < N,\ P > N,\ P < M,\ \sqrt{P} < \frac{M}{N}$ | $\frac{M}{N}$ | $\frac{N^3 P}{BM}$ | $\frac{N^2 P}{M}$ |
| $N \leq M,\ \frac{M}{N} < N,\ P > N,\ P < M,\ \sqrt{P} \in [\frac{M}{N}, N]$ | $\sqrt{P}$ | $\frac{N^2 P}{B\sqrt{P}}$ | $\frac{N^2 P}{\sqrt{P}\sqrt{M}}$ |
| $N \leq M,\ \frac{M}{N} < N,\ P > N,\ P < M,\ \sqrt{P} > N$ | $N$ | $\frac{N^2 P}{B\sqrt{M}}$ | $\frac{N^2 P}{M}$ |
| $N \leq M,\ \frac{M}{N} < N,\ P > N,\ P \geq M,\ \sqrt{P} \in [\frac{M}{N}, N]$ | $\sqrt{P}$ | $\frac{N^2 P}{B\sqrt{M}}$ | $\frac{N^2 P}{M}$ |
| $N \leq M,\ \frac{M}{N} < N,\ P > N,\ P \geq M,\ \sqrt{P} > N$ | $N$ | $\frac{N^2 P}{B\sqrt{M}}$ | $\frac{N^2 P}{M}$ |

Table 2.4: I/O and Seek complexities of the slab based algorithm on a square matrix

|  | $k$ | I/Os | Seeks |
|---|---|---|---|
| $N > M$ | $\sqrt{N}$ | $\frac{N^3}{B\sqrt{M}}$ | $\frac{N^3}{M}$ |
| $N \leq \sqrt{M}$ | $k \in [N, \frac{M}{N}]$ | $\frac{N^2}{B} + N$ | $N$ |
| $M^{2/3} < N \leq M$ | $\sqrt{N}$ | $\frac{N^3}{B\sqrt{N}}$ | $\frac{N^3}{\sqrt{N}\sqrt{M}}$ |
| $\sqrt{M} < N \leq M^{2/3}$ | $\frac{M}{N}$ | $\frac{N^4}{BM}$ | $\frac{N^3}{M}$ |

The I/O cost of the slab based QR-decomposition algorithm matches that of matrix multiplication in most of the cases. The best choice for slab width is not always the number of columns that fit in the main memory. Interestingly, a choice of $k = \sqrt{M}$ seems to minimise the seek complexity at $\frac{NPX}{M}$. However, it does not minimise the I/O cost.

We conclude that existing slab based implementations may not have to be completely redesigned for good Out-of-core performance. If some of the elementary matrix operations in them (for example, matrix multiplication here) are handled I/O efficiently, optimal I/O performances might be achievable for most choices of the problem parameters.

## 2.5 The Tile Approach

Slab based algorithms like those of the last section are not scalable: the larger the row size $N$ of the matrix, the smaller the width of the slab that can be fit in the main memory. An alternative that has been suggested is the tile approach: partition the matrix into a number of square blocks called "tiles", and process the tiles one at a time. It has been observed that the tile approach results in more scalable out-of-core algorithms for Cholesky decomposition [63, 93, 101] and QR-decomposition [64]. Tile based QR decomposition for multicore architecture is discussed in [28], using techniques similar to those of [64].

---

**Algorithm 2.4.** *QR decomposition using tile approach [64]*

**Input:** *An $N \times P$ matrix $A$.*

**Output:** *$A = QR$, where $R$ is an $N \times P$ upper triangular matrix and $Q$ is an $N \times N$ orthogonal matrix; $Q$ is not output explicitly, but is expressed as a product of terms of the form $I + YTY^T$, where $Y$ is an $N \times k$ matrix and $T$ is a $k \times k$ upper triangular matrix.*

---

*Choose $t = \Theta(\sqrt{M})$; Choose $k < t$;*
*Partition $A$ into $t \times t$ tiles;*
*Let $A_{ij}$ denote the tile that is in both the $i$-th tile row and $j$-th tile column;*
*$X = \min\{N, P\}$;*
**for** $i = 1$ **to** $X/t$ **do**
    *QR-decompose($A_{ii}$);*
    **for** $j = i + 1$ **to** $P/t$ **do**
        *QR-Multiply-from-left-1($A_{ii}, A_{ij}$);*
    **endfor**
    **for** $j = i + 1$ **to** $N/t$ **do**
        *QR-Update($A_{ii}, A_{ji}$);*
        **for** $l = i + 1$ **to** $P/t$ **do**
            *QR-Multiply-from-left-2($A_{ji}, A_{il}, A_{jl}$);*
        **endfor**
    **endfor**
**endfor**

---

## 2.5.1   The Algorithm

Here we analyse the QR-decomposition algorithm of [64] for its I/O complexity. A high level description of the algorithm is given in Algorithm 2.4.

Now, to analyse the algorithm, we assume that the input $N \times P$ matrix $A$ has already been "tile-transposed". (We define tile transposition of $A$ as the problem of permuting its elements so that each tile is available contiguously in column-major order.) Note that if $A$ is given in column-major order, then tile transposition can be done in $O(NP/B)$ I/Os and $O(NP/t)$ seeks, if the tile size $t \geq \max\{B, \sqrt{M}\}$, because a tile can be processed in $O(t^2/B)$ I/Os and $O(t)$ seeks. (The tile size is the number of columns (rows) in a tile.) Once we have found a $\sqrt{M} \times \sqrt{M}$ tiling, for any $t \leq \sqrt{M}$, a $t \times t$ tiling can be found in $O(NP/B)$ I/Os and $O(NP/M)$ seeks; read each $\sqrt{M} \times \sqrt{M}$ tile into the main memory, form a $t \times t$ tiling and write back. Putting the above two statements together, we have that if $\sqrt{M} > B$ (the tall cache assumption), for any $t$, a $t \times t$ tiling of $A$ can be found in $O(NP/B)$ I/Os and $O(NP/(t + \sqrt{M}))$ seeks.

We analyse for the following implementations, where $k < t$ is a parameter of the algorithm:

1. The single tile implementation, in which a single tile and three and a half narrow panels of size $t \times k$ occupy the main memory; i.e. $t^2 + 7kt/2 \leq M$.

2. The two tile implementation, in which two tiles together with two and a half narrow panels of size $t \times k$ occupy the main memory; i.e. $2t^2 + 5kt/2 \leq M$.

3. The three tile implementation, in which three tiles together with with a panel and a half of size $t \times k$ occupy the main memory; i.e. $3t^2 + 3kt/2 \leq M$. The half panel will hold the $T$ matrices and the full panel will provide space for computations.

The parameter $k$ can be chosen so as to minimise the number of I/Os. We now describe and analyse each procedure of Algorithm 2.4.

**Function QR-decompose($A_{ii}$)**

Assume that the diagonal tile $A_{ii}$ is divided into vertical slabs of width $k$ each. Let $r = t/k$. The invocation QR-decompose($A_{ii}$) finds matrices $Y_1, \ldots, Y_r, T_1, \ldots, T_r$ such that $R_{ii} = (I_t + Y_r T_r^T Y_r^T) \ldots (I_t + Y_1 T_1^T Y_1^T) A_{ii}$ is upper triangular. This is done as follows:

Read $A_{ii}$ into the main memory. Proceed in $r$ iterations. At the beginning of the $s$-th iteration, the subdiagonal elements of the first $(s-1)k$ columns of $A_{ii}$ are all zero. In the $s$-th iteration, calculate $Y_s$ and $T_s$ such that the subdiagonal elements of $(I_t + Y_s T_s^T Y_s^T) A_{ii}$ in the first $sk$ columns are all zero. Update $A_{ii}$ to $(I_t + Y_s T_s^T Y_s^T)A_{ii}$. Store $Y_s$ in the subdiagonal part of the $s$-th slab of $A_{ii}$.

When all slabs are processed, arrange the upper triangular part of $A_{ii}$ in row major order, and write $A_{ii}$ back to the disk.

The number of I/Os needed is $2t^2/B$; $A_{ii}$ is read and written; the $T$ matrices are kept in the main memory.

The number of seeks is $O(1)$ even for the one tile case.

**Function QR-Multiply-from-left-1**$(A_{ii}, A_{ij})$

If $A_{ii}$ QR-decomposes into $Q_{ii} R_{ii}$, then a pre-multiplication of $A$ by an orthogonal matrix obtained by replacing the $i$-th diagonal tile of $I_N$ by $Q_{ii}$ has the effect of making $A_{ii}$ upper triangular. This pre-multiplication also updates all tiles $A_{ij}$ with $j > i$.

The invocation QR-Multiply-from-left-1$(A_{ii}, A_{ij})$ achieves this update of $A_{ij}$ as follows:

Read $A_{ij}$ into the main memory. Use the $Y$ matrices in the lower triangular part of $A_{ii}$ and the $T$ matrices that remain in the main memory after the invocation "QR-decompose$(A_{ii})$" to update $A_{ij}$. When all updates are over, write $A_{ij}$ back in row major order. How the $Y$ matrices are accessed will depend on how many tiles fit in the main memory. If only one tile fits, then the $Y$ matrices are read in one after the other, and so the total number of I/Os is at most $2.5t^2/B$. If two tiles fit, we can assume that $A_{ii}$ resides in the main memory, and so the total number of I/Os will be at most $2t^2/B$.

The number of seeks is $O(1)$ even for the one tile case.

**Function QR-Update**$(A_{ii}, A_{ji})$

When $A_{ii}$ is upper triangular, the invocation QR-Update$(A_{ii}, A_{ji})$, finds a matrix $Q$ such that in $Q^T A$, every element of the $(j, i)$-th tile is zero, and $(i, i)$-th tile is upper triangular.

This is done over $r = t/k$ iterations. The $s$-th iteration processes $A_{ii}^{(s)}$, the $s$-th topmost horizontal slab of width $k$ of $A_{ii}$. Consider the matrix $C_s = \begin{pmatrix} A_{ii}^{(s)} \\ A_{ji} \end{pmatrix}$. Let $\begin{pmatrix} I_k \\ Y_s \end{pmatrix}$ and $T_s$ be matrices such that $\left( I_{k+t} + \begin{pmatrix} I_k \\ Y_s \end{pmatrix} T_s^T \begin{pmatrix} I_k & Y_s^T \end{pmatrix} \right) C_s$ has zeroes in the lowest $t$ rows of its $s$-th slab. Then, $\left( I_{2t} + \begin{pmatrix} 0_{(s-1)k} \\ I_k \\ 0_{t-sk} \\ Y_s \end{pmatrix} T_s^T \begin{pmatrix} 0_{(s-1)k} & I_k & 0_{t-sk} & Y_s^T \end{pmatrix} \right) \begin{pmatrix} A_{ii} \\ A_{ji} \end{pmatrix}$ has zeroes in the lower triangular part of its $s$-th slab. Set $\begin{pmatrix} A_{ii} \\ A_{ji} \end{pmatrix}$ to this matrix by updating $A_{ii}^{(s)}$ and $A_{ji}$. The $s$-th slab of $A_{ji}$ is now all zero. Therefore $Y_s$ can be stored there.

Let $Q_{ji}^T$ be the product

$$\left( I_{2t} + \begin{pmatrix} 0_{t-k} \\ I_k \\ Y_r \end{pmatrix} T_r^T \begin{pmatrix} 0_{t-k} & I_k & Y_r^T \end{pmatrix} \right) \ldots \left( I_{2t} + \begin{pmatrix} I_k \\ 0_{t-k} \\ Y_1 \end{pmatrix} T_1^T \begin{pmatrix} I_k & 0_{t-k} & Y_1^T \end{pmatrix} \right)$$

The above can be implemented as follows: Read $A_{ji}$. Read $A_{ii}$ one horizontal slab at a time. For each slab, update the slab, write it back and overwrite the corresponding vertical slab of $A_{ji}$ with the $Y$ matrix. The number of I/Os is $3t^2/B$, irrespective of whether one tile or two tiles fit in the main memory: one tile and a half are both read and written.

The number of seeks is $O(t/k)$ in the one tile case, and $O(1)$ in the two and three tile cases.

If both $A_{ii}$ and $A_{ij}$ are upper triangular (We will encountered such an invocation in Chapter 3.), then the number of I/Os is $2t^2/B$: two half tiles are both read and written. The number of seeks would be $O(1)$ even in the one tile case.

**QR-Multiply-from-left-2**$(A_{ji}, A_{il}, A_{jl})$

Here $i < j$ and $i < l$. Suppose $Q_{ji} \begin{pmatrix} R_{ii} \\ 0 \end{pmatrix}$ is a QR decomposition of $\begin{pmatrix} A_{ii} \\ A_{ji} \end{pmatrix}$ where $A_{ii}$ is upper triangular. Then,

$$Q_{ji}^T \begin{pmatrix} A_{ii} & A_{il} \\ A_{ji} & A_{jl} \end{pmatrix} = \begin{pmatrix} R_{ii} & B_{il} \\ 0 & B_{jl} \end{pmatrix}$$

where $\begin{pmatrix} B_{il} \\ B_{jl} \end{pmatrix} = Q_{ji}^T \begin{pmatrix} A_{il} \\ A_{jl} \end{pmatrix}$. Say, $Q_{ji}^T$ and $R_{ii}$ have been calculated by an invocation to QR-Update$(A_{ii}, A_{ji})$. At the end of the invocation the $Y$ matrices are stored in $A_{ji}$.

$B_{il}$ and $B_{jl}$ can be calculated as follows:

Read $A_{jl}$. Read a horizontal slab from $A_{il}$ and the corresponding vertical slab from $A_{ji}$. Update the horizontal slab and $A_{jl}$ using the vertical slab, and write the horizontal slab back. When all slabs have been processed, write $A_{jl}$ back. If only one tile fits in the main memory, then the number of I/Os is $5t^2/B$: $A_{jl}$ and $A_{il}$ is read and written; $A_{ji}$ is read.

If $A_{ji}$ is upper triangular (We will encounter such an invocation in Chapter 3.) then the number of I/Os would be $4.5t^2/B$.

If two tiles fit in the main memory, then $A_{ji}$ can be kept in the main memory, as $l$ varies, and so the number of I/Os is $4t^2/B$.

The number of seeks is $O(t/k)$ if one or two tiles fit in the main memory; $O(1)$ if three tiles fit in the main memory.

**I/O complexity**

The I/O complexity for the one tile implementation is

$$\frac{t^2}{B} \sum_{i=1}^{X/t} \left( 2 + \sum_{j=i+1}^{P/t} 2.5 + \sum_{j=i+1}^{N/t} \left( 3 + \sum_{l=i+1}^{P/t} 5 \right) \right)$$

$$= \frac{5}{6tB}(6NPX - 3NX^2 - 3PX^2 + 2X^3) + \frac{1}{4B}(2NX - X^2) + \frac{t}{12B}X$$

where $X = \min\{N, P\}$. In particular, when $N = P = X$, the I/O complexity is $\frac{5N^3}{3tB} + \frac{N^2}{4B} + \frac{tN}{12B}$ where $t \approx \sqrt{M}$.

The I/O complexity for the two (also, three) tile implementation is

$$\frac{t^2}{B} \sum_{i=1}^{X/t} \left( 2 + \sum_{j=i+1}^{P/t} 2 + \sum_{j=i+1}^{N/t} \left( 3 + \sum_{l=i+1}^{P/t} 4 \right) \right)$$

$$= \frac{4}{6tB}(6NPX - 3NX^2 - 3PX^2 + 2X^3) - \frac{1}{2B}(X^2 - 2NX) + \frac{t}{6B}X$$

In particular, when $N = P = X$, the I/O complexity is $\frac{4N^3}{3tB} + \frac{N^2}{2B} + \frac{tN}{6B}$ where $t \approx \sqrt{M/2}$ and $t \approx \sqrt{M/3}$ for two and three tile implementation respectively.

We also require a number of I/Os to write the $T$ matrices. We have $t/k$ number of $T$ matrices of size $k \times k$. Since the $T$ matrices are upper triangular, that amounts to roughly $k/2$ elements. There are $X/t$ number of single tile $QR$ decompositions and $N - i$ number of two tile $QR$ updates for $i = 1, \ldots, X/t - 1$. So I/Os required to handle the $T$ matrices is: $\sum_{i=1}^{X/t} \frac{kt}{2B} + \sum_{i=1}^{X/t-1} \frac{kt(N-i)}{2B} = \frac{k}{2tB}\left(NX - \frac{X^2}{2} + \frac{Xt}{2}\right)$

Clearly, from the above, the one tile implementation is the best of the three.

**Seek complexity**

If three tiles fit in the main memory, then the number of seeks is $O(\frac{NPX}{t^3})$. If at most two tiles fit, then it is $O(\frac{NPX}{kt^2})$.

## 2.5.2 An $O(1)$ factor improvement for the one tile implementation

We suggest a small modification to the above algorithm that can improve the number of I/Os by a constant factor for the one tile case.

After the invocation to "QR-decompose($A_{ii}$)", keep $A_{ii}$ in the main memory, while "QR-Multiply-from-left-1($A_{ii}, A_{ij}$)" is invoked for each $j > i$. Read a vertical slab of $A_{ij}$, update it and write it back in each iteration. Thus the invocation "QR-Multiply-from-left-1($A_{ii}, A_{ij}$)" would take only $\frac{2t^2}{B}$ I/Os. The updated $A_{ij}$ is now written in column major order. This version of the procedure, however, will incur $O(t/k)$ seeks in the one tile case.

This would necessitate a change in Procedure QR-Multiply-from-left-2, because now the invocation "QR-Multiply-from-left-2($A_{ji}, A_{il}, A_{jl}$)", would find that QR-Multiply-from-left-1($A_{ii}, A_{il}$) has left $A_{il}$ in column major order.

Suppose in QR-Multiply-from-left-2($A_{ji}, A_{il}, A_{jl}$), we keep $A_{ji}$ in the main memory, and bring a column each of $A_{il}$ and $A_{jl}$ into the main memory in each iteration. All the $Y$ matrices needed to update these columns are in the main memory now, and so are the columns. The updates of these columns are independent of those of the other columns. We can perform these updates and write the columns back. This reduces the number of I/Os performed by the invocation "QR-Multiply-from-left-2($A_{ji}, A_{il}, A_{jl}$)" to $\frac{4t^2}{B}$, even for a one tile implementation.

The I/O complexity, therefore, is

$$\frac{t^2}{B} \sum_{i=1}^{X/t} \left( 2 + \sum_{j=i+1}^{P/t} 2 + \sum_{j=i+1}^{N/T} \left( 3 + \sum_{l=i+1}^{P/T} 4 \right) \right)$$

$$= \frac{2}{3tB}(6NPX - 3NX^2 - 3PX^2 + 2X^3) + \frac{1}{2B}(2NX - X^2) + \frac{t}{6B}X$$

In particular, when $N = P = X$, the I/O complexity is $\frac{4N^3}{3tB} + \frac{N^2}{2B} + \frac{tN}{6B}$ where $t \approx \sqrt{M}$. The number of seeks is $O(\frac{NPX}{kt^2})$.

### 2.5.3 Comparison of seek times for slab and tile approaches

We compare the seek times of the slab based and tile based algorithms in Table 2.5. Order notation is omitted for brevity. Clearly, the tile approach is superior.

## 2.6 A Cache Oblivious Algorithm for $QR$ Decomposition

In this section, we present a cache oblivious algorithm for QR-decomposition with an I/O complexity of $O(\frac{N^3}{B\sqrt{M}} + \frac{N^2}{B})$, asymptotically the same as that of the algorithm discussed in Section 2.5.

Table 2.5: Comparison of seeks needed by the slab and tile base algorithms

| | $N \le \sqrt{M}$ | $\sqrt{M} < N \le M^{2/3}$ | $M^{2/3} < N \le M$ | $N > M$ |
|---|---|---|---|---|
| slab | $N$ | $\frac{N^3}{M}$ | $\frac{N^3}{\sqrt{NM}}$ | $\frac{N^3}{M}$ |
| 3 tiles | $\frac{N^2}{M}$ | $\frac{N^3}{M\sqrt{M}}$ | $\frac{N^3}{M\sqrt{M}}$ | $\frac{N^3}{M\sqrt{M}}$ |
| 1 or 2 tiles | $\frac{N^2}{M}$ | $\frac{N^3}{M\kappa}$ | $\frac{N^3}{M\kappa}$ | $\frac{N^3}{M\kappa}$ |

A cache oblivious QR-decomposition algorithm based on Givens rotations is proposed in [53]. That algorithm consists of two mutually recursive functions, $f$ and $e$, where the function $f$ QR-decomposes a block and the function $e$ annihilates a block using an upper-triangular block as the pivoting block, and assumes that the input matrix is given in Morton order (also called bit-interleaved order or Z-order). Our algorithm uses Householder transformations, and is simpler in that it does not use mutual recursion. Also, we do not assume any input representation. However, if the input is in Morton order, our algorithm will run efficiently even without the "tall cache assumption" [54].

Our algorithm uses the divide-and-conquer technique. First we present the algorithm for square matrices, and generalise it later. The algorithm consists of two functions "CO-QR-decompose" and "CO-QR-Update". "CO-QR-decompose" QR-decomposes the input matrix using Householder transformations and returns the orthogonal transformation $Q$ and upper triangular $R$. "CO-QR-Update" takes as input two blocks $A$ and $B$ of size $N \times N$ each, where $A$ is upper triangular, and returns an orthogonal matrix $Q \in \mathbb{R}^{N \times N}$ and upper triangular matrix $R \in \mathbb{R}^{N/2 \times N/2}$ so that $\begin{pmatrix} A \\ B \end{pmatrix} = Q \begin{pmatrix} R \\ 0 \end{pmatrix}$. The functions "CO-QR-decompose" and "CO-QR-Update" are described in Algorithm 2.5 and Algorithm 2.6 respectively.

---

**Algorithm 2.5.** *CO-QR-decompose(A)*

**Input:** *An $N \times N$ matrix $A$.*

**Output:** *An orthogonal matrix $Q \in \mathbb{R}^{N \times N}$ such that $A = QR$, where $R \in \mathbb{R}^{N \times N}$ is upper triangular and overwrites $A$.*

---

| | |
|---|---|
| *Step-1* | **if $N = 1$ then return** [1]; |
| *Step-2* | *Partition $A$ into four tiles of equal size:* $\begin{pmatrix} A_{00} & A_{01} \\ A_{10} & A_{11} \end{pmatrix}$; |
| *Step-3* | $Q_1 = CO\text{-}QR\text{-}decompose(A_{00})$; |
| *Step-4* | *Let* $\hat{Q}_1 = \begin{pmatrix} Q_1 & 0 \\ 0 & I_{N/2} \end{pmatrix}$ |
| *Step-5* | $Q_2 = CO\text{-}QR\text{-}Update(A_{00}, A_{10})$; |
| *Step-6* | *Let* $Q_3 = \hat{Q}_1 Q_2$; |
| *Step-7* | $\begin{pmatrix} A_{01} \\ A_{11} \end{pmatrix} = Q_3^T \begin{pmatrix} A_{01} \\ A_{11} \end{pmatrix}$; |
| *Step-8* | $Q_4 = CO\text{-}QR\text{-}decompose(A_{11})$; |
| *Step-9* | **Return** $Q_3 \cdot \begin{pmatrix} I_{N/2} & 0 \\ 0 & Q_4 \end{pmatrix}$; |

---

## 2.6.1   I/O Analysis

Let $S(N)$ denote the worst case I/O complexity of an invocation "CO-QR-Update$(A, B)$", where $A$ and $B$ are $N/2 \times N/2$ matrices. The invocation spawns four recursive invocations each on a pair of $N/4 \times N/4$ matrices, and 28 multiplications of two $N/4 \times N/4$ matrices. (See steps 7, 8 and 13 of Algorithm 2.6.) If $N < \sqrt{M}$, then $S(N) = \Theta(N^2/B)$. For $N \geq \sqrt{M}$, we have the following recurrence relation:

$$S(N) = 4S(N/2) + 28V(N/4) \tag{2.1}$$

where $V(N)$ denotes the I/O complexity of multiplying two $N \times N$ matrices. When $N \in (\sqrt{M}, \ \sqrt{M}/2]$,   $S(N) = \Theta(N^2/B) = \Theta(\frac{N^3}{B\sqrt{M}})$; that is, there is a constant $d$ such that $S(N) \leq \frac{dN^3}{B\sqrt{M}}$. Moreover, there is a constant $a$ such that $V(N) \leq \frac{aN^3}{B\sqrt{M}}$ [54]. For $N \geq \sqrt{M}$, if $d \geq 7a/8$, we can use induction to claim that $S(N) \leq \frac{dN^3}{B\sqrt{M}}$, because then

$$S(N) \leq \frac{4d(N/2)^3}{B\sqrt{M}} + \frac{28a(N/4)^3}{B\sqrt{M}} \leq \frac{dN^3}{B\sqrt{M}} \tag{2.2}$$

---

**Algorithm 2.6.** *CO-QR-Update(A,B)*

**Input:** $A, B \in \mathbb{R}^{N/2 \times N/2}$ *matrices, A is upper triangular.*

**Output:** $\begin{pmatrix} A \\ B \end{pmatrix} = Q \begin{pmatrix} R \\ 0 \end{pmatrix}$, *where* $R \in \mathbb{R}^{N/2 \times N/2}$ *is upper triangular, and* $Q \in \mathbb{R}^{N \times N}$ *is orthogonal matrix.*

---

*Step-1*     **if** $N = 2$, **then** *compute the Householder transformation Q such that* $Q^T A$
           *is of the form* $\begin{pmatrix} r \\ 0 \end{pmatrix}$;
           *Let* $A = \begin{pmatrix} r \\ 0 \end{pmatrix}$; **Return** $Q$;

*Step-2*     *Partition the matrices A and B into* $\begin{pmatrix} A \\ B \end{pmatrix} = \left( \begin{array}{c|c} A_{00} & A_{01} \\ 0 & A_{11} \\ \hline B_{20} & B_{21} \\ B_{30} & B_{31} \end{array} \right)$,
           *where $A_{00}$ and $A_{11}$ are upper triangular and* $A_{**}, B_{**} \in \mathbb{R}^{N/4 \times N/4}$;

*Step-3*     $Q_1 = CO\text{-}QR\text{-}Update(A_{00}, B_{20})$;
           *Partition $Q_1$ into* $\begin{pmatrix} Q_{00}^1 & Q_{01}^1 \\ Q_{10}^1 & Q_{11}^1 \end{pmatrix}$ *where* $Q_{**}^1 \in \mathbb{R}^{N/4 \times N/4}$;

*Step-4*     *Let* $\hat{Q}_1 = \begin{pmatrix} Q_{00}^1 & 0 & Q_{01}^1 & 0 \\ 0 & I_{N/4} & 0 & 0 \\ Q_{10}^1 & 0 & Q_{11}^1 & 0 \\ 0 & 0 & 0 & I_{N/4} \end{pmatrix}$;

*Step-5*     $Q_2 = CO\text{-}QR\text{-}Update(A_{00}, B_{30})$;
           *Partition $Q_2$ into* $\begin{pmatrix} Q_{00}^2 & Q_{01}^2 \\ Q_{10}^2 & Q_{11}^2 \end{pmatrix}$ *where* $Q_{**}^2 \in \mathbb{R}^{N/4 \times N/4}$;

*Step-6*     *Let* $\hat{Q}_2 = \begin{pmatrix} Q_{00}^2 & 0 & 0 & Q_{01}^2 \\ 0 & I_{N/4} & 0 & 0 \\ 0 & 0 & I_{N/4} & 0 \\ Q_{10}^2 & 0 & 0 & Q_{11}^2 \end{pmatrix}$;

*Step-7*     *Let* $Q_3 = \hat{Q}_1 \cdot \hat{Q}_2$;

*Step-8*     *Set* $\begin{pmatrix} A_{01} \\ A_{11} \\ B_{21} \\ B_{31} \end{pmatrix} = Q_3 \cdot \begin{pmatrix} A_{01} \\ A_{11} \\ B_{21} \\ B_{31} \end{pmatrix}$;

*Step-9*     $Q_4 = CO\text{-}QR\text{-}Update(A_{11}, B_{21})$;
           *Partition $Q_4$ into* $\begin{pmatrix} Q_{00}^4 & Q_{01}^4 \\ Q_{10}^4 & Q_{11}^4 \end{pmatrix}$ *where* $Q_{**}^4 \in \mathbb{R}^{N/4 \times N/4}$;

*Step-10*     *Let* $\hat{Q}_4 = \begin{pmatrix} I_{N/4} & 0 & 0 & 0 \\ 0 & Q_{00}^4 & Q_{01}^4 & 0 \\ 0 & Q_{01}^4 & Q_{11}^4 & 0 \\ 0 & 0 & 0 & I_{N/4} \end{pmatrix}$;

*Step-11*     $Q_5 = CO\text{-}QR\text{-}Update(A_{11}, B_{31})$;
           *Partition $Q_5$ into* $\begin{pmatrix} Q_{00}^5 & Q_{01}^5 \\ Q_{10}^5 & Q_{11}^5 \end{pmatrix}$ *where* $Q_{**}^5 \in \mathbb{R}^{N/4 \times N/4}$;

*Step-12*     *Let* $\hat{Q}_5 = \begin{pmatrix} I_{N/2} & 0 & 0 & 0 \\ 0 & Q_{00}^5 & 0 & Q_{01}^5 \\ 0 & 0 & I_{N/2} & 0 \\ 0 & Q_{10}^5 & 0 & Q_{11}^5 \end{pmatrix}$;

*step-13*     **Return** $Q = Q_3 \cdot \hat{Q}_4 \cdot \hat{Q}_5$;

Similarly, let $T(N)$ denote the worst case I/O complexity of an invocation "CO-QR-decompose$(A)$", where $A$ is a $N \times N$ matrix. If $N < \sqrt{M}$, then $T(N) = \Theta(N^2/B)$. For $N \geq \sqrt{M}$, we have the following recurrence relation:

$$T(N) = 2T(N/2) + S(N) + 8V(N/2) \tag{2.3}$$

See Algorithm 2.5. When $N \in (\sqrt{M}, \ \sqrt{M}/2]$, $T(N) = \Theta(N^2/B) = \Theta(\frac{N^3}{B\sqrt{M}})$; thus, there is a constant constant $c$ such that $T(N) \leq \frac{cN^3}{B\sqrt{M}}$. For $N \geq \sqrt{M}$, if $c \geq 4(d+a)/3 \geq 5a/2$, we can use induction to claim that $T(N) \leq \frac{cN^3}{B\sqrt{M}}$, because then

$$T(N) \leq \frac{2c(N/2)^3}{B\sqrt{M}} + \frac{dN^3}{B\sqrt{M}} + \frac{8a(N/2)^3}{B\sqrt{M}} \leq \frac{cN^3}{B\sqrt{M}} \tag{2.4}$$

## 2.6.2 The General Case

Now we extend the algorithm for general matrices. Assume that the input is an $N \times P$ matrix $A$.

If $N < P$, then divide $A$ vertically into two slabs $A_1$ and $A_2$, $A = (A_1 \ A_2)$, so that $A_1$ has $N$ columns. Perform a QR decomposition of $A_1$, $A_1 = QR$, using "CO-QR-decompose". Then compute the product $Q^T A_2$. The I/O complexity of this procedure is

$$T'(N, \ P) = T(N) + V'(N, \ N, \ P - N)$$

where $V'(P, Q, R)$ is the I/O complexity of multiplying a $P \times Q$ matrix and a $Q \times R$ matrix. If $N < \sqrt{M}$, then the I/O complexity is $T'(N, \ P) = O(NP/B)$. If $N \geq \sqrt{M}$ then, $T'(N, \ P) = O\left(\frac{N^3}{B\sqrt{M}}\right) + O\left(\frac{N^2(P-N)}{B\sqrt{M}}\right) = O\left(\frac{N^2 P}{B\sqrt{M}}\right)$

If $N > P$, then partition $A$ into $P \times P$ blocks from the top; $A = \begin{pmatrix} A_1 \\ \vdots \\ A_{N/P} \end{pmatrix}$ where each $A_i$ is a $P \times P$ matrix. Perform a QR-decomposition of $A_1$ by invoking "CO-QR-decompose". Then, for $1 < i \leq N/P$, perform a QR-update of $A_1$ and $A_i$ using "CO-QR-Update". The

I/O complexity of this procedure is

$$T'(N,\ P) = T(P) + (N/P - 1)S(P)$$

If $P < \sqrt{M}$, then both $T(P)$ and $S(P)$ are $O(P^2/B)$. Therefore, $T'(N,\ P) = O\left(\frac{NP}{B}\right)$. If $P \geq \sqrt{M}$, then both $T(P)$ and $S(P)$ are $O(\frac{P^3}{B\sqrt{M}})$; $T'(N,\ P) = O\left(\frac{NP^2}{B\sqrt{M}}\right)$.

# Chapter 3

# Banded Hessenberg Reductions and Related Problems

## 3.1 Introduction

Computing of eigenvalues and singular values of matrices have wide-ranging applications in engineering and computational sciences such as control theory, vibration analysis, electric circuits, signal processing, pattern recognition, numerical weather prediction and information technology, to name a few [6, 33, 65, 120].

The eigenvalues of a square matrix are typically computed in a two stage process. In the first stage the matrix is reduced, through a sequence of orthogonal similarity transformations (OSTs), to a condensed form (Hessenberg form, if the matrix is nonsymmetric, and tridiagonal form, otherwise), and in the second stage an iterative method called the $QR$ algorithm is applied to the condensed form [58,115]. The reduction of a nonsymmetric $N \times N$ matrix to upper Hessenberg form as well as a symmetric $N \times N$ matrix to tridiagonal form using OSTs takes $O(N^3)$ flops. Each iteration of the QR algorithm takes $O(N^2)$ and $O(N)$ flops respectively, when applied on Hessenberg and tridiagonal matrices.

The singular values of a matrix are also typically computed in a similar two stage process. First the matrix is reduced to bidiagonal form using orthogonal equivalence transformations (OETs); this takes $O(N^3)$ flops. Then the modified QR algorithm [57] is applied to the bidiagonal matrix. Each iteration of the QR algorithm takes $O(N)$ flops.

(An OST on a square matrix $A \in \mathbb{R}^{N \times N}$ transforms $A$ into $Q^T A Q$ using an orthogonal matrix $Q$. An OET on a matrix $A \in \mathbb{R}^{N \times P}$ transforms $A$ into $Q^T A P$ using orthogonal matrices $Q$ and $P$. OSTs and OETs preserve eigenvalues.)

Thus, in each case, the first stage of the computation, which costs $O(N^3)$ flops, forms a bottleneck. Each of these algorithms also needs $O(N^3/B)$ I/Os to perform its $O(N^3)$ operations, as we shall see in Chapter 4.

The traditional Hessenberg, tridiagonal, and bidiagonal reduction algorithms based on Householders or rotators are rich in V-V and V-M operations, and not in M-M operations [58, 115]. Householder based sequential and parallel algorithms for Hessenberg, tridiagonal and bidiagonal reductions using the slab approach have been proposed [45, 46, 89]. But even in these, V-M operations dominate the performance: after the reduction of each column of a slab, the rest of the matrix needs to be read in to update the next column before it will be ready to be reduced. This is because of the two sidedness of the transformations involved. This also makes it difficult to design tile based versions of these algorithms. To the best of our knowledge, no tile based algorithm has been proposed for the above direct reductions.

Due to the above reasons, it has been proposed that the reduction in the first stage be split into two steps [17, 19, 21, 77, 80], the first reducing the matrix to a block condensed form (banded Hessenberg, symmetric banded or banded, as is relevant) [11, 61, 81], and the second further reducing it to Hessenberg, tridiagonal or bidiagonal form [21, 77, 78, 80]. Almost all operations of the first step are performed using M-M operations. Though the modified first stage takes more flops, its richness in M-M operations, makes it more efficient on machines with multiple levels of memory. Usually these reductions are imple-mented using relatively expensive two sided orthogonal transformations rather than the inexpensive single sided Gauss elimination, because orthogonal transformations guarantee stability [102, 115].

We now present an overview of the results in this chapter.

### 3.1.1  Reduction of a nonsymmetric matrix to banded Hessenberg form

Suppose the input is a nonsymmetric matrix $A \in \mathbb{R}^{N \times N}$, and it is to be reduced to banded upper Hessenberg form $H_t$ of bandwidth $t$. This could be done using an OST: construct an orthogonal matrix $Q \in \mathbb{R}^{N \times N}$ such that $H_t = Q^T A Q$ [17,58,81,83]. A slab based sequential algorithm [58], and a parallel algorithm for message passing multicomputers [17] are known. Tile based algorithms for multicore architectures using Householders and Givens rotations are presented in [81, 83].

We study the slab based algorithm of [58], and analyse it for its I/O complexity for varying values of $M$ and $N$, for a given value of $t$. This algorithm assumes that the slab size $k = t$. We generalise this into an algorithm with $k$ not necessarily the same as $t$. We find that the algorithm performs the best when $k = \min\{t, \sqrt{M}\}$.

We also investigate the I/O complexity of some tile based algorithms [27,64,90].

These algorithms, typically choose $t$ as the tile size, and also at any time, maintain only a small number of tiles in the main memory. If $t \ll \sqrt{M}$, then most of the main memory stays unused. We propose an algorithm that allows the number of tiles kept in the main memory to depend on the tile size. We also propose an algorithm for $t > \sqrt{M}$, which partitions the matrix into tiles of size $\sqrt{M/3}$; this gives a constant factor improvement over the choice of $t$ as the tile size. We conclude that banded Hessenberg reduction can be achieved in $O(N(N-t)^2/(B\min\{t, \sqrt{M}\}))$ I/Os using the tile approach, when $(N-t) = \Theta(N)$.

### 3.1.2   Reduction of a symmetric matrix to symmetric banded form

Suppose the input symmetric matrix $A \in \mathbb{R}^{N \times N}$, reduces to symmetric banded form using a two stage approach called successive band reduction (SBR) [20–22]. Here we consider the first stage of SBR.

Suppose the input is a symmetric matrix $A \in \mathbb{R}^{N \times N}$, and it is to be reduced to symmetric banded form $T_t$ of bandwidth $2t + 1$. This could be done using an OST: construct an orthogonal matrix $Q \in \mathbb{R}^{N \times N}$ such that $T_t = Q^T A Q$ [17, 58, 81, 83]. In symmetric banded reduction, due to symmetry only the lower or upper banded portion needs to be stored.

The problem is well studied in parallel and out-of-core contexts [10, 11, 55, 60, 119]. All these algorithms are slab based. The performance of these algorithms depends upon various factors like data layout, the final bandwidth, and the slab width. Usually the slab width $k$ is assumed to be same as the desired bandwidth $t$ [11, 19, 55, 60].

We look at the out-of-core algorithm of [60] in particular. Parallel version of this algorithm appears in [8, 18]. A parallel algorithm using PLAPACK was proposed in [119] and also implemented in [10, 11]; these assume that $k$ is not necessarily equal to $t$.

The block-tridiagonal divide-and-conquer (BD&C) algorithm of [56] computes eigenvalues of banded matrices directly instead of further reducing it to tridiagonal form. See [10, 11, 55] for further details on BD&C. These works also discuss the symmetric band reduction. In particular, in [10] it is mentioned that $k$ should always be less than or equal $t$ for optimum performance. The reduction of a banded form to tridiagonal form is discussed in [21, 55, 72, 77, 95, 97].

In this chapter, we recall the standard the slab based algorithm for the problem. Our tile based algorithms for banded Hessenberg reduction work for symmetric band reduction too, and perform about half as many I/Os.

### 3.1.3   Reduction to banded form

Suppose the input matrix is $A \in \mathbb{R}^{N \times P}$, and it is to be reduced to banded form $B_t$ of bandwidth $t$. This could be done using an OET: construct orthogonal matrices $U \in \mathbb{R}^{N \times N}$ and $V \in \mathbb{R}^{P \times P}$ such that $B_t = U^T A V$ [61, 82, 83]. If $N \geq P$ then $B_t$ would be in upper banded form, otherwise in lower banded form.

A slab based parallel algorithm in the context of distributed memory for the two stage reduction is described in [61]. An algorithm for the first stage on a multicore architecture is given in [82, 83]. The second stage is discussed in [78, 80]. See Chapter 4 for details.

We describe slab and tile based algorithms for the problem and analyse them for their I/O complexities, as in the cases of banded Hessenberg and symmetric band reductions.

### 3.1.4   Organisation of this Chapter

The rest of the chapter is organised as follows: Section 2 discusses the reduction of a nonsymmetric matrix to banded Hessenberg form using the slab approach. In Section 3, we recall and propose tile approach algorithms for banded Hessenberg reduction. Reduction of symmetric matrix to banded form using the slab approach is discussed in Section 4. In Section 5, we propose tile based reduction algorithms for symmetric band reduction. The reduction of a general matrix to upper banded form using the slab approach is given in Section 6. In Section 7, we describe some tile based algorithms for the reduction of a general matrix to upper banded form.

## 3.2   Reduction of a Nonsymmetric Matrix to Banded Hessenberg Form using the Slab Approach

Let $A \in \mathbb{R}^{N \times N}$ be the nonsymmetric matrix that is to be reduced to banded upper Hessenberg form $H_t$ of bandwidth $t$.

### 3.2.1   Reduction with a slab width of $t$

The algorithm of [58], when $N - t > t$, partitions the matrix into slabs of width $t$, and then proceeds in $N/t - 1$ iterations, each of which reduces a slab using block QR factorisations and then updates the rest of the matrix from both sides using aggregated Householders [23]. The algorithm is illustrated below:

Consider the first slab of $A$:

$$A = \begin{bmatrix} \overset{t}{A_{11}} & \overset{N-t}{A_{12}} \\ \hline A_{21} & A_{22} \end{bmatrix} \begin{matrix} t \\ \\ N-t \end{matrix}$$

Perform a QR factorisation on $A_{21}$ such that $A_{21} = Q_{21}R_{21}$. Then an OST with $Q = \begin{pmatrix} I_t & 0 \\ 0 & Q_{21} \end{pmatrix}$ gives:

$$Q^T A Q = \left( \begin{array}{c|c} A_{11} & A_{12}Q_{21} \\ \hline R_{21} & Q_{21}^T A_{22} Q_{21} \end{array} \right)$$

If $Q$ is in WY-representation, this involves computing a product of the form $(I + WY^T)^T A (I + WY^T)$. Now the first slab is in banded Hessenberg form. Repeat this $N/t - 1$ times, and the matrix reduces to banded upper Hessenberg form $H_t$:

$$H_t = \begin{pmatrix} H_{11} & H_{12} & \cdots & \cdots & H_{1\frac{N}{t}} \\ H_{21} & H_{22} & \cdots & \cdots & H_{2\frac{N}{t}} \\ 0 & H_{32} & \ddots & \cdots & H_{3\frac{N}{t}} \\ \vdots & \vdots & \ddots & \ddots & \vdots \\ 0 & 0 & \cdots & H_{\frac{N}{t}(\frac{N}{t}-1)} & H_{\frac{N}{t}\frac{N}{t}} \end{pmatrix}$$

where each $H_{ij}$ is a $t \times t$ tile; $H_{ij}$ is a full tile when $i \le j$, upper triangular when $i = j + 1$, and all zero otherwise. Thus, $H_t$ is banded upper Hessenberg with bandwidth $t$.

The I/O complexity of this algorithm can obtained from that of Case-1 in Algorithm 3.1, by substituting $k = t$.

If $N - t \leq t$ then $A$ is partitioned as follows:

$$A = \begin{bmatrix} \overset{N-t}{A_{11}} & \overset{t}{A_{12}} \\ \hline A_{21} & A_{22} \end{bmatrix} \begin{matrix} t \\ \\ N-t \end{matrix}$$

Perform a QR factorisation on $A_{21}$ such that $A_{21} = Q_{21}R_{21}$. Then an OST with $Q = \begin{pmatrix} I_t & 0 \\ 0 & Q_{21} \end{pmatrix}$ gives $Q^T A Q$ which is in the desired banded Hessenberg form. The I/O complexity of this is as given in Table 3.1. **(Refer to the table in Chapter 2)**.

Table 3.1: The case of $(N - t) < t$. The number of I/Os for QR decomposition, matrix multiplication, and the total are given in columns titled QRD, MM and Total respectively

| Conditions on $M$ & $N - t$ | QRD | MM | Total |
|---|---|---|---|
| $N - t > M$ | $\frac{(N-t)^3}{\sqrt{M}B}$ | $\frac{N(N-t)^2}{\sqrt{M}B}$ | $\frac{N(N-t)^2}{\sqrt{M}B}$ |
| $N - t \leq \sqrt{M}$ | $\frac{(N-t)^2}{B}$ | $\frac{N(N-t)}{B}$ | $\frac{N(N-t)}{B}$ |
| $M^{2/3} < N - t \leq M$ | $\frac{(N-t)^3}{\sqrt{N-t}B}$ | $\frac{N(N-t)^2}{\sqrt{M}B}$ | $\frac{(N-t)^3}{\sqrt{N-t}B} + \frac{N(N-t)^2}{\sqrt{M}B}$ |
| $\sqrt{M} < N - t \leq M^{2/3}$ | $\frac{(N-t)^4}{MB}$ | $\frac{N(N-t)^2}{\sqrt{M}B}$ | $\frac{(N-t)^4}{MB} + \frac{N(N-t)^2}{\sqrt{M}B}$ |

### 3.2.2   Reduction with a slab width not necessarily equal to $t$

In the above algorithm, if the slabs to be QR decomposed are very small compared to the main memory ($(N - it) \times t \ll M$), then the main memory would seem under utilised; a larger slab width might be appropriate. Also, if the slab is too big to fit in the main memory ($(N - it) \times t \gg M$) then the QR factorisations and subsequent updates are to performed out-of-core; a smaller slab width might do better. We present an out-of-core algorithm (see Algorithm 3.1) that uses a slab width of $k$ not necessarily the same as $t$.

Algorithms of a similar vein have been proposed for the reduction of a full symmetric matrix to symmetric banded form [10, 11, 119], and it has been observed [10] that choosing

---

**Algorithm 3.1.** *Banded Hessenberg reduction using the slab approach*

---

**Input:** *An $N \times N$ matrix $A$, bandwidth $t$ and algorithmic block size $k$.*

**Output:** *$H = Q^T A Q$, $H$ is $N \times N$ banded upper Hessenberg with bandwidth $t$, a number of Householder aggregates with $Q$ as their product.*

---

**Case-1:** *$k \leq t$. For ease of exposition, assume that $k$ divides $t$*
    **for** $i = 1$ **to** $(N - t)/t$ **do**
        *Let $g = (i - 1)t$;*
        **for** $j = 1$ **to** $t/k$ **do**
            *Let $h = (j - 1)k$;*
            *QR-Decompose($A[g + h + t + 1 : N, \ g + h + 1 : \ g + h + k]$);*
            *Let $Q_{ij} = (I + Y_{ij} T_{ij} Y_{ij}^T)$ be the resulting compact WY representation of $Q$;*
            **Update the rest of the matrix:**
            *Left-multiply $A[g + h + t + 1 : N, \ g + h + k + 1 : N]$ with $Q_{ij}$;*
            *Right-multiply $A[1 : N; \ g + h + t + 1 : N]$ with $Q_{ij}$;*
        **endfor***;*
    **endfor***;*
**Case-2:** *$k > t$. For ease of exposition, assume that $t$ divides $k$;*
    **for** $i = 1$ **to** $(N - t)/k$ **do**
        *Let $g = (i - 1)k$;*
        *Let $y$ denote the range $g + t + 1 : N$;*
        *Let $z$ denote the range $g + k + 1 : N$;*
        *$\hat{A}[y, \ y] = A[y, \ y]$;*
        **for** $j = 1$ **to** $k/t$ **do**
            *Let $h = (j - 1)t$;*
            *QR-Decompose($A[g + h + t + 1 : N, \ g + h + 1 : \ g + h + t]$);*
            *Let $Q_{ij} = (I + Y_{ij} T_{ij} Y_{ij}^T)$ be the resulting compact WY representation of $Q$;*
            *Let $Y_i = Y_{ij}$ if $j = 1$ and $Y_i = (Y_i \ Y_{ij})$ otherwise;*
            *Let $T_i = T_{ij}$ if $j = 1$ and $T_i = \begin{pmatrix} T_i & T_i Y_i^T Y_{ij} T_{ij} \\ 0 & T_{ij} \end{pmatrix}$ otherwise;*
            *Let $x$ denote the range $g + h + t + 1 : \ g + h + 2t$;*
            **If** $(j \neq k/t)$ **do**
                **Update the next $t$ columns of the panel $i$:**
                *Compute $B_{ij} = \hat{A}[y, \ x] + \hat{A} Y_i T_i \ \left( Y_i^T[1 : h + t, \ x] \right)$;*
                *Compute $A[y, \ x] = (I + Y_i T_i Y_i^T) B_{ij}$;*
            **end if***;*
        **end for***;*
        **Update the rest of the matrix:**
        *Right-multiply $A[1 : g + t, \ y]$ with $(I + Y_i T_i Y_i^T)$;*
        *Compute $A[y : z] = \hat{A}[y : z] + \hat{A} Y_i T_i \ \left( Y_i^T[1 : k, z] \right)$;*
        *Left-multiply $A[y : z]$ with $(I + Y_i T_i Y_i^T)$;*
    **endfor***;*

---

a value of at most $t$ for $k$ will do better than otherwise. To the best of our knowledge, no such out-of-core algorithm has been proposed for banded Hessenberg reduction. Case-2 of Algorithm 3.1 has been inspired by the parallel algorithm of [119], while Case-1 is a generalisation of the algorithm described in Subsection 3.2.1 [58].

Algorithm 3.1 divides $A$ into vertical slabs of $k$ columns each. Given the values of $M$, $N$ and $t$, the I/O complexity of the algorithm depends upon the choice of $k$. We analyse the two cases of $t < k$ and $t \geq k$ separately, with an intent of finding the choice of $k$ that would minimise the number of I/Os.

### 3.2.2.1   Case 1: $k \leq t$

For $1 \leq i \leq (N-t)/t$ and $1 \leq j \leq t/k$, in the $(i,j)$-th iteration there are a QR decomposition of an $(N-g-h-t) \times k$ matrix, a matrix multiplication chain of dimensions $(N-g-h-t, k, k, N-g-h-t, N-g-h-k)$, and a matrix multiplication chain of dimensions $(N, N-g-h-t, k, k, N-g-h-t)$, where $g = (i-1)t$ and $h = (j-1)k$. Let $\alpha$, $\beta$ and $\gamma$ denote the I/O complexities of these operations, in that order. Clearly, $\gamma \geq \beta$, in all cases. As $i$ varies from 1 to $(N-t)/t$ and $j$ varies from 1 to $t/k$, $(N-g-h-t)$ takes on values $lk$ for $(N-t)/k \geq l \geq 1$. The slab fits in the memory when $(N-g-h-t)k = lk^2 \leq M$; that is, $l \leq M/k^2$.

The following table gives the asymptotic I/O complexity of the $l$-th iteration, for $l \geq 1$. We omit the $O$-notation for brevity. **(Refer to the table in Chapter 2.)**

Let $c = \sqrt{M}/k$.

If $c \geq 1$, then the table simplifies as follows:

Summing over the iterations, the I/O complexity is as given in Table 3.4. For example,

Table 3.2: $\alpha$, $\gamma$ and their sum in the $l$-th iteration for various values of $l$

| | $\alpha$ | $\gamma$ | $\alpha + \gamma$ |
|---|---|---|---|
| $1 \le l \le \frac{B}{k}$ | $k$ | $\frac{Nlk}{B}$ | $\frac{Nlk}{B}$ |
| $\frac{B}{k} < l \le c^2$ | $\frac{lk^2}{B}$ | $\frac{Nlk}{B}$ | $\frac{Nlk}{B}$ |
| $c^2 < l \le c^2\sqrt{k}$ | $\frac{l^2k^4}{BM}$ | $\left(1+\frac{1}{c}\right)\frac{Nlk}{B}$ | $\left(1+\frac{1}{c}+\frac{lk^3}{NM}\right)\frac{Nlk}{B}$ |
| $c^2\sqrt{k} < l \le c^2k$ | $\frac{lk^3}{B\sqrt{k}}$ | $\left(1+\frac{1}{c}\right)\frac{Nlk}{B}$ | $\left(1+\frac{1}{c}+\frac{k^2}{N\sqrt{k}}\right)\frac{Nlk}{B}$ |
| $c^2k < l$ | $\frac{lk^3}{B}\left(\frac{1}{\sqrt{k}}+\frac{1}{\sqrt{M}}\right)$ | $\left(1+\frac{1}{c}\right)\frac{Nlk}{B}$ | $\left(1+\frac{1}{c}+\frac{k^2}{N\sqrt{k}}+\frac{k^2}{N\sqrt{M}}\right)\frac{Nlk}{B}$ |

Table 3.3: $\alpha$, $\gamma$ and their sum in the $l$-th iteration for various values of $l$, when $c \ge 1$

| | $\alpha$ | $\gamma$ | $\alpha + \gamma$ |
|---|---|---|---|
| $1 \le l \le \frac{B}{k}$ | $k$ | $\frac{Nlk}{B}$ | $\frac{Nlk}{B}$ |
| $\frac{B}{k} < l \le c^2$ | $\frac{lk^2}{B}$ | $\frac{Nlk}{B}$ | $\frac{Nlk}{B}$ |
| $c^2 < l \le c^2\sqrt{k}$ | $\frac{l^2k^4}{BM}$ | $\frac{Nlk}{B}$ | $\left(1+\frac{lk^3}{NM}\right)\frac{Nlk}{B}$ |
| $c^2\sqrt{k} < l$ | $\frac{lk^3}{B\sqrt{k}}$ | $\frac{Nlk}{B}$ | $\left(1+\frac{k^2}{N\sqrt{k}}\right)\frac{Nlk}{B}$ |

when $1 \le \frac{(N-t)}{k} \le c^2$, $\sum_{l=1}^{(N-t)/k} \alpha + \gamma = \sum_{l=1}^{(N-t)/k} Nlk/B = N(N-t)^2/Bk$.

Table 3.4: The I/O complexity for Case 1 and $c \ge 1$, under various conditions

| Condition | I/Os | Condition paraphrased |
|---|---|---|
| $1 \le \frac{(N-t)}{k} \le c^2$ | $\frac{N(N-t)^2}{Bk}$ | $k \le \frac{M}{(N-t)}$ |
| $c^2 < \frac{(N-t)}{k} \le c^2\sqrt{k}$ | $\frac{N(N-t)^2}{Bk} + \frac{k(N-t)^3}{BM} - \frac{M^2}{Bk^2}$ | $\frac{M}{(N-t)} < k \le \frac{M^2}{(N-t)^2}$ |
| $c^2\sqrt{k} < \frac{(N-t)}{k}$ | $\frac{N(N-t)^2}{Bk} + \frac{k(N-t)^2-M^2}{B\sqrt{k}}$ | $\frac{M^2}{(N-t)^2} < k$ |

Consider the following five statements: (The blank is to be filled in by the phrases listed below to obtain the five respective statements.) There exist values $M$, $N$ and $t$ with $t \leq \sqrt{M}$, (which makes it possible to choose $k = t$ with $c = \frac{\sqrt{M}}{k} = \frac{\sqrt{M}}{t} \geq 1$ satisfied) and ——————— is less than the cost of $k = t$.

1. $t > \frac{M^2}{(N-t)^2}$ and so that the cost of $k = \frac{M}{(N-t)}$

2. $t > \frac{M^2}{(N-t)^2}$ and so that the cost of the best $k$ in $(\frac{M}{N-t}, \frac{M^2}{(N-t)^2}]$

3. $t > \frac{M^2}{(N-t)^2}$ and so that the cost of some $k$ in $(\frac{M^2}{(N-t)^2}, t)$

4. $\frac{M}{N-t} < t \leq \frac{M^2}{(N-t)^2}$ and so that the cost of $k = \frac{M}{(N-t)}$

5. $\frac{M}{N-t} < t \leq \frac{M^2}{(N-t)^2}$ and so that the cost of some $k$ in $(\frac{M}{N-t}, t)$

We claim that each of these statements is false. Proofs by contradiction follow:

Define $a = \log_M t$, $b = \log_M(N - t)$. Then $a, b > 0$, $t = M^a$, $(N - t) = M^b$ and $N = M^a + M^b$. Thus, $t \leq \sqrt{M} \Rightarrow a \leq 1/2$. For Statements 1, 2 and 3, therefore, $t > \frac{M^2}{(N-t)^2} \Rightarrow a > 2 - 2b \Rightarrow b > 1 - a/2 \Rightarrow b > 3/4$. So $N = (M^a + M^b) \approx M^b$.

For Statement 1, when $k = M/(N - t)$ the cost is $\frac{N(N-t)^2}{Bk} = NM^{3b-1}/B$. When $k = t$, the cost is $\leq \frac{N(N-t)^2}{Bk} + \frac{k(N-t)^2}{B\sqrt{k}} = (NM^{2b-a} + M^{2b+a/2})/B$. But, $NM^{3b-1} < M^{2b+a/2} \Rightarrow 4b - 1 < 2b + a/2 \Rightarrow (a/2 + 1) > 2b \geq 3/2 \Rightarrow a > 1$. Contradiction.

The cost for $k$ in $(\frac{M}{N-t}, \frac{M^2}{(N-t)^2}]$ is lower bounded by: $\frac{N(N-t)^2}{Bk} - \frac{M^2}{Bk^2}$ which is minimised at $k = \frac{M^2}{(N-t)^2} = M^{2-2b}$. For Statement 2, the cost at $k = M^{2-2b}$ is at most $(M^{5b-2} - M^{4b-2})/B \approx M^{5b-2}/B$. When $k = t$, the cost is, as before, $(NM^{2b-a} + M^{2b+a/2})/B$. But, $M^{5b-2} < M^{2b+a/2} \Rightarrow 5b - 1 < 2b + a/2 \Rightarrow (a/2 + 2) > 3b \geq 9/4 \Rightarrow a > 1/2$. Contradiction.

For Statement 3, the cost at $t$ is $(NM^{2b-a} + M^{a/2+2b} - M^{2-a/2})/B$. The cost at $M^c$, $\frac{M^2}{(N-t)^2} < M^c < t$ is $(NM^{2b-c} + M^{c/2+2b} - M^{2-c/2})/B$. But, $NM^{2b}(M^{-a} - M^{-c}) + M^{2b}(M^{a/2} - M^{c/2}) - M^2(M^{-a/2} - M^{-c/2}) > 0 \Rightarrow (2 - (a+c)/2 > 3b - a/2 - c) \vee (2b > 3b - a/2 - c) \Rightarrow b < 3/4$. Contradiction.

For statements 4 and 5, $M^{1-b} < M^a \leq M^{2-2b} \Rightarrow (1-b) < a \leq 2-2b \Rightarrow b > 1/2$. That is, $a \leq 1/2 < b$.

Consider Statement 4. When $k = M/(N-t)$ the cost is $\frac{N(N-t)^2}{Bk} = NM^{3b-1}/B$. When $k = t$, the cost is $\leq \frac{N(N-t)^2}{Bt} + \frac{t(N-t)^3}{BM} = (NM^{2b-a} + M^{a+3b-1})/B$. But, $NM^{3b-1} < M^{a+3b-1} \Rightarrow N < M^a$. Contradiction.

For Statement 5, the cost at $t$ is $(NM^{2b-a} + M^{a+3b-1} - M^{2-2a})/B$. The cost at $M^c$, $\frac{M}{(N-t)} < M^c < t \leq \frac{M^2}{(N-t)^2}$, is $(NM^{2b-c} + M^{c+3b-1} - M^{2-2c})/B$. But, $NM^{2b}(M^{-a} - M^{-c}) + M^{3b-1}(M^a - M^c) - M^2(M^{-2a} - M^{-2c}) > 0 \Rightarrow a + c > 1 \lor b < (2-c)/3$. As $c < a \leq 1/2$, $a + c > 1$ cannot be. On the other hand, $b < (2-c)/3 \Rightarrow \frac{M}{(N-t)} = M^{(1-b)} > M^{(1+c)/3} \Rightarrow c > (1+c)/3 \Rightarrow c > 1/2$; contradiction.

What we have shown is that when $t \leq \sqrt{M}$, $t$ is a better choice for $k$ than any of the smaller values. An analogous proof shows that if $\sqrt{M} < t$, then $\sqrt{M}$ is better than any smaller value.

If $c < 1$, (that is, $k > \sqrt{M}$) then Table 3.2 simplifies as follows:

Table 3.5: $\alpha$, $\gamma$ and their sum in the $l$-th iteration for various values of $l$, when $c < 1$

|  | $\alpha$ | $\gamma$ | $\alpha + \gamma$ |
|---|---|---|---|
| $1 \leq l \leq c^2\sqrt{k}$ | $\frac{l^2 k^4}{BM}$ | $\frac{Nlk}{cB}$ | $\left(\frac{1}{c} + \frac{lk^3}{NM}\right)\frac{Nlk}{B}$ |
| $c^2\sqrt{k} < l \leq c^2 k$ | $\frac{lk^3}{B\sqrt{k}}$ | $\frac{Nlk}{cB}$ | $\left(\frac{1}{c} + \frac{k^2}{N\sqrt{k}}\right)\frac{Nlk}{B}$ |
| $c^2 k < l$ | $\frac{lk^3}{B}\left(\frac{1}{\sqrt{k}} + \frac{1}{\sqrt{M}}\right)$ | $\frac{Nlk}{cB}$ | $\left(\frac{1}{c} + \frac{k^2}{N\sqrt{k}} + \frac{k^2}{N\sqrt{M}}\right)\frac{Nlk}{B}$ |

Choosing $c < 1$ we would have $\sqrt{M} < k \leq t$. As the simplified tables above show, any such choice would be worse than the choice of $k = \sqrt{M}$ in the $c \geq 1$ case discussed above. That is, if $\sqrt{M} < t$, then $\sqrt{M}$ is better than any larger value of $k \leq t$.

Table 3.6: The I/O complexity for Case 1 and $c < 1$, under various conditions

| Condition | I/Os | Condition paraphrased |
|---|---|---|
| $1 \leq \frac{(N-t)}{k} \leq c^2\sqrt{k}$ | $\frac{N(N-t)^2}{B\sqrt{M}} + \frac{k(N-t)^3}{BM}$ | $k \leq \frac{M^2}{(N-t)^2}$ |
| $c^2\sqrt{k} < \frac{(N-t)}{k} \leq c^2k$ | $\frac{N(N-t)^2}{B\sqrt{M}} + \frac{k(N-t)^2-M^2}{B\sqrt{k}}$ | $\frac{M^2}{(N-t)^2} < k$ and $(N-t) \leq M$ |
| $c^2k < \frac{(N-t)}{k}$ | $\frac{N(N-t)^2}{B\sqrt{M}} + \frac{k(N-t)^2-M^2}{B\sqrt{k}} + \frac{k(N-t)^2-kM^2}{B\sqrt{M}}$ | $\frac{M^2}{(N-t)^2} < k$ and $(N-t) > M$ |

If $k$ is to be at most $t$, choosing it as $\min\{t, \sqrt{M}\}$ is the best.

### 3.2.2.2    Case 2: $k > t$

For $1 \leq i \leq (N-t)/k$ and $1 \leq j \leq k/t$, in the $(i,j)$-th iteration of the inner loop there are

- a QR decomposition of an $(N - g - h - t) \times t$ matrix,

- a matrix multiplication chain of dimensions $(h, N - g - t, t, t)$,

- a matrix multiplication chain of dimensions $(N - g - t, N - g - t, h + t, h + t, t)$, and

- a matrix multiplication chain of dimensions $(N - g - t, h + t, h + t, N - g - t, t)$,

where $g = (i - 1)k$ and $h = (j - 1)t$. Let $\alpha$, $\beta$, $\gamma$ and $\delta$ denote the I/O complexities of these operations, in that order. For $1 \leq i \leq (N-t)/k$, in the $i$-th iteration of the outer loop there is a

- a matrix multiplication chain of dimensions $(g + t, N - g - t, k, k, N - g - t)$,

- a matrix multiplication chain of dimensions $(N - g - t, N - g - t, k, k, N - g - k)$, and

- a matrix multiplication chain of dimensions $(N - g - t, k, k, N - g - t, N - g - k)$.

Let $\mu$, $\psi$ and $\phi$ denote the I/O complexities of these operations, in that order.

Let $c = \sqrt{M}/t$. Proceeding as in the analysis of case 1, we find that as $i$ varies from 1 to $(N-t)/k$ and $j$ varies from 1 to $k/t$, $(N-g-h-t)$ takes on values $lt$ for $(N-t)/t \geq l \geq 1$, and therefore $\alpha$ in the $l$-th iteration is as given in the table below: (We omit the order notation for brevity.)

As $0 \leq g \leq N-t-k$, we have that $t \leq g+t \leq N-k$, $N-t \geq N-g-t \geq k$ and

Table 3.7: $\alpha$ in the $l$-th iteration for various values of $l$, when $k > t$

|  | $\alpha$ |
|---|---|
| $1 \leq l \leq \frac{B}{t}$ | $t$ |
| $\frac{B}{t} < l \leq c^2$ | $\frac{lt^2}{B}$ |
| $c^2 < l \leq c^2\sqrt{t}$ | $\frac{l^2 t^4}{BM}$ |
| $c^2\sqrt{t} < l \leq c^2 t$ | $\frac{lt^3}{B\sqrt{t}}$ |
| $c^2 t < l$ | $\frac{lt^3}{B}\left(\frac{1}{\sqrt{t}} + \frac{1}{\sqrt{M}}\right)$ |

$N-k \geq N-g-k \geq t$. Therefore,

$\sum_{i=1}^{(N-t)/k} \mu(i)$: $O\left(\left(\frac{k}{\sqrt{M}}+1\right) \sum_{i=1}^{(N-t)/k} \frac{(g+t)(N-g-t)}{B}\right)$,

$\sum_{i=1}^{(N-t)/k} \psi(i)$: $O\left(\left(\frac{k}{\sqrt{M}}+1\right) \sum_{i=1}^{(N-t)/k} \frac{(N-g-t)^2}{B}\right)$, and

$\sum_{i=1}^{(N-t)/k} \phi(i)$: $O\left(\left(\frac{k}{\sqrt{M}}+1\right) \sum_{i=1}^{(N-t)/k} \frac{(N-g-k)(N-g-t)}{B}\right)$

$\sum_{i=1}^{(N-t)/k} \mu(i)+\psi(i)+\phi(i) \leq O\left(\left(\frac{k}{\sqrt{M}}+1\right) \sum_{i=1}^{(N-t)/k} \frac{N(N-g-t)}{B}\right) = O\left(\frac{N(N-t)^2}{kB}\left(\frac{k}{\sqrt{M}}+1\right)\right)$

Similarly, $\sum_{i=1}^{(N-t)/k} \sum_{j=1}^{k/t} \gamma(i,j) = O\left(\sum_{i=1}^{(N-t)/k} \sum_{j=1}^{k/t} \frac{(N-g-t)^2}{B}\left(\frac{t}{\sqrt{M}}+1\right)\right)$.

Also, $\sum_{i=1}^{(N-t)/k} \sum_{j=1}^{k/t} \beta(i,j)+\delta(i,j) = O\left(\sum_{i=1}^{(N-t)/k} \sum_{j=1}^{k/t} \frac{(N-g-t)(h+t)}{B}\left(\frac{t}{\sqrt{M}}+1\right)\right)$

As $N-g-t \geq h+t$, we have:

$$\sum_{i=1}^{(N-t)/k} \sum_{j=1}^{k/t} \beta(i,j)+\gamma(i,j)+\delta(i,j) \leq O\left(\left(\frac{t}{\sqrt{M}}+1\right) \sum_{i=1}^{(N-t)/k} \sum_{j=1}^{k/t} \frac{(N-g-t)^2}{B}\right)$$

$$= O\left(\left(\frac{t}{\sqrt{M}}+1\right)\frac{k}{t}\sum_{i=1}^{(N-t)/k}\frac{(N-g-t)^2}{B}\right) = O\left(\left(\frac{(N-t)^3}{tB}\left(\frac{t}{\sqrt{M}}+1\right)\right)\right)$$

The total cost of all matrix multiplications is: $O\left(\frac{N(N-t)^2}{B}\left(\frac{1}{\sqrt{M}}+\frac{1}{t}+\frac{1}{k}\right)\right)$.

The total cost of QR-decomposition is:

Table 3.8: The total cost of QR-decomposition, under various conditions

| Condition | Cost of QR decomposition |
|---|---|
| $1 \leq \frac{(N-t)}{t} \leq \frac{B}{t}$ | $N-t$ |
| $\frac{B}{t} < \frac{(N-t)}{t} \leq c^2$ | $\frac{(N-t)^2}{B}+B$ |
| $c^2 < \frac{(N-t)}{t} \leq c^2\sqrt{t}$ | $\frac{t(N-t)^3}{BM}$ |
| $c^2\sqrt{t} < \frac{(N-t)}{t} \leq c^2 t$ | $\frac{t(N-t)^2-M^2}{B\sqrt{t}}$ |
| $c^2 t < \frac{(N-t)}{t}$ | $\frac{t(N-t)^2-M^2}{B\sqrt{t}} + \frac{t(N-t)^2-tM^2}{B\sqrt{M}}$ |

As can be seen, the I/O complexity, the sum of the above two, is independent of the choice of $k$. But, when $t \leq \sqrt{M}$, a choice of $k = t$, and when $t > \sqrt{M}$, a choice of $k = \sqrt{M}$ would give lower costs than these.

Thus, we conclude that the slab based algorithm does the best when $k$ is chosen as $\min\{\sqrt{M}, t\}$.

## 3.3  Reduction to Banded Hessenberg Form using the Tile Approach

In this section, we discuss the reduction of a nonsymmetric $N \times N$ matrix $A$ into banded Hessenberg form using the tile approach. Suppose $t$ is the desired bandwidth. Without loss of generality, we assume that $N$ is a multiple of $t$. Partition the matrix $A$ into tiles of size $t \times t$; there are $N/t$ tile rows and $N/t$ tile columns.

A tile based parallel algorithm for message passing multicomputers for banded upper Hessenberg reduction was proposed in [17]. Recently, tile based algorithms have been designed in the multicore environment using both Householders and Givens rotations [81, 83].

We assume that $A$ has already been "tile-transposed". (We define tile transposition of $A$ as the problem of permuting its elements so that each tile is available contiguously in column-major order.) As we showed in Chapter 2, if $\sqrt{M} > B$ (the tall cache assumption), for any $t$, a $t \times t$ tiling of $A$ can be found in $O(N^2/B)$ I/Os and $O(N^2/(t + \sqrt{M}))$ seeks.

## 3.3.1   An Algorithm for $t < \sqrt{M}$

First we consider an algorithm derived from the parallel algorithms of [17]; it is described in Algorithm 3.2. We use the narrow panel techniques of [27,64,90]: QR decompose narrow panels of size $t \times k$, aggregate its $k$ Householders for each panel, and apply the aggregates from the left and right one by one, by accessing narrow panels of rows and columns.

Now, the functions invoked by the algorithm are described, and analysed for their I/O complexity. We consider three cases where one, two and three tiles fit in the main memory, respectively.

The functions "QR-decompose", "QR-Update", "QR-Multiply-from-left-1", and "QR-Multiply-from-left-2" are described in Chapter 2.

**Function QR-Multiply-from-right-1**$(A_{ji}, A_{lj})$

If $A_{ji}$ QR-decomposes into $Q_{ji}R_{ji}$, then a pre-multiplication of $A$ by an orthogonal matrix $Q^T$ obtained by replacing the $j$-th diagonal tile of $I_N$ by $Q_{ji}^T$ has the effect of making

---

**Algorithm 3.2.** *Banded Hessenberg reduction using the tile approach*

**Input:** *A nonsymmetric matrix* $A \in \mathbb{R}^{N \times N}$ *and a parameter* $t < \sqrt{M}$

**Output:** *A banded Hessenberg matrix of bandwidth* $t$.

---

**for** $i = 1$ **to** $N/t - 1$ **do**
　**for** $j = i + 1$ **to** $N/t$ **do**
　　*QR-decompose*$(A_{ji})$;
　　**for** $k = i + 1$ **to** $N/t$ **do**
　　　*QR-Multiply-from-left-1*$(A_{ji}, A_{jk})$;
　　**endfor**;
　　**for** $l = 1$ **to** $N/t$ **do**
　　　*QR-Multiply-from-right-1*$(A_{ji}, A_{lj})$;
　　**endfor**;
　**endfor**;
　**for** $j = N/t$ **to** $i + 2$ **step** $-1$ **do**
　　*QR-Update*$(A_{j-1\,i}, A_{ji})$;
　　**for** $k = i + 1$ **to** $N/t$ **do**
　　　*QR-Multiply-from-left-2*$(A_{ji}, A_{j-1\,k}, A_{jk})$;
　　**endfor**;
　　**for** $l = 1$ **to** $N/t$ **do**
　　　*QR-Multiply-from-right-2*$(A_{ji}, A_{lj-1}, A_{lj})$;
　　**endfor**;
　**endfor**;
**endfor**;

---

$A_{ji}$ upper triangular. In order to complete the similarity transformation, $A$ has to be postmultiplied by $Q$; that is, every tile $A_{lj}$ has to be post-multiplied by $Q_{ji}$.

The invocation QR-Multiply-from-right-1$(A_{ji}, A_{lj})$ achieves this update of $A_{lj}$ as follows: Bring $A_{ji}$ (that is, the $Y$ matrix) into the main memory, and update horizontal slabs of $A_{lj}$ of width $k$ one at a time from right using the $Y$ matrices, and the $T$ matrices that remain in the memory after the invocation QR-decompose$(A_{ji})$. How $A_{lj}$ is accessed will depend on the number of tiles that fit in the main memory. If only one tile fits, then the row slabs of $A_{lj}$ are read, updated and written one after the other. If two tiles fit, then $A_{lj}$ can be kept in the main memory. So, the total number of I/Os will be at most $2t^2/B$, irrespective of the number of tiles that fit in the main memory.

The number of seeks is $O(t/k)$ for the one tile case, and $O(1)$ for the two and three tile cases.

**Function QR-Multiply-from-right-2**$(A_{ji}, A_{lh}, A_{lj})$

Here $h < j$. Suppose $Q_{hji} \begin{pmatrix} R_{hi} \\ 0 \end{pmatrix}$ is a QR decomposition of $\begin{pmatrix} A_{hi} \\ A_{ji} \end{pmatrix}$. Then, in order to complete the similarity transformation, $Q_{hji}$ has to be applied from the right to the $h$-th and $j$-th block columns of tiles. Let

$$\begin{pmatrix} B_{lh} & B_{lj} \end{pmatrix} = \begin{pmatrix} A_{lh} & A_{lj} \end{pmatrix} Q_{hji}.$$

Say, $Q_{hji}^T$ and $R_{hi}$ have been calculated by an invocation to QR-Update$(A_{hi}, A_{ji})$. At the end of the invocation the $Y$ matrices are stored in $A_{ji}$.

$B_{lh}$ and $B_{lj}$ can be calculated as follows:

Suppose the $Y$ (that is, $A_{ji}$) and $T$ matrices are kept in the main memory. Repeatedly, read the next horizontal slabs from $A_{lh}$ and $A_{lj}$, update them using the $Y$ and $T$ matrices, and write them back. Irrespective of the number of tiles that fit in the main memory, the number of I/Os is $4t^2/B$: $A_{lj}$ and $A_{lh}$ are read and written; $A_{ji}$ is read in only once for the entire column update. In every invocation in Algorithm 3.2, $A_{ji}$ would be upper triangular and therefore the number of additional I/Os per column update would be only $0.5t^2/B$.

The number of seeks is $O(t/k)$ if one or two tiles fit in the main memory; $O(1)$ if three tiles fit in the main memory.

## I/O Complexity

We maintain all tiles in row major order. Note that the multiplications from the left and right are not done the same way. This asymmetry allows us to dispense with all tile transpositions that would have been necessary otherwise.

The I/O complexity of the one tile implementation is:

$$\frac{t^2}{B} \sum_{i=1}^{N/t-1} \left[ \sum_{j=i+1}^{N/t} \left( 2.5 + \sum_{k=i+1}^{N/t} 2.5 + \sum_{l=1}^{N/t} 2 \right) + \sum_{j=i+2}^{N/t} \left( 2.5 + \sum_{k=i+1}^{N/t} 4.5 + \sum_{l=1}^{N/t} 4 \right) \right] = \frac{32N^3}{6tB} + \frac{-61.5N^2 + 14.5Nt - 15t^2}{6B}$$

The I/O complexities of two and three tile implementations are:

$$\frac{t^2}{B} \sum_{i=1}^{N/t-1} \left[ \sum_{j=i+1}^{N/t} \left( 2 + \sum_{k=i+1}^{N/t} 2 + \sum_{l=1}^{N/t} 2 \right) + \sum_{j=i+2}^{N/t} \left( 2.5 + \sum_{k=i+1}^{N/t} 4 + \sum_{l=1}^{N/t} 4 \right) \right] = \frac{5N^3}{tB} + \frac{-10N^2 + 3Nt - 2t^2}{B}$$

Note that here the desired bandwidth $t$ decides how many tiles fit in the memory; the implementation to be chosen is the one that fits as many tiles as possible.

If we were to proceed as in [64], then the right updates (with "QR-Multiply-from-right-1" and "QR-Multiply-from-right-2") in the one and two tile cases would involve reading the $Y$ matrix afresh every time a tile is to be updated. In addition to that, after the completion of each update, the unreduced tiles would need to be transposed, costing adding extra I/Os. Henceforth all right updates in this thesis will be done in our way, and not the one of [64].

In the single tile implementation, a single tile and three and a half narrow panels of size $t \times k$ occupy the main memory; i.e. $t^2 + 7kt/2 \leq M$. In the two tile implementation, two tiles together with two and a half narrow panels of size $t \times k$ occupy the main memory; i.e. $2t^2 + 5kt/2 \leq M$. In the three tile implementation, three tiles together with with a panel and a half of size $t \times k$ occupy the main memory; i.e. $3t^2 + 3kt/2 \leq M$. The half panel will hold the $T$ matrices and the full panel will provide space for computations. The parameter $k$ is usually kept small; $1 \leq k \ll t$.

We also require a number of I/Os to deal with the $T$ matrices. Whether we use the one, two or three tiles approach, we have $t/k$ number of $T$ matrices of size $k \times k$ each. Since the $T$ matrices are upper triangular, that amounts to roughly $tk/2$ elements. There are $N/t - i$ number of single tile QR factorisations and $N/t - i - 1$ number of two tile QR updates for $i = 1, \ldots, N/t - 1$. So I/Os required to handle the $T$ matrices is:

$$\sum_{i=1}^{N/t-1} \frac{kt(N/t - i) + kt(N/t - i - 1)}{2B} = O\left(\frac{N(N - t)k}{tB}\right)$$

**Seek Complexity**

If only one or two tiles fit in the main memory, then the number of seeks needed is $O(\frac{N^2(N-t)}{kt^2})$. If three tiles fit, then it is $O(\frac{N^2(N-t)}{t^3})$.

## 3.3.2   An $O(1)$ factor improvement in I/Os

A small modification to the above algorithm can improve the number of I/Os by a constant factor. Some of the invocations "QR-decompose($A_{ji}$)", $1 \leq i < N/t$ and $i < j \leq N/t$, and their subsequent left and right updates in Algorithm 3.2 can be avoided, as described in [81, 83] for multicore architectures, if instead, we, for $1 \leq i < N/t$, QR factorize the subdiagonal tile $A_{i+1,i}$ and then annihilate the lower subdiagonal tiles by invoking "QR-Update($A_{i+1,i}, A_{ji}$)", for $j > i+1$; here $A_{ji}$ is a full tile instead of an upper triangular tile as in Algorithm 3.2. The other steps of the algorithm remain unchanged. See Algorithm 3.3.

The functions invoked here are the same as those in Algorithm 3.2.

**I/O Complexity**

The I/O complexity of the one tile implementation is:

$$\frac{t^2}{B} \sum_{i=1}^{N/t-1} \left[ 2 + \sum_{k=i+1}^{N/t} 2.5 + \sum_{l=1}^{N/t} 2 + \sum_{j=i+2}^{N/t} \left( 3.5 + \sum_{k=i+1}^{N/t} 5 + \sum_{l=1}^{N/t} 4 \right) \right] = \frac{22N^3}{6tB} + \frac{-35N^2 + 5Nt + 9t^2}{6B}$$

The I/O complexity of the two tile implementation is:

$$\frac{t^2}{B} \sum_{i=1}^{N/t-1} \left[ 2 + \sum_{k=i+1}^{N/t} 2 + \sum_{l=1}^{N/t} 2 + \sum_{j=i+2}^{N/t} \left( 3.5 + \sum_{k=i+1}^{N/t} 4 + \sum_{l=1}^{N/t} 4 \right) \right] = \frac{20N^3}{6tB} + \frac{-31.5N^2 + 2.5Nt + 9t^2}{6B}$$

---

**Algorithm 3.3.** *Banded Hessenberg reduction using the tile approach, an $O(1)$ factor improvement*

**Input:** *A nonsymmetric matrix $A \in \mathbb{R}^{N \times N}$ and a parameter $t$*

**Output:** *A banded Hessenberg matrix of bandwidth $t$*

---

**for** $i = 1$ **to** $N/t - 1$ **do**
    *QR-decompose*$(A_{i+1\,i})$;
    **for** $j = i + 1$ **to** $N/t$ **do**
        *QR-Multiply-from-left-1*$(A_{i+1\,i},\ A_{i+1\,j})$;
    **endfor**;
    **for** $k = 1$ **to** $N/t$ **do**
        *QR-Multiply-from-right-1*$(A_{i+1\,i},\ A_{k\,i+1})$;
    **endfor**;
    **for** $j = i + 2$ **to** $N/t$ **do**
        *QR-Update*$(A_{i+1\,i}, A_{j\,i})$;
        **for** $k = i + 1$ **to** $N/t$ **do**
            *QR-Multiply-from-left-2*$(A_{j\,i}, A_{i+1\,k}, A_{j\,k})$;
        **endfor**;
        **for** $l = 1$ **to** $N/t$ **do**
            *QR-Multiply-from-right-2*$(A_{j\,i}, A_{l\,i+1}, A_{l\,j})$;
        **endfor**;
    **endfor**;
**endfor**;

---

The cost of the three tile case is asymptotically the same as that of the two tile case.

QR-decomposition is invoked only once for each $i = 1, \ldots, N/t - 1$. So the number of I/Os required to handle the $T$ matrices is:

$$\sum_{i=1}^{N/t-1} \frac{kt + kt(N/t - i - 1)}{2B} = O\left(\frac{N(N-t)k}{tB}\right)$$

**Seek Complexity**

If only one or two tiles fit in the main memory, then the number of seeks needed is $O(\frac{N^2(N-t)}{kt^2})$. If three tiles fit, then it is $O(\frac{N^2(N-t)}{t^3})$.

### 3.3.3   When $r > 3$ tiles fit in the main memory

So far, in the algorithms of this section, we assumed a presence of at most three tiles in the main memory. Particularly when $t = o(\sqrt{M})$, the main memory remain unused throughout the reduction limiting the performance. If more tiles would fit in the main memory, we could improve on memory utilisation and thus reduce the number of I/Os.

We now present an algorithm that assumes that $r > 3$ tiles fit in the main memory; that is, $rt^2 \leq M$. See Algorithm 3.4. This is similar to Algorithm 3.3, except in that it QR updates $r/2 - 1$ (instead of one) subdiagonal tiles at a time, and multiplies $r/2$ (instead of two) tiles at a time.

---

**Algorithm 3.4.** *Banded Hessenberg reduction using $r > 3$ tiles in the main memory*

**Input:** *A nonsymmetric matrix $A \in \mathbb{R}^{N \times N}$ and a parameter $t$*

**Output:** *A banded Hessenberg matrix of bandwidth $t$*

---
$s = r/2 - 1;$
**for** $i = 1$ **to** $N/t - 1$ **do**
    *QR-decompose*$(A_{i+1\,i});$
    **for** $j = i + 1$ **to** $N/t$ **do**
        *QR-Multiply-from-left-1*$(A_{i+1\,i}, A_{i+1\,j});$
    **endfor**;
    **for** $k = 1$ **to** $N/t$ **do**
        *QR-Multiply-from-right-1*$(A_{i+1\,i}, A_{k\,i+1});$
    **endfor**;
    **for** $j = 1$ **to** $\lceil (N/t - (i+1))/s \rceil$ **do**
        Let $x_1, \ldots, x_s$ respectively be $i + (j-1)s + 2, \ldots, i + js + 1;$
        *QR-Update-n*$(A_{i+1\,i}, A_{x_1\,i}, \ldots, A_{x_s\,i});$
        **for** $k = i + 1$ **to** $N/t$ **do**
            *QR-Multiply-from-left-n*$(A_{i+1\,k}, A_{x_1\,k}, \ldots, A_{x_s\,k});$
        **endfor**;
        **for** $l = 1$ **to** $N/t$ **do**
            *QR-Multiply-from-right-n*$(A_{l\,i+1}, A_{l\,x_1}, \ldots, A_{l\,x_s});$
        **endfor**;
    **endfor**;
**endfor**;

---

The functions "QR-decompose", "QR-Multiply-from-left-1", and "QR-Multiply-from-right-1" are the same as in Algorithm 3.3.

**Function QR-Update-n**$(A_{i+1\,i}, A_{x_1\,i}, \ldots, A_{x_s\,i})$

The invocation of "QR-Update-n$(A_{i+1\,i}, A_{x_1\,i}, \ldots, A_{x_s\,i})$" annihilates the $s = r/2 - 1$ tiles $A_{x_1\,i}, \ldots, A_{x_s\,i}$ from the current block column $i$ by taking the upper triangular tile $A_{i+1,i}$ as pivot, overwrites the $st \times t$ sized $Y$ matrix in the corresponding $s$ annihilated tiles. The $Y$ and the upper triangular $t \times t$ sized $T$ matrices are kept in memory. So the I/O complexity is $2st^2/B$: read and write $s$ tiles. Seeks: $O(s)$.

**Function QR-Multiply-from-left-n**$(A_{i+1\,k}, A_{x_1\,k}, \ldots, A_{x_s\,k})$

This function updates the $s + 1$ tiles $A_{i+1\,k}, A_{x_1\,k}, \ldots, A_{x_s\,k}$ from the left using the in-core $Y$ and $T$ matrices. So the I/O complexity is $rt^2/B$: read and write $(s + 1)$ tiles. Seeks: $O(s)$.

**QR-Multiply-from-right-n**$(A_{l\,i+1}, A_{l\,x_1}, \ldots, A_{l\,x_s})$

This function updates $s$ tiles $A_{l\,i+1}, A_{l\,x_1}, \ldots, A_{l\,x_s}$ from right using the in-core $Y$ and the $T$ matrices. So the I/O complexity is $rt^2/B$: read and write $(s + 1)$ tiles. Seeks: $O(s)$.

**I/O Complexity**

The I/O complexity of the algorithm is:

$$\frac{t^2}{B} \sum_{i=1}^{N/t-1} \left[ 2 + \sum_{k=i+1}^{N/t} 2 + \sum_{l=1}^{N/t} 2 + \sum_{j=1}^{\lceil \frac{N/t-(i+1)}{s} \rceil} \left( 2s + \sum_{k=i+1}^{N/t} r + \sum_{l=1}^{N/t} r \right) \right] = \frac{5rN^3}{6stB} + \frac{(-15r+24s)N^2 + (10r-24s)}{6sB}$$

When $r$ is large this is $\frac{10N^3}{6tB} + o\left(\frac{N^3}{tB}\right)$

"QR-Update-n$(A_{i+1\,i}, A_{x_1\,i}, \ldots, A_{x_s\,i})$" is invoked $\lceil \frac{N/t-(i+1)}{s} \rceil$ times for $i = 1, \ldots, N/t-1$. So I/Os required to handle the $T$ matrices is:

$$\sum_{i=1}^{N/t-1} \frac{t^2 + t^2 \lceil \frac{N/t-(i+1)}{s} \rceil}{2B} = O\left( \frac{N(N-t)}{rB} \right)$$

As can be seen from the I/O complexity this algorithm does better than the other algorithms discussed in the previous sections. Also it can be observed that if $rt^2 < M$, i.e., the entire block column fits in memory then the algorithm behaves like the slab approach algorithm of Section 3.2.1.

As expected this algorithm reduces the I/O cost by a factor of two over Algorithm 3.3.

**Seek Complexity**

The number of seeks is $O\left( \frac{N(N-t)^2}{t^3} \right)$, as in the three tile implementations seen earlier.

## 3.3.4   An algorithm for $t > \sqrt{M}$

Without loss of generality, assume that $t$ is a multiple of $w = \sqrt{M/3}$. Divide the input matrix $A$ into $w \times w$ tiles; three $w \times w$ tiles can fit in the main memory. Then apply a modified version of Algorithm 3.3 (see Algorithm 3.5). In Algorithm 3.5 the QR-decomposition of the $i$-th block column starts from the tile $A_{i+w\,i}$ instead of $A_{i+1\,i}$. Also the block columns reduced are 1 to $(N-t)/w$, instead of 1 to $N/t-1$.

The functions invoked here are the same as those in Algorithm 3.3.

**I/O Complexity**

Assume that three tiles fit in the main memory.

The I/O complexity of the algorithm is:

$$\frac{w^2}{B} \sum_{i=1}^{\frac{N-t}{w}} \left[ 2 + \sum_{k=i+1}^{N/w} 2 + \sum_{l=1}^{N/w} 2 + \sum_{j=i+1+t/w}^{N/w} \left( 3.5 + \sum_{k=i+1}^{N/w} 4 + \sum_{l=1}^{N/w} 4 \right) \right]$$

---

**Algorithm 3.5.** *Banded Hessenberg reduction for* $t > \sqrt{M}$

---

**Input:** *A nonsymmetric matrix* $A \in \mathbb{R}^{N \times N}$ *and a parameter* $t > \sqrt{M}$
**Output:** *A banded Hessenberg matrix of bandwidth* $t$

---

*Let* $w = \sqrt{M/3}$ *and* $s = t/w$;
**for** $i = 1$ **to** $\frac{N}{w} - s$ **do**
  *QR-decompose*($A_{i+si}$);
  **for** $j = i + 1$ **to** $N/w$ **do**
    *QR-Multiply-from-left-1*($A_{i+si}$, $A_{i+sj}$);
  **endfor**;
  **for** $k = 1$ **to** $N/w$ **do**
    *QR-Multiply-from-right-1*($A_{i+si}$, $A_{ki+s}$);
  **endfor**;
  **for** $j = s + i + 1$ **to** $N/w$ **do**
    *QR-Update*($A_{i+si}, A_{ji}$);
    **for** $k = i + 1$ **to** $N/w$ **do**
      *QR-Multiply-from-left-2*($A_{ji}, A_{i+sk}, A_{jk}$);
    **endfor**;
    **for** $l = 1$ **to** $N/w$ **do**
      *QR-Multiply-from-right-2*($A_{ji}, A_{li+s}, A_{lj}$);
    **endfor**;
  **endfor**;
**endfor**;

---

$$= \frac{(20N^3 - 12N^2 t + 4Nt^2 + 12t^3)}{6wB} + \frac{-19.5N^2 - Nt - 3.5t^2 - 0.5(N-t)w}{6B}$$

Note that when $t = w$, Algorithms 3.3 and 3.5 have the same I/O complexity.

The number of I/Os needed to deal with the $T$ matrices is

$$\sum_{i=1}^{\frac{N-t}{w}} \frac{w^2}{2B} + \sum_{i=1}^{\frac{N-t}{w}-1} \sum_{j=i+1+t/w}^{\frac{N}{w}} \frac{w^2}{2B} = O\left(\frac{N(N-t)}{B}\right)$$

**Seek Complexity**

The number of seeks is: $O\left(\frac{N^3}{M\sqrt{M}}\right)$.

## 3.4 Reduction of a Symmetric Matrix to Symmetric Banded Form using the Slab Approach

Reduction of a symmetric matrix $A \in \mathbb{R}^{N \times N}$ to symmetric banded form with a semi band width $t < N$, is usually the first step of tridiagonal reduction [20–22]. This is usually done by finding an orthogonal transformation $Q$ such that $Q^T A Q$ is in the desired form. Due to symmetry only the upper (or lower) banded portion needs to be stored. Symmetric band reduction has received a lot of attention [10, 11, 55, 60, 119].

### 3.4.1 Reduction with a slab width of $t$

It is typical to reduce the matrix is by assuming that the algorithmic slab size $k$ is same as the desired bandwidth [11, 19, 55, 60]. Algorithms with this assumption have been implemented on parallel machines [8, 18]. An out-of-core algorithm with this assumption in the context of symmetric generalised eigenvalue problem is given in [60].

In this section we discuss the out-of-core algorithm of [60] which proceeds as follows. (See Algorithm 3.6.)

Let the input symmetric matrix $A \in \mathbb{R}^{N \times N}$ be partitioned into tiles of size $t \times t$, and let $X$ be the first slab (block column) of $A$ excluding the diagonal tile.

$$A = \begin{pmatrix} A_{11} & A_{21}^T & \cdots & A_{\frac{N}{t}1}^T \\ A_{21} & A_{22} & \cdots & A_{\frac{N}{t}2}^T \\ \vdots & \vdots & \ddots & \vdots \\ A_{\frac{N}{t}1} & A_{\frac{N}{t}2} & \cdots & A_{\frac{N}{t}\frac{N}{t}} \end{pmatrix} \quad X = \begin{pmatrix} A_{21} \\ A_{31} \\ \vdots \\ A_{N1} \end{pmatrix}$$

QR factorise $X$ and aggregate the $t$ Householder transformations into a matrix $Q$ in $WY$ representation [23].

$$Q = \begin{pmatrix} I_t & 0 \\ 0 & I + WY^T \end{pmatrix}$$

Apply $Q$ on $A$ from both sides. This reduces the first block column and block row of $A$, and modifies all other tiles of $A$:

$$Q^T A Q = \begin{pmatrix} A_{11} & R_{21}^T & 0 & \cdots & 0 \\ R_{21} & \tilde{A}_{22} & \tilde{A}_{32}^T & \cdots & \tilde{A}_{\frac{N}{t}2}^T \\ 0 & \tilde{A}_{32} & \tilde{A}_{33} & \cdots & \tilde{A}_{\frac{N}{t}3}^T \\ \vdots & \vdots & \vdots & \ddots & \vdots \\ 0 & \tilde{A}_{\frac{N}{t}2} & \tilde{A}_{\frac{N}{t}3} & \cdots & \tilde{A}_{\frac{N}{t}\frac{N}{t}} \end{pmatrix}$$

Let $A^{(2)}$ and $\tilde{A}^{(2)}$ be the intersections of block rows and columns numbered from 2 to $N/t$, respectively, before and after the transformation. Then

$$\tilde{A}^{(2)} = (I + WY^T)^T A^{(2)} (I + WY^T)$$

Expanding,

$$\tilde{A}^{(2)} = A^{(2)} + YW^T A^{(2)} + A^{(2)} WY^T + YW^T A^{(2)} WY^T \qquad (3.1)$$

Letting $S = A^{(2)}W$ and $V = W^T A^{(2)} W$ and substituting for $S$ and $V$,

$$\tilde{A}^{(2)} = A^{(2)} + YS^T + SY^T + YVY^T \qquad (3.2)$$

That is,

$$\tilde{A}^{(2)} = A^{(2)} + Y(S^T + (1/2)VY^T) + (S + (1/2)YV)Y^T. \qquad (3.3)$$

Let $T = S + (1/2)YV$, and we have:

$$\tilde{A}^{(2)} = A^{(2)} + YT^T + TY^T$$

Recurse with $\tilde{A}^{(2)}$.

The asymptotic I/O complexity of the algorithm is same as that of the banded Hessenberg reduction described in Section 3.2.1; use $k = t$. Due to symmetry some computations can be avoided. (Note that the use of the $YTY^T$ representation in one the $WY^T$ representation in the other for Householder aggregates do not cause a difference in the I/O complexities.)

---

**Algorithm 3.6.** *Reduction to symmetric banded form [60]*

---

**Input:** *An $N \times N$ symmetric matrix $A$ and a parameter $t$.*
**Output:** *A symmetric banded matrix of bandwidth $t$, orthogonally similar to $A$.*

---

**for** $k = 1, N/t - 1$ **do**
   **for** $j = k + 1, N/t$ **do**
     *Read in $A_{jk}$ and store in $X_j$;*
   **endfor**;
   *Compute the QR factorisation of $X$ accumulating the matrices $W, Y,$ and $R$;*
   *Write $R$ over block $A_{k+1k}$;*
   *Write $W$ and $Y$ to secondary storage;*
   *Compute $S = A^{(k+1)}W$ with $S$ overwriting $W$;  /* Can't overwrite $W$, needed next line */*
   *Compute $V = W^T S$ with $V$ overwriting $R$;*
   *Compute $T = S - 1/2YV$ with $T$ overwriting $W$;*
   **for** $j = k + 1, N/t$ **do**
     **for** $i = j, N/t$ **do**
       *Read in $A_{ij}$;*
       *$A_{ij} = A_{ij} - Y_i T_j^T - T_i Y_j^T$;*
       *Write out $A_{ij}$;*
     **endfor**;
   **endfor**;
**endfor**;

---

### 3.4.2   Reduction with a slab width not necessarily equal to $t$

Algorithm 3.6, chooses the slab width $k = t$. The parallel algorithms of $[11, 119]$ generalise by assuming that $k$ could be different from $t$. The algorithm of $[119]$ is similar to Algorithm 3.1 for banded Hessenberg reduction. The analysis is essentially the same. We observed in our discussions on the banded Hessenberg reduction that $k$ should not exceed $t$. It is applicable here too. The same observation is made in $[10]$.

## 3.5    Reduction to Symmetric Banded Form using the Tile Approach

Suppose the input symmetric matrix $A \in \mathbb{R}^{N \times N}$ has been partitioned into tiles of size $t \times t$. Without loss of generality, assume that $N$ is an integer multiple of $t$. The algorithms of Section 3.3.2 work as they are.

Some improvements in I/O are possible though. When $A$ is to be transformed into orthogonally similar $Q^T A Q$, and $Q^T$, from left, applies to a set $S$ of rows, then $Q$ from the right applies to the corresponding set of columns. Thus, only the elements which are on the intersection of the concerned rows and columns need be updated from the right. So some of the right updates in those algorithms can avoided, thus saving I/O as well as flops.

The algorithms when analysed keeping the above in mind, expectedly, are found to require about one half the number of I/Os and seeks as in the banded Hessenberg reduction. We omit the details.

## 3.6    Reduction to Upper Banded Form using the Slab Approach

The reduction of a general matrix to banded form using orthogonal equivalence transformations is discussed in [61, 82, 83]. Given a matrix $A \in \mathbb{R}^{N \times P}$, $N \geq P$, and a desired bandwidth $t$, our aim is to find orthogonal transformations $U \in \mathbb{R}^{N \times N}$ and $V \in \mathbb{R}^{P \times P}$ such that $B_t = U^T A V$ is in upper banded form of bandwidth $t$; i.e., $B_t(i, j) = 0$ whenever $|i - j| > t$. If $N < P$, we reduce the matrix to a lower banded form. It is easy to generalise these algorithms to get reduction into a matrix with $m$ upper and $t$ lower diagonals; we will not discuss this further. Also, we recall only the reduction to upper banded form ($N \geq P$); the reduction to lower banded form ($N < P$) is analogous.

In this section, we discuss Householder based upper banded reduction ($N \geq P$) using the slab approach. Two factorisation methods, QR and LQ, are interleaved from the left and right respectively to reduce the matrix to banded form. A slab based parallel algorithm for the problem was given in [61]; an external memory implementation of that is discussed here. See Algorithm 3.7 for a description.

In the algorithm the notations $A_s^{(RF)}$ and $A_s^{(LF)}$ stand for the submatrices which have to be QR and LQ factorised respectively in the $s$-th iteration. Similarly the notations $A_s^{(RU)}$ and $A_s^{(LU)}$ stand for the submatrices that have to be updated by orthogonal matrices $U$ and $V$ from the left and right respectively in the $s$-th iteration. The computation proceeds as follow:

Let the remaining unreduced portion $\tilde{A}$ of $A$ after $(s-1)$ iterations be represented as $(A_s^{(RF)}, A_s^{(RU)})$, where $A_s^{(RF)} \in \mathbb{R}^{[N-(s-1)t] \times t}$ and $A_s^{(RU)} \in \mathbb{R}^{[N-(s-1)t] \times (P-st)}$ are submatrices. Then QR factorise $A_s^{(RF)}$ into $Q_s R_s$ using the slab based algorithm, and aggregate the $t$ Householders into a matrix $Q_s$ using, say, the $I + YTY^T$ representation [23]. Apply $Q_s^T$ from the left:

$$Q_s^T \tilde{A} = \left[ \begin{array}{cc} R_s & Q_s^T A_s^{(RU)} \end{array} \right]$$

Now the $s$-th block column is reduced. We next reduce the first block row of $Q_s^T A_s^{(RU)}$ to banded form using LQ factorisation. Again partition the updated $\tilde{A}$ as follows:

$$\tilde{A} = \begin{array}{c} \\ \left[ \begin{array}{cc} \overset{t}{R'_s} & \overset{P-st}{A_s^{(LF)}} \\ 0 & A_s^{(LU)} \end{array} \right] \begin{array}{c} t \\ N-st \end{array}$$

Now LQ factorise $A_s^{(LF)} = L_s P_s^T$. Apply $P_s$ from right,

$$\tilde{A} P_s = \begin{array}{c} \\ \left[ \begin{array}{cc} \overset{t}{R'_s} & \overset{P-st}{L_s} \\ 0 & A_s^{(LU)} P_s \end{array} \right] \begin{array}{c} t \\ N-st \end{array}$$

Now the $s$-th block row is reduced to upper banded form. Continue this another $P/t - 1 - s$ times to reduce the matrix to upper bidiagonal form.

---

**Algorithm 3.7.** *Reduction to banded form [61]*

---

**Input:** *An $N \times P$ ($N \geq P$) matrix and a parameter $t$.*
**Output:** *An orthogonally equivalent upper banded matrix of bandwidth $t$.*

---

$i = 0$;
**while** $i < P$ **do**
    $s = i/t + 1$;
    *Define $A_s^{(RF)} \equiv A[i+1:N; \; i+1:i+t]$;*
    *Define $A_s^{(RU)} \equiv A[i+1:N; \; i+t+1:P]$;*
    $A_s^{(RF)} = Q_s R_s$; /*QR decomposition */
    $A_s^{(RU)} \leftarrow Q_s^T A_s^{(RU)}$; /*update*/
    **if** $i + t < p$ **do**
        *Define $A_s^{(LF)} \equiv A[i+1:i+t; \; i+t+1:P]$;*
        *Define $A_s^{(LU)} \equiv A[i+t+1:N; \; i+t+1:P]$;*
        $A_s^{(LF)} = L_s P_s^T$; /*LQ decomposition */
        $A_s^{(LU)} \leftarrow A_s^{(LU)} P_s$; /*update*/
    **endif**;
    $i = i + t$;
**endwhile**;

---

The LQ factorisations are analogous to the QR factorisations, except that the Householder vectors are stored in row major order in the zeroed upper triangular portion of the matrix. The left updates are analogous to the ones in QR factorisation. (See Chapter 2.) The right updates are similar to the ones in banded Hessenberg reduction but use orthogonal transformations different from the left side. Therefore the I/O complexity of the reduction is dominated by the I/O complexity of QR decomposition. The details of the analysis are hence omitted.

## 3.7   Reduction to Upper Banded Form using the Tile Approach

In this section, we discuss band reduction using the tile approach. Parallel algorithms using the tile approach were proposed in [17,61]. Parallel reductions using the tile approach have been proposed for multicore architectures too [82, 83].

## 3.7.1   An algorithm for $t < \sqrt{M}$

Suppose the $N \times P$, $N \geq P$ matrix has been partitioned into tiles of size $t \times t$. Without loss of generality, assume that $N$ and $P$ are integer multiples of $t$ and denote, $X = N/t$ and $Y = P/t$, that is the matrix is partitioned into $X \times Y$ tiles. The reduction is described in Algorithm 3.8.

---

**Algorithm 3.8.** *Reduction to banded form [82, 83]*

---

**Input:** *An $N \times P$ ($N \geq P$) matrix and a parameter t.*
**Output:** *An orthogonally equivalent upper banded matrix with bandwidth t.*

---

*Let $X = N/t$ and $Y = P/t$;*
**for** $i = 1$ *to Y* **do**
   *QR-decompose($A_{ii}$ ); /\* QR factorisation\*/*
     **for** $j = i + 1$ *to Y* **do**
       *QR-Multiply-from-left-1($A_{ii}, A_{ij}$);*
     **endfor***;*
     **for** $k = i + 1$ *to X* **do**
       *QR-Update($A_{ii}, A_{ki}$);*
       **for** $j = i + 1$ *to Y* **do**
         *QR-Multiply-from-left-2($A_{ki}, A_{ij}, A_{kj}$);*
       **endfor***;*
     **endfor***;*
  **if** $i < Y$ **then**
   *LQ-decompose($A_{ii+1}$ ); /\* LQ factorisation\*/*
   **for** $j = i + 1$ *to X* **do**
     *LQ-Multiply-from-right-1($A_{ii+1}, A_{ji+1}$);*
   **endfor***;*
   **for** $k = i + 2$ *to Y* **do**
     *LQ-Update($A_{ii+1}, A_{ik}$);*
     **for** $j = i + 1$ *to X* **do**
       *LQ-Multiply-from-right-2($A_{ik}, A_{ji+1}, A_{jk}$);*
     **endfor***;*
   **endfor***;*
  **endif***;*
**endfor***;*

---

The functions "QR-decompose", "QR-Update", "QR-Multiply-from-left-1", and "QR-Multiply-from-left-2" invoked here in Algorithm 3.8, are the same as those in Algorithm 3.3.

**Function LQ-decompose**$(A_{i\,i+1})$

Assume that the tile $A_{i\,i+1}$ is divided into horizontal slabs of width $k$ each. Let $r = t/k$. The invocation LQ-decompose$(A_{i\,i+1})$ finds matrices $Y_1, \ldots, Y_r, T_1, \ldots, T_r$ such that $L_{i\,i+1} = A_{i\,i+1}(I_t + Y_1 T_1^T Y_1^T) \ldots (I_t + Y_r T_r^T Y_r^T)$ is lower triangular. This is done as follows:

Read $A_{i\,i+1}$ into the main memory. Proceed in $r$ iterations. At the beginning of the $s$-th iteration, the superdiagonal elements of the first $(s-1)k$ rows of $A_{i\,i+1}$ are all zero. In the $s$-th iteration, calculate $Y_s$ and $T_s$ such that the superdiagonal elements of $A_{i\,i+1}(I_t + Y_s T_s^T Y_s^T)$ in the first $sk$ rows are all zero. Update $A_{i\,i+1}$ to $A_{i\,i+1}(I_t + Y_s T_s^T Y_s^T)$. Store $Y_s$ in the superdiagonal part of the $s$-th slab of $A_{i\,i+1}$.

When all slabs are processed, arrange the lower triangular part of $A_{i\,i+1}$ in column major order, and write $A_{i\,i+1}$ back to the disk.

The number of I/Os needed is $2t^2/B$; $A_{i\,i+1}$ is read and written; the $T$ matrices are kept in the main memory.

The number of seeks is $O(1)$ even for the one tile case.

**Function LQ-Multiply-from-right-1**$(A_{i\,i+1}, A_{j\,i+1})$

If $A_{i\,i+1}$ LQ-decomposes into $L_{i\,i+1}Q_{i\,i+1}$, then a post-multiplication of $A$ by an orthogonal matrix $Q$ obtained by replacing the $(i+1)$-st diagonal tile of $I_N$ by $Q_{i\,i+1}$ has the effect of making $A_{i\,i+1}$ lower triangular. In order to complete the orthogonal equivalence transformation, $A$ has to be postmultiplied by $Q$; that is, every tile $A_{j\,i+1}$ has to be post-multiplied by $Q_{i\,i+1}$.

The invocation LQ-Multiply-from-right-1($A_{i\,i+1}$, $A_{j\,i+1}$) achieves this update of $A_{ji+1}$ as follows: Bring $A_{i\,i+1}$ (that is, the $Y$ matrix) into the main memory, and update horizontal slabs of $A_{j\,i+1}$ of width $k$ one at a time from right using the $Y$ matrices, and the $T$ matrices that remain in the memory after the invocation LQ-decompose($A_{i\,i+1}$). How $A_{ji+1}$ is accessed will depend on the number of tiles that fit in the main memory. If only one tile fits, then the row slabs of $A_{j\,i+1}$ are read, updated and written one after the other. If two tiles fit, then $A_{j\,i+1}$ can be kept in the main memory. So, the total number of I/Os will be at most $2t^2/B$, irrespective of the number of tiles that fit in the main memory.

The number of seeks is $O(t/k)$ for the one tile case, and $O(1)$ for the two and three tile cases.

**Function LQ-Update**($A_{i\,h}$, $A_{ik}$)

Here $A_{i\,h}$ is lower triangular. The invocation LQ-Update($A_{i\,h}$, $A_{ik}$)), finds a matrix $Q$ such that every element of the $(i,k)$-th tile of $AQ^T$ is zero. This is done over $r = t/k$ iterations. The $s$-th iteration processes $A_{i\,h}^{(s)}$, the $s$-th leftmost vertical slab of width $k$ of $A_{i\,h}$. Consider the matrix $C_s = \left(\begin{array}{cc} A_{i\,h}^{(s)} & A_{ik} \end{array}\right)$.

Let $\left(\begin{array}{c} I_k \\ Y_s \end{array}\right)$ and $T_s$ be matrices such that $C_s \cdot \left( I_{k+t} + \left(\begin{array}{c} I_k \\ Y_s \end{array}\right) T_s^T \left(\begin{array}{cc} I_k & Y_s^T \end{array}\right) \right)$ has zeroes in the rightmost $t$ columns of its $s$-th horizontal slab.

Then, $\left(\begin{array}{cc} A_{i\,h} & A_{ik} \end{array}\right) \cdot \left( I_{2t} + \left(\begin{array}{c} 0_{(s-1)k} \\ I_k \\ 0_{t-sk} \\ Y_s \end{array}\right) T_s^T \left(\begin{array}{cccc} 0_{(s-1)k}^T & I_k & 0_{t-sk}^T & Y_s^T \end{array}\right) \right)$ has zeroes in the upper triangular part of its $s$-th horizontal slab. Set $\left(\begin{array}{c} A_{ih}A_{ik} \end{array}\right)$ to this matrix by updating $A_{i\,h}^{(s)}$ and $A_{ik}$. The $s$-th slab $A_{ik}$ is now all zero. Therefore $Y_s$ can be stored there.

Let $Q_{ik}^T$ be the product

$$\left( I_{2t} + \left(\begin{array}{c} 0_{t-k} \\ I_k \\ Y_1 \end{array}\right) T_1^T \left(\begin{array}{ccc} 0_{t-k} & I_k & Y_1^T \end{array}\right) \right) \ldots \left( I_{2t} + \left(\begin{array}{c} I_k \\ 0_{t-k} \\ Y_r \end{array}\right) T_r^T \left(\begin{array}{ccc} I_k & 0_{t-k} & Y_r^T \end{array}\right) \right)$$

The above can be implemented as follows: Read $A_{ik}$. Read $A_{ih}$ one vertical slab at a time. For each slab, update the slab, write it back and overwrite the corresponding horizontal slab of $A_{ik}$ with the $Y$ matrix. The number of I/Os is at most $3t^2/B$, irrespective of whether one tile or two tiles fit in the main memory: one tile and a half are both read and written.

The number of seeks is $O(t/k)$ for the one tile case, and $O(1)$ for the two and three tile cases.

**Function LQ-Multiply-from-right-2**$(A_{ik}, A_{j\,i+1}, A_{jk})$

Here $i + 1 < k$. Suppose $\begin{pmatrix} L_{i\,i+1} & 0 \end{pmatrix}.Q_{i\,i+1\,k}$ is an LQ decomposition of $\begin{pmatrix} A_{i\,i+1} & A_{ik} \end{pmatrix}$. Then, in order to complete the orthogonal equivalence transformations, $Q_{i\,i+1\,k}$ has to be applied from the right to the $(i+1)$-st and $k$-th block columns of tiles. Let

$$\begin{pmatrix} B_{j\,i+1} & B_{jk} \end{pmatrix} = \begin{pmatrix} A_{j\,i+1} & A_{jk} \end{pmatrix} Q^T_{i\,i+1\,k}.$$

Say, $Q_{i\,i+1\,k}$ and $L_{i\,i+1}$ have been calculated by an invocation to LQ-Update$(A_{i\,i+1}, A_{ik})$. At the end of the invocation the $Y$ matrices are stored in $A_{ik}$.

$B_{j\,i+1}$ and $B_{jk}$ can be calculated as follows:

Suppose the $Y$ (that is, $A_{ik}$) and $T$ matrices are kept in the main memory. Repeatedly, read the next horizontal slabs from $A_{j\,i+1}$ and $A_{jk}$, update them using the $Y$ and $T$ matrices, and write them back. Irrespective of the number of tiles that fit in the main memory, the number of I/Os is $4t^2/B$: $A_{j\,i+1}$ and $A_{jk}$ are read and written; $A_{ik}$ is read in only once for the entire column update. In every invocation in Algorithm 3.8, the number of additional I/Os per column update would be only $0.5t^2/B$: to read $A_{ik}$.

The number of seeks is $O(t/k)$ if one or two tiles fit in the main memory; $O(1)$ if three tiles fit in the main memory.

**I/O Complexity**

The I/O complexity for the one tile implementation is:

$$\frac{t^2}{B}\left[\sum_{i=1}^{P/t}\left(2+\sum_{k=i+1}^{P/t}2.5+\sum_{k=i+1}^{N/t}\left(3+\sum_{j=i+1}^{P/t}5\right)\right)+\sum_{i=1}^{P/t-1}\left(2+\sum_{l=i+1}^{N/t}2+\sum_{k=i+2}^{P/t}\left(3+\sum_{j=i+1}^{N/t}4\right)\right)\right]$$
$$=\frac{9NP^2}{2tB}-\frac{3P^3}{2tB}+o\left(\frac{NP^2}{tB}\right)$$

The I/O complexities for the two and three tile implementations are:

$$\frac{t^2}{B}\left[\sum_{i=1}^{P/t}\left(2+\sum_{k=i+1}^{P/t}2+\sum_{k=i+1}^{N/t}\left(3+\sum_{j=i+1}^{P/t}4\right)\right)+\sum_{i=1}^{P/t-1}\left(2+\sum_{l=i+1}^{N/t}2+\sum_{k=i+2}^{P/t}\left(3+\sum_{j=i+1}^{N/t}4\right)\right)\right]$$
$$=\frac{4NP^2}{tB}-\frac{4P^3}{3tB}+o\left(\frac{NP^2}{tB}\right)$$

The number of I/Os to deal with the $T$ matrices is asymptotically the same as that of the Algorithm 3.3.

**Seek Complexity**

If one tile or two tiles are available then number of seeks is $O\left(\frac{NP^2}{kt^2}\right)$. But if three tiles are available then it is $O\left(\frac{NP^2}{t^3}\right)$.

### 3.7.2 When $r > 3$ tiles fit in the main memory

Here we propose an algorithm that keeps $r$ tiles in the main memory at a time, where $rt^2 \leq M$. This is similar to Algorithm 3.4 and is described in Algorithm 3.9.

The functions "QR-decompose", "QR-Multiply-from-left-1", and "QR-Multiply-from-left-n" invoked here are the same as in Algorithm 3.4. Similarly the functions "LQ-decompose" and "LQ-Multiply-from-right-1" are the same as in Algorithm 3.8.

---

**Algorithm 3.9.** *Reduction to banded form when $r > 3$ tiles fit in the main memory*

**Input:** *An $N \times P$ ($N \geq P$) matrix $A$ and a parameter $t < \sqrt{M}$.*
**Output:** *An upper banded matrix orthogonally equivalent to $A$.*

---

*Let $s = r/2 - 1$;*
*Let $X = N/t$ and $Y = P/t$;*
**for** $i = 1$ *to* $Y$ **do**
  *QR-decompose($A_{ii}$);*
  **for** $j = i + 1$ *to* $Y$ **do**
    *QR-Multiply-from-left-1($A_{ii}$, $A_{ij}$);*
  **endfor**;
  **for** $k = 1$ *to* $\lceil (X - i)/s \rceil$ **do**
    *Let $x_1, \ldots, x_s$ respectively be $i + (k-1)s + 1$, $\ldots$, $i + ks$;*
    *QR-Update-n($A_{ii}, A_{x_1 i}, \ldots, A_{x_s i}$);*
    **for** $j = i + 1$ *to* $Y$ **do**
      *QR-Multiply-from-left-n($A_{ij}, A_{x_1 j}, \ldots, A_{x_s j}$);*
    **endfor**;
  **endfor**;
  **if** $i < Y$ **then**
    *LQ-decompose($A_{i\,i+1}$); /* LQ factorisation*/*
    **for** $j = i + 1$ *to* $X$ **do**
      *LQ-Multiply-from-right-1($A_{i\,i+1}$, $A_{j\,i+1}$);*
    **endfor**;
    **for** $k = 1$ *to* $\lceil (Y - i - 1)/s \rceil$ **do**
    *Let $y_1, \ldots, y_s$ respectively be $i + (k-1)s + 2, \ldots, i + ks + 1$;*
    *LQ-Update-n($A_{i\,i+1}, A_{i\,y_1}, \ldots, A_{i\,y_s}$);*
      **for** $j = i + 1$ *to* $X$ **do**
        *LQ-Multiply-from-right-n($A_{j\,i+1}, A_{j\,y_1}, \ldots, A_{j\,y_s}$);*
      **endfor**;
    **endfor**;
  **endif**;
**endfor**;

---

**Function LQ-Update-n**($A_{i\,i+1}, A_{i\,y_1}, \ldots, A_{i\,y_s}$)

The invocation of "LQ-Update-n($A_{i\,i+1}, A_{i\,y_1}, \ldots, A_{i\,y_s}$)" annihilates the $s = r/2 - 1$ super diagonal tiles $A_{i\,y_1}, \ldots, A_{i\,y_s}$ from the current block row $i$ by taking the lower triangular tile $A_{i\,i+1}$ as pivot, overwrites the $st \times t$ sized $Y$ matrix in the corresponding $s$ annihilated tiles. The $Y$ and the upper triangular $t \times t$ sized $T$ matrices are kept in memory. So the I/O complexity is $2st^2/B$: read and write $s$ tiles. Seeks: $O(s)$.

**LQ-Multiply-from-right-n**$(A_{j\,i+1}, A_{j\,y_1}, \ldots, A_{j\,y_s})$

This function updates $s$ tiles $(A_{j\,i+1}, A_{j\,y_1}, \ldots, A_{j\,y_s})$ from right using the in-core $Y$ and the $T$ matrices. So the I/O complexity is $rt^2/B$: read and write $(s+1)$ tiles. Seeks: $O(s)$.

### I/O Complexity

The I/O complexity of the Algorithm 3.9 is:

$$\frac{t^2}{B}\left(\sum_{i=1}^{P/t}\left(2+\sum_{k=i+1}^{P/t}2+\sum_{k=1}^{\left\lceil\frac{N/t-i}{s}\right\rceil}\left(2s+\sum_{j=i+1}^{P/t}r\right)\right)+\sum_{i=1}^{P/t-1}\left(2+\sum_{l=i+1}^{N/t}2+\sum_{k=1}^{\left\lceil\frac{P/t-i-1}{s}\right\rceil}\left(2s+\sum_{j=i+1}^{N/t}r\right)\right)\right)$$
$$=\frac{2NP^2}{tB}-\frac{2P^3}{3tB}+o\left(\frac{NP^2}{tB}\right)$$

The I/Os to deal with the $T$ matrices is analogous to Algorithm 3.4 and is $O(NP/rB)$.

### Seek Complexity

Since all the $Y, T$ and the tiles to be updates are brought into memory by full tiles instead of narrow panels, so the number of seeks is $O\left(\frac{NP^2}{t^3}\right)$.

## 3.7.3   Reduction to upper banded form for tile size $t > \sqrt{M}$

The techniques used in Section 3.3.4 also can be employed here to improve the I/O. See Algorithm 3.10 for details. All the functions invoked here are the same as those in Algorithm 3.8.

### I/O Complexity

The I/O complexity of the algorithm is:

$$\frac{w^2}{B}\left[\sum_{i=1}^{P/w}\left(2+\sum_{k=i+1}^{P/w}2+\sum_{k=i+1}^{N/w}\left(3+\sum_{j=i+1}^{P/w}4\right)\right)+\sum_{i=1}^{P/w-s}\left(2+\sum_{l=i+1}^{N/w}2+\sum_{k=i+s+1}^{P/w}\left(3+\sum_{j=i+1}^{N/w}4\right)\right)\right]$$
$$=\frac{4NP^2}{wB}-\frac{4P^3}{3wB}+o\left(\frac{NP^2}{wB}\right)$$

---

**Algorithm 3.10.** *Reduction to banded form for $t > \sqrt{M}$*

---

**Input:** *An $N \times P$ ($N \geq P$) matrix $A$ and a parameter $t > \sqrt{M}$.*
**Output:** *An upper banded matrix orthogonally equivalent to $A$.*

---

*Let $w = \sqrt{M/3}$ and $s = t/w$;*
**for** $i = 1$ *to* $\frac{P}{w}$ **do**
  *QR-decompose($A_{ii}$);*
  **for** $l = i + 1$ *to* $P/w$ **do**
    *QR-Multiply-from-left-1($A_{ii}$, $A_{il}$);*
  **endfor**;
  **for** $j = i + 1$ *to* $N/w$ **do**
    *QR-Update($A_{ii}$, $A_{ji}$);*
    **for** $k = i + 1$ *to* $P/w$ **do**
      *QR-Multiply-from-left-2($A_{ji}$, $A_{ik}$, $A_{jk}$);*
    **endfor**;
  **endfor**;
  **if** $i \leq P/w - s$ **do**
    *LQ-decompose($A_{ii+s}$);*
    **for** $j = i + 1$ *to* $N/w$ **do**
      *LQ-Multiply-from-right-1($A_{ii+s}$, $A_{ji+s}$);*
    **endfor**;
    **for** $k = i + s + 1$ *to* $P/w$ **do**
      *LQ-update($A_{ii+s}$, $A_{ik}$);*
      **for** $l = i + 1$ *to* $N/w$ **do**
        *LQ-Multiply-from-right-2($A_{ik}$, $A_{li+s}$, $A_{lk}$);*
      **endfor**;
    **endfor**;
  **endif**;
**endfor**;

---

The I/Os to deal with the $T$ matrices is as follows:

$$\frac{w^2}{B} \sum_{i=1}^{\frac{P}{w}} \left( 1 + \sum_{l=i+1}^{N/w} 2 \right) + \frac{w^2}{B} \sum_{i=1}^{(P-t)/w} \left( 1 + \sum_{k=i+1+t/w}^{P/w} 1 \right) = O(NP/B)$$

**Seek Complexity**

The number of seeks is: $O\left( \frac{NP^2}{M\sqrt{M}} \right)$.

# Chapter 4

# Reduction from Banded Hessenberg Form to Hessenberg Form and Related Problems

## 4.1 Introduction

We briefly recapitulate from the Introduction to Chapter 3: The eigenvalues and singular values are typically computed in a two stage process. In the first stage the input matrix is reduced, through a sequence of OSTs or OETs to a condensed form and in the second stage an iterative method called the $QR$ algorithm is applied to the condensed form. All known algorithms for the first stage need $O(N^3/B)$ I/Os and $O(N^3)$ flops. Therefore, the first stage is usually the performance bottleneck. It has been proposed that the reduction in the first stage be split into two steps [17,19,21,77,80], the first reducing the matrix to a banded Hessenberg, symmetric banded or nonsymmetric banded form, as is relevant [11,61,81], and the second further reducing it to Hessenberg, tridiagonal or bidiagonal form [21,77,78,80].

In Chapter 3 we discussed the first step of the first stage. In this chapter we study the second step.

Unblocked Hessenberg reductions were known even in 1960s [85]. Blocked algorithms that try to incorporate M-M operations have been invented since then [45–47,89]. As we shall show, these algorithms run in $O(N^3/B)$ I/Os on full matrices. If the input matrix is in banded Hessenberg form, these algorithms are not suitable, as they don't exploit the sparsity in the input. After a few steps of these algorithms, the matrix becomes full with the fill-ins created, resulting $O(N^3)$ flops [21, 77, 78] and $O(N^3/B)$ I/Os. To the best of our knowledge, no efficient algorithm has been proposed to reduce a banded Hessenberg matrix to Hessenberg form.

For the tridiagonal reduction, if the input matrix is full, the conventional Householder or rotation based unblocked algorithms are useful [7, 32, 58, 115]. Blocked algorithms are also known [45]. Parallel algorithms are described in [46, 66]. These and their straight forward adaptations all run in $O(N^3)$ flops and $O(N^3/B)$ I/Os. Again, if the matrix is in symmetric banded form, a few steps of these algorithms will fill the matrix with bulges. Sequential and parallel algorithms for the tridiagonalisation of symmetric banded matrices have been known [20–22, 55, 72, 77, 95, 97].

For the bidiagonal reduction, if the input matrix $A \in \mathbb{R}^{N \times P}$ is full, the conventional algorithms use Householders to reduce columns and rows, one by one [58, 115]. A Householder based blocked algorithm that performs about half of its work in M-M is described in [45]. A bidiagonalisation using one-sided orthogonal transformations was proposed in [92]; its stability was later improved in [14, 24]. These algorithms have an I/O complexity of $O(NP\min(N, P)/B)$. Again, if the input matrix is banded, a few steps of these algorithms will fill the matrix with bulges, and so these are not suitable for such inputs. Sequential and parallel algorithms for reducing banded matrices to bidiagonal form were proposed in [78].

We now present an overview of the results in this chapter.

We recall an unblocked and a blocked direct Hessenberg reduction algorithm, and show that they both have an I/O complexity of $O(N^3/B)$. For the BH-H reduction we propose two algorithms. The first one requires $O(N^3/B)$ I/Os, but the second one, an extension of the first using tight packing of bulges requires only $O\left(\frac{t^2N^2}{B} + \frac{N^3}{tB}\right)$ I/Os when $t$, the bandwidth, is at most $\sqrt{M}$. Combining this with the results of Chapter 3, we show that the Hessenberg reduction can be performed in $O(N^3/\sqrt{M}B)$ I/Os, when $N$ is sufficiently large. This matches the known lower bound on data movement for the problem [13].

We recall unblocked and blocked tridiagonal reductions, and show that they have an I/O complexity of $O(N^3/B)$. For the symmetric banded form to tridiagonal form reduction, we show, the known algorithms take $O(N^2t/B)$ I/Os. Combining this with the results of Chapter 3, we show that the tridiagonal reduction can be performed in $O(N^3/\sqrt{M}B)$ I/Os, when $N$ is sufficiently large. This matches the known lower bound on the data movement for the problem [13].

We recall unblocked and a blocked bidiagonal reductions, and show that they have an I/O complexity of $O(NP\min\{N,P\}/B)$, when applied on an $N \times P$ matrix. For the banded to bidiagonal reduction, we show, the known algorithms take $O(\min\{N,P\}^2t)/B)$ I/Os, when $t$ is the bandwidth. Combining this with the results of Chapter 3, we show that the bidiagonal reduction can be performed in $O\left(\frac{\min\{N,P\}.NP}{\sqrt{M}B}\right)$ I/Os when the matrix is sufficiently large. This matches the known lower bound on the data movent for the problem [13].

## 4.1.1   Organisation of this Chapter

In Section 2, we study some of the direct Hessenberg/tridiagonal/bidiagonal reductions, and their I/O performances. In Section 3, we present our new BH-H reductions. Tridiagonal reductions are discussed in Section 4, and Section 5 is a study of bidiagonal reductions.

## 4.2    Preliminaries

The reductions of matrices to Hessenberg or tridiagonal form (the latter if it is symmetric, the former otherwise) or bidiagonal form have traditionally been performed without constructing an intermediate banded form. We study such direct reductions in this section.

### 4.2.1    Direct Hessenberg reduction

For the reduction of a nonsymmetric matrix to Hessenberg form, unblocked and blocked algorithms have been known.

#### 4.2.1.1    Unblocked direct Hessenberg reduction

Let $A \in \mathbb{R}^{N \times N}$ be a nonsymmetric matrix. Find Householder transformations $Q_1, Q_2, \ldots,$ $Q_{N-2}$ such that each $Q_i$ is designed to introduce zeroes below the first subdiagonal in the $i$-th column of the matrix $Q_{i-1}^T \cdots Q_1^T A Q_1 \cdots Q_{i-1}$ [32, 45, 46, 58, 115].

Then $H = Q_{N-2}^T \cdots Q_1^T A Q_1 \cdots Q_{N-2}$ is a Hessenberg matrix.

For a Householder $Q = (I - uu^T)$, where $\|u\|_2 = 1/\sqrt{2}$, $Q^T A Q = (I - uu^T)^T A (I - uu^T) = A - uu^T A - Auu^T + uu^T Auu^T = A - uv^T - (Au - uv^T u)u^T = A - uv^T - wu^T$, where $v^T = u^T A$ and $w = Au - uv^T u$. Hence Algorithm 4.1.

---

**Algorithm 4.1.**  *Unblocked direct Hessenberg reduction*

---

**Input:** *An $N \times N$ matrix $A$*

**Output:** *$A$ is overwritten with a Hessenberg matrix similar to $A$*

---

**for** $k = 1, \cdots, N - 2$ **do**
   *Find a Householder vector $u$, $\|u\|_2 = 1/\sqrt{2}$ to zero the elements*
   *below the subdiagonal in the $k$-th column of $A$;*
   *$v^T = u^T A$;*
   *$w = Au - (v^T u)u$;*
   *update $A = A - uv^T - wu^T$;*
**endfor**

---

As shown in [89], the total flops required by the above algorithm is $\sum_{k=1}^{N-2} 4N(N-k) +$ $\sum_{k=1}^{N-2} 4(N-k)^2 \approx 10N^3/3$. The computations of $u^T A$ and $Au - (v^T u)u$ require $A$ in column major order and row major order respectively. Therefore, if $N > B$, the algorithm, as it is presented, needs $O(N^3)$ I/Os. However, a matrix transposition in each iteration can bring the cost down to $O(N^3/B)$ I/Os.

To access each column/row if the matrix is in row/column major order takes $N$ seeks. So the seek complexity of the algorithm is $O(N^3)$ if the algorithm is implemented as it is presented. But with a matrix transposition in each iteration, the seek complexity would be $O(N^3/\sqrt{M} + N^2)$, which is $O(N^3/\sqrt{M})$ as $N > \sqrt{M}$.

The main drawback of this algorithm is that all its operations are V-M or V-V. Blocking has been shown to improve on this I/O performance [32, 45, 46].

### 4.2.1.2 Blocked direct Hessenberg reduction

A blocked version of the above algorithm has been suggested [32, 45, 46] using aggregated Householder transformations [23, 96]. We describe and analyse this algorithm now.

Let $A$ be the input matrix, and let $A_1 = A$. Find a Householder vector $u_1$ with $\|u_1\|_2 = 1/\sqrt{2}$ to reduce the first column of $A_1$. Let $v_1 = A_1^T u$ and $w_1 = A_1 u_1 - u_1 u_1^T A_1 u_1$.

Initialise $U = [u_1]$, $V = [v_1]$, and $W = [w_1]$.

Then $A_2 = Q_{u_1}^T A_1 Q_{u_1} = A_1 - u_1 v_1^T - w_1 u_1^T = A_1 - UV^T - WU^T$.

The second column of $A_2$ can be computed like this: $a_2 = A_1[*, 2] - UV^T[*, 2] - WU^T[*, 2]$. Find Householder vector $u_2$ of euclidian norm $1/\sqrt{2}$ for reducing $a_2$. Let $v_2 = A_2^T u_2 = (A_1 - UV^T - WU^T)^T u_2$ and $w_2 = A_2 u_2 - u_2 v_2^T u_2 = (A_1 - UV^T - WU^T)u_2 - u_2 v_2^T u_2$. Then $A_3 = Q_{u_2}^T A_2 Q_{u_2} = A_2 - u_2 v_2^T - w_2 u_2^T = A_1 - UV^T - WU^T - u_2 v_2^T - w_2 u_2^T = A_1 - \begin{pmatrix} U & u_2 \end{pmatrix} \begin{pmatrix} V^T \\ v_2^T \end{pmatrix} - \begin{pmatrix} W & w_2 \end{pmatrix} \begin{pmatrix} U^T \\ u_2^T \end{pmatrix}$. Set $U = \begin{pmatrix} U & u_2 \end{pmatrix}$, $V = \begin{pmatrix} V & v_2 \end{pmatrix}$ and $W = \begin{pmatrix} W & w_2 \end{pmatrix}$.

That is, $A_3$ can be computed without explicitly computing $A_2$; $a_2$ is enough. If we continue like this for $k$ columns, the aggregates $U$, $V$ and $W$ can be used to update the rest of the matrix as in:

$$A_k = A_1 - UV^T - WU^T \tag{4.1}$$

We would have computed $A_k$ without explicitly computing $A_2, \ldots, A_{k-1}$. See Algorithm 4.2 below:

---

**Algorithm 4.2.** *Slab based direct Hessenberg reduction [46]*

---

**Input:** *Matrix $A$ of size $N \times N$.*
**Output:** *$A$ is overwritten with a similar Hessenberg matrix.*

---

*Let $\hat{A} = A$*
**for** $i = 1$ **to** $N/k$ **do**
   *Find a Householder vector $u$ with $\|u\|_2 = 1/\sqrt{2}$ to reduce*
     $A[(i-1)k+2:N; \ (i-1)k+1]$ *to form $(*, 0, \ldots, 0)^T$;*
   *Compute $v = A^T u$ and $w = Au - u(v^T u)$;*
   *Initialise $U = [u]$, $V = [v]$, $W = [w]$;*
   **for** $j = 2, \cdots, k$ **do**
     *Let $a_j = A[*, j] - UV^T[*, j] - WU^T[*, j]$;*
     *Find a Householder vector $u$ of size $N$ with $\|u\|_2 = 1/\sqrt{2}$ to reduce*
       $a_j[(i-1)k+j+1:N]$ *to form $(*, 0, \ldots, 0)^T$;*
     *Compute $v = A^T u - VU^T u - UW^T u$;*
     *Compute $w = Au - UV^T u - WU^T u - uv^T u$;*
     *$U = [U \ u]$; $V = [V \ v]$; $W = [W, \ w]$;*
   **endfor***;*
   **for** $j = 1, \ldots k$ **do**
     *$A[*; \ (i-1)k+j] = a_j$;*
   **endfor***;*
   *$\hat{A}[*; \ ik+1:N] = \hat{A}[*; \ ik+1:N] - UV^T[*; \ ik+1:N] - WU^T[*; \ ik+1:N]$;*
**endfor***;*

---

As shown in [89], the number of flops computed with matrix vector products is approximately, $2\sum_{k=1}^{N-k} N(N-k) \approx N^3$, while the total number of flops remains nearly the same as that in the unblocked algorithm. That is, this algorithm spends about 30% of its work in level-2 operations and about 70% of work in level-3 operations. But that does not result in an asymptotic improvement in I/O complexity. If $A^T$ is computed once in the beginning of each $i$-loop, then the total number of blocks moved by the algorithm is $O(N^3/B)$. If $N \leq M$ and $k = \Theta(M/N)$, then the number of seeks is $O(\frac{N^3}{k\sqrt{M}} + \frac{N^3}{M})$. If $N > M$ and $k \leq \sqrt{M}$, then the number of seeks is $O(\frac{N^3}{k\sqrt{M}} + \frac{N^2 k^2}{M})$. If $N > M$ and $M \geq k > \sqrt{M}$, then the number of seeks is $O(\frac{N^3}{M} + \frac{N^2 k^2}{M})$. Though most of the computations are performed using M-M operations, M-V and V-V operations still dominate the I/O.

## 4.2.2 Direct tridiagonal reduction

Let $A \in \mathbb{R}^{N \times N}$ be a symmetric matrix. Find Householder transformations $Q_1, \ldots, Q_{N-2}$ such that each $Q_i$ is designed to introduce zeroes below the first subdiagonal in the $i$-th column of the matrix $Q_{i-1}^T \cdots Q_1^T A Q_1 \cdots Q_{i-1}$ [58,115]. Then $T = Q_{N-2}^T \cdots Q_1^T A Q_1 \cdots Q_{N-2}$ is a tridiagonal matrix.

For a Householder $Q = (I - u u^T)$, where $\|u\|_2 = 1/\sqrt{2}$, $Q^T A Q = (I - u u^T)^T A (I - u u^T) = A - u v^T - v u^T$, where $v = Au - (1/2) u u^T A u$. Thus, an algorithm similar to the unblocked Hessenberg reduction works for tridiagonal reduction. Due to symmetry, only the lower triangular part of the matrix needs to be updated. The number of flops and I/Os would be approximately half of that of the Hessenberg reduction.

A blocked algorithm analogous to Algorithm 4.2 can be devised for tridiagonal reduction too. The observations made there apply here too. The asymptotic complexity remains the same; that is, $O(N^3/B)$.

The asymptotic seek complexity remains same as the Hessenberg reduction.

## 4.2.3 Direct bidiagonal reduction

Let $A \in \mathbb{R}^{N \times P}$, and $N \geq P$. Find Householder transformations $Q_{u_P} \ldots Q_{u_1}, Q_{v_1}, \ldots, Q_{v_{P-2}}$ such that for $1 \leq i \leq P-2$, $Q_{u_i}$ is designed to introduce zeroes below the diagonal in the $i$-th column of the matrix $Q_{u_{i-1}}^T \cdots Q_1^T A Q_{v_1} \cdots Q_{v_{i-1}}$, and $Q_{v_i}$ is designed to introduce zeroes to the right of the first superdiagonal in the $i$-th row of the matrix $Q_{u_i}^T \cdots Q_1^T A Q_{v_1} \cdots Q_{v_{i-1}}$. For $P - 1 \leq i \leq P$, $Q_{u_i}$ introduces zeroes below the diagonal in the $i$-th column of $Q_{u_{i-1}}^T \cdots Q_1^T A Q_{v_1} \cdots Q_{v_{P-2}}$.

Let $U = Q_{u_1}, \ldots, Q_{u_P}$ and $V = Q_{v_1}, \ldots, Q_{v_{P-2}}$. Then $B = U^T A V$ is upper bidiagonal [58, 61, 115].

With the necessary matrix transpositions thrown in, the algorithm can be made to run in $O(NP^2/B)$ I/Os and $O(NP^2/\sqrt{M} + NP)$ seeks. The cases of $N < P$, and lower bidiagonal form can be handled similarly.

If the Householders are aggregated and applied in lots, the computation can be made richer in M-M operations. We now discuss one such algorithm [32, 45].

If $u$ and $v$ are two Householder vectors, then

$$(I - uu^T)A(I - vv^T) = A - uw^T - yv^T$$

where

$$y = Av, \quad h = A^T u \text{ and } w = h - (u^T y)v$$

Therefore, if we can compute $u_j$ and $v_j$ independently of each other, after reducing the $(j - 1)$-th column and row of the matrix, then the Algorithm 4.3 works:

The computation of $v_j$ would seem to depend on $u_j$. In $j$-th step of the algorithm, the $j$-th row is altered by the left multiplication with $Q_{u_j}$, and $Q_{v_j}$ is to reduce this altered $j$-th row. Similarly, application of $Q_{v_j}$ from the right alters the $(j + 1)$-st column of the matrix, which is to be reduced by $Q_{u_{j+1}}$.

But observe that $Q_{v_1} \ldots Q_{v_{P-2}}$ is precisely the sequence of Householders that would transform the symmetric matrix $A^T A$ to a tridiagonal form $T$ in the algorithms of Section 4.2.2. That is, if $V = Q_{v_1} \ldots Q_{v_{P-2}}$, then $V^T A^T A V = B^T B$ is tridiagonal.

---

**Algorithm 4.3.** *Blocked Direct Bidiagonal Reduction [45]*

---

**Input:** $A \in \mathbb{R}^{N \times P}$ *with* $N \geq P$, *and* $k$, *the slab width.*

**Output:** *A is overwritten with a bidiagonal matrix orthogonally equivalent to A.*

---

*Let* $r = \lceil (P-2)/k \rceil$;
**for** $i = 1$ **to** $r$ **do**
     $s = (i-1)k + 1$;
     $U = Y = [\,]$;     /* $U$ *and* $Y$ *have* $N$ *rows each* */
     $V = W = X = [\,]$;     /* $V$, $W$, *and* $X$ *have* $P$ *rows each* */
     **for** $j = s$ **to** $\min\{P-2, s+k-1\}$ **do**
         *Let* $a_j = A[*, j] - UW^T[*, j] - YV^T[*, j]$;
         *Compute* $u_j$ *to reduce* $a_j$;
         *Compute* $v_j$ *by invoking Algorithm 4.4 with* $A_s = A[s:N; \; s:P]$, $V$, $X$ *and* $j$;
         $y_j = (A - UW^T - YV^T)v_j$;    $h_j = (A - UW^T - YV^T)^T u_j$;
         $w_j = h_j - (u_j^T y_j)v_j$;
         $U = (U, u_j)$; $V = (V, v_j)$; $W = (W, w_j)$; $Y = (Y, y_j)$;
     **endfor**;
     $A[1:N; \; s+1:s+k-1] = (a_{s+1}, \ldots, a_{s+k-1})$;
     $A[s+k:N; \; s+k:P] = (A - UW^T - YV^T)[s+k:N; \; s+k:P]$;
**endfor**;
*Compute* $u_{P-1}$ *and* $u_P$;
*update* $A[P-1:N; \; P-1:P] = Q_{u_P}Q_{u_{P-1}}A[P-1:N; \; P-1:P]$;

---

Hence $v_j$ can be computed as follows:

---

**Algorithm 4.4.** *Computation of* $v_j$ *for Algorithm 4.3 [45]*

---

**Input:** *A submatrix* $A_s = A[s:P; \; s:N]$ *of* $A$, $V$, $X$ *and* $j$.

**Output:** *A Householder vector* $v_j$ *such that* $Q_{v_j}$, *when applied from the right, zeroes the elements to the right of the superdiagonal in the* $j$-th *row of* $A_s$.

---

     $z_j = A_s^T A_s[*, j] - XV^T[*, j] - VX^T[*, j]$
     *Compute the Householder vector* $v_j$ *from* $z_j$
     $t_j = (A_s^T A_s - V_{j-1}X_{j-1}^T - X_{j-1}V_{j-1}^T)v_j$;
     $x_j = t_j - \frac{1}{2}(t_j^T v_j)v_j$
     $X_j = (X_{j-1}, x_j)$

---

If $A^T$ is computed once in the beginning of each $i$-loop, then the total number of blocks moved by the algorithm is $O(NP^2/B)$. If $N \leq M$ and $k = \Theta(M/N)$, then the number of seeks is $O(\frac{NP^2}{k\sqrt{M}} + \frac{NP^2}{M})$. If $N > M$ and $k \leq \sqrt{M}$, then the number of seeks is $O(\frac{NP^2}{k\sqrt{M}} + \frac{NPk^2}{M})$. If $N > M$ and $M \geq k > \sqrt{M}$, then the number of seeks is $O(\frac{NP^2}{M} + \frac{NPk^2}{M})$.

## 4.3 Reduction from Banded Hessenberg Form to Hessenberg Form using Householders

The direct Hessenberg reductions described in Section 4.2.1 can be used, but after a few steps of those, the matrix at hand becomes full, necessitating, on the whole, $O(N^3)$ flops and $O(N^3/B)$ I/Os. Algorithms that can exploit the sparsity of the matrix can do better. In this section, we propose two such algorithms.

Let $A$ be the input matrix. Let it be of dimensions $N \times N$ and bandwidth $t$.

### 4.3.1 A bulge chasing algorithm

A reduction from symmetric banded form to tridiagonal form is presented in [77]. We adapt this.

Annihilate the unwanted subdiagonal elements of the band, column by column, using Householders. To reduce the $i$-th column, for $1 \leq i \leq N - 2$, construct a Householder and apply it from the left and right, respectively, to rows and columns numbered $i + 1$ to $i + t$. This will create a triangular $t \times t$ bulge below the band in columns $i + 1$ to $i + t$, i.e., at positions $A[i + t + 1 : i + 2t; \; i + 1 : i + t]$. Chase the first column of the bulge along the band, and out of the matrix. On the completion of the chasing, non-overlapping harmless $(t-1) \times (t-1)$ bulges will lie below the band all along the matrix. The $i$-th column is done. Now reduce the $(i + 1)$-st column. Each $t \times t$ bulge created now will subsume the harmless $(t-1) \times (t-1)$ bulges created earlier while processing the $i$-th column. See Algorithm 4.5 for a formal description. The first sweep of the algorithm is illustrated in Figure 4.1.

---

**Algorithm 4.5.** *Banded Hessenberg form to Hessenberg form using Householders*

**Input:** *A banded Hessenberg matrix $A \in \mathbb{R}^{N \times N}$ of bandwidth $t$.*
**Output:** *A Overwritten with a similar Hessenberg matrix.*

---

**for** $i = 1, \ldots, N - 2$ **do**
    /*At this point, $A[1 : i - 1, 1 : i - 1]$ is a Hessenberg matrix. The band elements
    of the $i$-th column are handled in this iteration*/
    Let $X = \min\{i + t, N\}$;
    Find a Householder $Q$ such that $Q^T.A[i + 1 : X, i]$ has form $(*, 0, \ldots, 0)^T$.
    Let $k = \lceil \frac{N-i}{t} \rceil$;
    **for** $j = 0, \ldots, k - 1$ **do**
        Let $Y = \min\{i + (j + 1)t, N\}$;
        Multiply $A[i + jt + 1 : Y; \ i + jt + 1 : N]$ from the left with $Q^T$.
        Let $Z = \min\{i + (j + 2)t, N\}$;
        Multiply $A[1 : Z; i + jt + 1 : Y]$ from the right with $Q$.
        **if**$(j < k - 1)$ **do**
            Find a Householder $Q$ such that $Q^T.A[(i + 1) + (j + 1)t : Z; \ i + jt + 1]$
            has form $(*, 0, \ldots, 0)^T$.
        **endif**
    **endfor**
**endfor**

---

The householder vectors can be stored in the lower triangular part of the matrix. All the Householders generated from the annihilation of the $i$-th column and subsequent bulge chasing is stored below the subdiagonal of the $i$-th column one after another.

Zeroing of the $i$-th column needs $O(N^2/B)$ blocks of data to be moved. So the total number of blocks moved by the algorithm is $O(N^3/B)$.

The algorithm involves $O(N^2/t)$ multiplications from the left and right with Householders. Without loss of generality, assume that the matrix is in column major order. Then, for its right updates, the algorithm requires a total of $O(N^2/t)$ seeks if $Nt \leq M$ (an entire panel of $t$ columns fits in the main memory and can be read in with a seek), $O(N^3t/M)$ seeks if $t \leq M < Nt$ ($M$ elements can be read in with $t$ seeks), and $O(N^3)$ seeks if $t > M$. For its left updates the algorithm requires a total of $O(N^3/t)$ seeks in all cases.

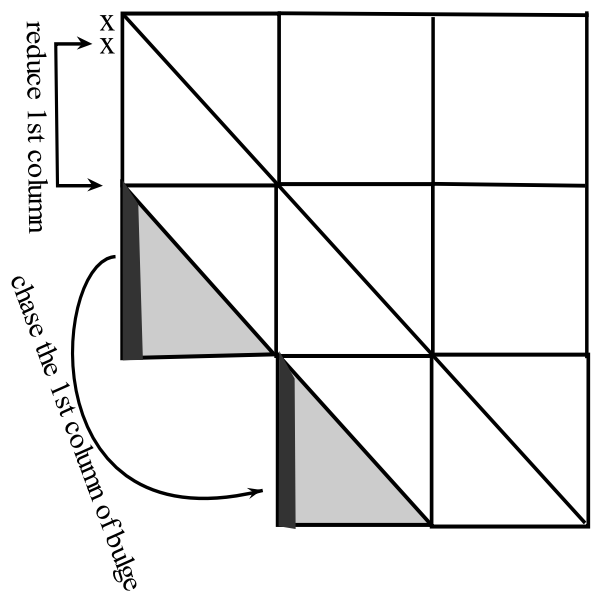

Figure 4.1: Reduction of the first column and the corresponding bulge chasing steps

A small modification to the algorithm can improve its seek complexity. Defer the left application of the $Q$ matrices to the horizontal panels. In particular, while reducing the $i$-th row, repeatedly, compute a Householder $Q$, read the vertical panel on which it is to be applied, update the panel using all the pending left updates (the Householder vectors for these are available in the $i$-th column), and finally apply $Q$ on the panel from the right. This avoids the need for accessing the horizontal panels. So the overall asymptotic seek complexity of the algorithm would be $O(N^2/t)$ if $Nt \leq M$, $O(N^3 t/M)$ if $t \leq M < Nt$ and $O(N^3)$ if $t > M$.

## 4.3.2   An asymptotic improvement

The above reduction does asymptotically no better than the direct Hessenberg reduction. Nevertheless, a similar algorithm is popular for tridiagonal and bidiagonal reductions. The main drawback of the above algorithm is that to reduce a single column, the whole of the portion to its right is read in and written back.

We can improve on this if we defer chasing the bulges. We make the observation that it is enough to chase the bulges a short distance to make the next few columns ready for processing, and use it to achieve an asymptotic improvement. We maintain a box about the column that is being reduced, and chase the bulges to the bottom right corner of the box. When a number of columns are reduced we would have lined up an equal number of bulges at the bottom right corner of the box in a tightly packed sequence. While doing this, we need to update with the Householders only the portion inside the box. After lining up an appropriate number of bulges, we use the aggregate of the Householders to update the portion outside the box at one go.

On a banded upper Hessenberg matrix $A \in \mathbb{R}^{N \times N}$ of bandwidth $t$, the algorithm proceeds as follows:

Divide the matrix into slabs of $t$ columns each. Process the slabs one by one. Begin with the leftmost slab.

Let $w$ be $2t^2 - t$. Define the "small box" as the intersection of columns $l = 1$ to $r = 1 + w$ (left and right) and rows $a = 2$ to $b = 1 + w + t$ (above and below). Thus, it has $w + 1$ columns and $w + t$ rows. For $h \geq 1$, let the $h$-th large box be the intersection of columns $l(h) = 1 + t + (h - 1)k$ to $r(h) = 1 + w + hk$ and rows $a(h) = i + 2t + (h - 1)k$ to $b(h) = 1 + w + t + hk$. Thus, it has $k + w - t + 1$ columns and rows. ($k$ is a parameter to be chosen later.)

Process the columns of the first slab one by one. Begin with the first column. Find a Householder $Q_1^{(1)}$ s.t. $Q_1^{(1)} * A[2 : 1+t; 1]$ is of the form $(* \; 0 \cdots 0)^T$. Multiply $A[2 : 1+t; 1 : r]$ and $A[a : 1 + 2t; 2 : 1 + t]$ with $Q_1^{(1)} = (Q_1^{(1)})^T$ from the left and right respectively. Note that only the portion inside the small box is involved in these multiplications. A bulge forms in columns $2 : 1 + t$ at rows $2 + t : 1 + 2t$.

Chase the leftmost column of this bulge: Find a Householder $Q_2^{(1)}$ s.t. $Q_2^{(1)} * A[2 + t :$

$1 + 2t, 2]$ is of the form $(* \, 0 \cdots \, 0)^T$. Multiply $A[2 + t : 1 + 2t; 2 : r]$ from the left, and $A[a : 1 + 3t; 2 + t : 1 + 2t]$ from the right with $Q_2^{(1)} = (Q_2^{(1)})^T$. A bulge forms in columns $2 + t : 1 + 2t$ at rows $2 + 2t : 1 + 3t$.

Chase the leftmost column of this new bulge similarly, and continue like that till the right boundaries of the latest bulge and the small box coincide. This happens after the $(2t - 1)$-th Householder. Nail the bulge in its present position.

The nailed bulge is unreduced and hence spans $t$ columns, whereas each of the other (harmless) bulges has lost one column and so spans $t - 1$ columns.

Now process the second column, and chase the bulges similarly, until the latest bulge is $t-1$ columns away from the nailed bulge of the first column. This involves finding of $(2t-3)$ Householders. Note that the bulges created while processing the second column subsume, when they are as yet unreduced, the harmless bulges left behind by the processing of the first column.

Process the $t$ columns of the slab like this. When all are done, we will be left with $t$ nailed bulges spanning columns $t + 1$ to $2t^2 - t + 1$, and any two consecutive nailed bulges will be separated by a harmless bulge that spans $t - 1$ columns.

Note that the slab and the interspersed sequence of the nailed and harmless bulges are tightly packed inside the small box.

The Householders found till now have been applied only to the relevant portions of the small box. Now we form groups of these Householders, aggregate each group into an $(I + YTY^T)$ representation and apply from the left on the portion to the right of the small box, and apply from the right on the portion above the small box. (The details of the aggregation are given later.)

Next consider the first large box. The nailed bulges of the small box now occupy the top left corner of this box. Move them into the bottom right corner, one by one, starting with the rightmost bulge, and arrange them there in a similar tightly packed manner. This process would leave harmless bulges each spanning $t - 1$ columns in the small box. In particular, columns $t + 2$ to $2t$ hold the harmless remnant of the bulge introduced by the reduction of column $t$. But column $t + 1$ is now without a bulge, and so is ready for processing. The Householders found during this process are applied on the fly only to the relevant portions of the first large box. Now group them, aggregate the groups and apply the aggregates on portions outside the first large box. (The details of the aggregation are given later.)

We repeat the process for the remaining large boxes in a left to right order, until the nailed bulges are all within the last $k + w$ columns of the matrix.

Next we chase bulges from the bottom right corner of the last large box, and out of the matrix. The matrix is now identical to the one produced by the first $t$ iteration of Algorithm 4.5 applied on the original input.

Now the first slab has been processed. Continue with remaining slabs, until at most $k + w - t + 1$ rightmost columns of the matrix remain to be reduced. Then invoke Algorithm 4.5 on the unreduced square submatrix at the bottom right corner.

Now the matrix will be in Hessenberg form.

A more formal description of the algorithm follows:

---

**Algorithm 4.6.** *Banded Hessenberg form to Hessenberg form using bulge packing*

---

**Input:** *A banded upper Hessenberg matrix $A \in \mathbb{R}^{N \times N}$ of bandwidth $t$.*

**Output:** *A is overwritten with a similar Hessenberg matrix.*

---

*Let $w = 2t^2 - t$;*

**for** *($i = 1$; $i \leq \lceil (N - k - w + t - 1)/t \rceil t$; $i = i + t$)* **do**

    *Let the small box be the intersection of columns $l(i) = i$ to $r(i) = i + w$ and*

*rows $a(i) = i + 1$ to $b(i) = i + w + t$.*

*Thus, it has $w + 1$ columns and $w + t$ rows.*

**for** ($j = i$;  $j \leq i + t - 1$;  $j++$) **do**    /* *process the $j$-th column* */

   *Initialise $G_1 \ldots G_{2(t-1)+1}$ to identity matrices of appropriate dimensions.*

   *Find Householder $Q_1^{(j)}$ s.t. $Q_1^{(j)} * A[j+1 : j+t; j]$ has form $(* \, 0 \cdots 0)^T$.*

   *Aggregate $Q_1^{(j)}$ into $G_1$.*

   *Multiply $A[j+1 : j+t; j : r(i)]$ and $A[a(i) : b(i); j+1 : j+t]$ with*

      *$Q_1^{(j)}$ from the left and right respectively.*

   *A bulge forms in columns $j+1 : j+t$ and rows $j+t+1 : j+2t$.*

   *Let $x_j = 2 * (t - (j - i + 1)) + 1$. Then $x_j - 1$ is the number of times*

   *this bulge will be chased before it's fixed in the small box.*

   **for** ($z = 2$;  $z \leq x_j$;  $z++$) **do**

      *Find $Q_z^{(j)}$ s.t. $Q_z^{(j)} * A[j + (z-1)t + 1 : j + zt, j + (z-2)t + 1]$*

         *has form $(* \, 0 \cdots 0)^T$. Aggregate $Q_z^{(j)}$ into $G_z$.*

      *Multiply $A[j + (z-1)t + 1 : j + zt; j + (z-2)t + 1 : r(i)]$ from the left, and*

      *$A[a(i) : j + (z+1)t; j + (z-1)t + 1 : j + zt]$ from the right with $Q_z^{(j)}$.*

      *A bulge forms in columns $j + (z-1)t + 1 : j + zt$ and*

      *rows $j + zt + 1 : j + (z+1)t$. See Remark below.*

   **endfor** /* *Nail the bulge at the present position* */

**endfor**

*Multiply $A[a(i) : b(i) - t; r(i) + 1 : N]$ (right of the small box) from the left,*

*and multiply $A[1 : a(i) - 1; l(i) : r(i)]$ (above the small box) from*

*the right with $(G_1 \ldots G_{2(t-1)+1})^T$ and $(G_{2(t-1)+1} \ldots G_1)$ respectively.*

*See Algorithm 4.7*

*Let the h-th large box be the intersection of*

*columns $l(i, h) = i + t + (h-1)k$ to $r(i, h) = i + w + hk$ and*

*rows $a(i, h) = i + 2t + (h-1)k$ to $b(i, h) = i + w + t + hk$.*

*Thus, it has $k + w - t + 1$ columns and rows.*

*$k$ is a parameter to be chosen later.*

**for** $(h = 1; \;\; r(i, h) \leq N - t; \;\; h++)$ **do**

    /* move the bulges to the bottom right of the h-th large box */

    **for** $(j = i; \;\; j \leq i + t - 1; \;\; j++)$ **do**

       *Initialise $G_{hk/t+2t-1} \ldots G_{(h-1)k/t+2}$ to identity matrices of appropriate dimensions.*

       $x_j = 2 * (t - (j - i + 1)) + 1;$

       **for** $(p = 1; \;\; p \leq k/t; \;\; p++)$ **do**

          *Let $z = p + (h-1)k/t + x_j$.    /* for each column j we pick up the*

          *z-count from where the previous box (large or small) left off */*

          *Find $Q_z^{(j)}$ s.t. $Q_z^{(j)} * A[j + (z-1)t + 1 : j + zt; \; j + (z-2)t + 1]$*

          *has form $(* \, 0 \ldots 0)^T$. Aggregate $Q_z^{(j)}$ with $G_z$.*

          *Multiply $A[j + (z-1)t + 1 : j + zt; \; j + (z-2)t + 1 : r(i, h)]$*

          *from the left, and $A[a(i, h) : j + (z+1)t; \; j + (z-1)t + 1 : j + zt]$*

          *from the right with $Q_z^{(j)}$. A bulge forms in columns*

          *$j + (z-1)t + 1 : j + zt$ and rows $j + zt + 1 : j + (z+1)t$.*

      **endfor**

    **endfor**

*The smallest and largest values assigned to z in the above are*

*$(h-1)k/t + 2$ and $hk/t + 2t - 1$, when $(p = 1, j = i + t - 1)$ and*

*$(p = k/t, j = i)$ respectively. See Algorithm 4.8.*

*Multiply $A[a(i,h) : b(i,h) - t; \; r(i,h) + 1 : \; N]$ (right of the h-th large box)*

*from the left, and multiply $A[1 : a(i,h) - 1; \; l(i,h) : \; r(i,h)]$ (above the h-th large box)*

*from the right with $(G_{(h-1)k/t+2} \ldots G_{hk/t+2t-1})^T$*

*and $(G_{hk/t+2t-1} \ldots G_{(h-1)k/t+2})$ respectively.*

*See Algorithm 4.8.*

**endfor**

*Chase the bulges that are now at the bottom right corner of the $(h-1)$-th large*

*box out of the matrix. (We omit the details.)*

**endfor**

*Invoke Algorithm 4.5 on $A[N + 1 - \lceil \frac{(N-k-w+t)}{t} \rceil t : N; \; N + 1 - \lceil \frac{(N-k-w+t)}{t} \rceil t : N]$.*

*Aggregate the Householders into groups after all t columns have been reduced*

*Update corresponding columns of submatrix*

*$A[1 : N - \lceil \frac{(N-k-w+t)}{t} \rceil t; \; N + 1 - \lceil \frac{(N-k-w+t)}{t} \rceil t : N]$ from the right.*

---

**Remark:** The bulge that forms in columns $j + (z-1)t + 1 : j + zt$ and rows $j + zt + 1 :$ $j + (z+1)t$ subsumes any harmless elements that may have been present there. Columns $j + (z-1)t + 1 : j + zt$ cannot have elements below this bulge; column $j + zt$ has been cleared of any bulge before proceeding to this step. A snapshot of the matrix after applying the bulge introduction phase to the first slab $(t = 3)$ is shown in Figure 4.2. In the figure, the submatrix $A[i + 1 : i + 2t^2; \; i : i + 2t^2 - t] = A[2 : 19; \; 1 : 16]$ is updated from both the left and right, the submatrix $A[i + 1 : i + 2t^2 - t; \ i + w + 1 : N] = A[2 : 16; \ 17 : N]$ is updated from the left and the submatrix $A[1 : i; \ i + 1 : i + w] = A[1 : 1; \ 2 : 16]$ is updated from the right.

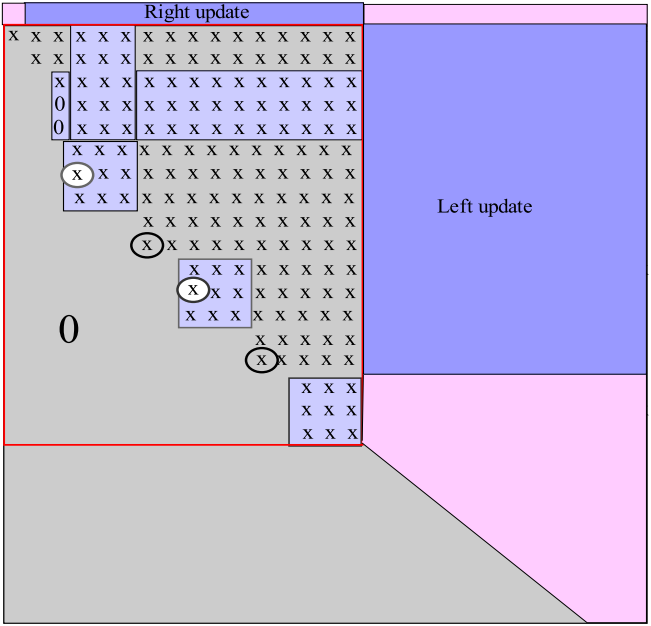

Figure 4.2: Introduction of $t = 3$ tightly packed bulges of size $3 \times 3$ after reducing first 3 columns.

### 4.3.2.1 Aggregation: the bulge introduction phase

Let $Q_1^{(j)}$ be the Householder that reduces the $j$-th column, and for $z > 1$, let $Q_z^{(j)}$ be the Householder that reduces the first column of the $z$-th bulge induced by $j$-th column. $Q_z^{(j)}$ acts along dimensions $j + 1 + (z - 1)t$ to $j + zt$.

Consider the processing of the slab beginning at column $i$.

The Householders are constructed and applied inside the small box in the following order: $Q_1^{(i)} \ldots Q_{x_i}^{(i)} Q_1^{(i+1)} \ldots Q_{x_i-2}^{(i+1)} \ldots Q_1^{(i+t-1)}$. But while updating the relevant portions outside the small box they need not be applied in the same order. We can aggregate them into different groups using a technique proposed in [78], because of the following

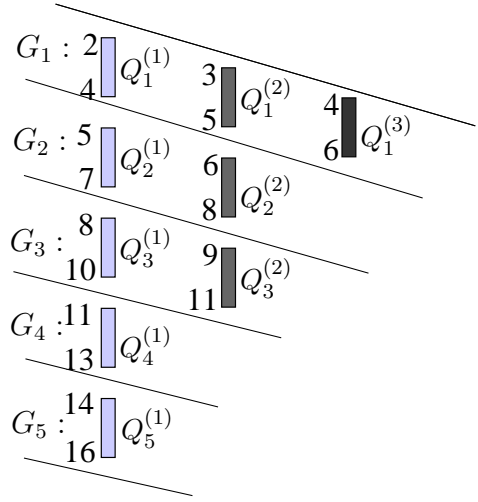

Figure 4.3: Householders constructed to introduce $m = 3$ tightly packed bulges after reducing first block(first 3) column.

observations: (See Figure 4.2 and Figure 4.3.)

1. $Q_*^{(j)}$'s, which reduce the $j$-th column and the bulges induced by it, act along disjoint dimensions. That means, if $Q_*^{(j)}$'s are to be applied on a submatrix only from one side (the left or right), then they can be applied in any order.

2. The dimensions of $Q_z^{(j)}$ and $Q_y^{(j-1)}$ overlap iff $y = z$ or $y = z + 1$. Therefore, $Q_z^{(j)}$ can be applied only after $Q_z^{(j-1)}$ and $Q_{z+1}^{(j-1)}$.

3. $Q_z^{(*)}$'s are to be applied in the order of their construction. Let $G_z$ be the aggregation of $Q_z^{(*)}$'s in the order of construction.

4. If the $G_z$'s are applied in the decreasing order of $z$, then the above two requirements will be met.

The computing of $(G_1 \dots G_{2(t-1)+1})^T A[a(i) : b(i) - t; \ r(i) + 1 : \ N]$ and $A[1 : a(i) - 1; \ l(i) : r(i)](G_{2(t-1)+1} \dots G_1)$ can be done as follows. Note that $x_i = 2(t-1) + 1$, $G_{x_i} = Q_{x_i}^{(1)}$ and $G_{x_i-1} = Q_{x_i-1}^{(1)}$. In a slab only the first column is subjected to more than $x_i - 2$ chases.

$G_{x_i-2}$ is an aggregation of $Q^{(1)}_{x_i-2}$ and $Q^{(2)}_{x_i-2}$. $G_{x_i-3}$ is an aggregation of $Q^{(1)}_{x_i-3}$ and $Q^{(2)}_{x_i-3}$. In a slab only the first two columns are subjected to more than $x_i - 4$ chases. Continuing like this, we find that for $1 \leq z \leq x_i$, if $q = x_i - z + 1$, then $1 + \lfloor (q-1)/2 \rfloor$ is the number of Householders aggregated into $G_z$, which therefore operates on $y = t + \lfloor (q-1)/2 \rfloor$ rows and columns.

| Group | Householders | $y$ | left end of the range |
|---|---|---|---|
| $G_1$ | $Q^{(i)}_1, Q^{(i+1)}_1, \ldots, Q^{(i+t-2)}_1, Q^{(i+t-1)}_1$ | $2t-1$ | $i+1$ |
| $G_2$ | $Q^{(i)}_2, Q^{(i+1)}_2, \ldots, Q^{(i+t-2)}_2$ | $2t-2$ | $i+t+1$ |
| $G_3$ | $Q^{(i)}_3, Q^{(i+1)}_3, \ldots, Q^{(i+t-2)}_3$ | $2t-2$ | $i+2t+1$ |
| $\vdots$ | | | |
| $G_{2r}$ | $Q^{(i)}_{2r}, Q^{(i+1)}_{2r}, \ldots, Q^{(i+t-r-1)}_{2r}$ | $2t-r-1$ | $i+(2r-1)t+1$ |
| $G_{2r+1}$ | $Q^{(i)}_{2r+1}, Q^{(i+1)}_{2r+1}, \ldots, Q^{(i+t-r-1)}_{2r+1}$ | $2t-r-1$ | $i+2rt+1$ |
| $\vdots$ | | | |
| $G_{2t-2}$ | $Q^{(i)}_{2t-2}$ | $t$ | $i+(2t-3)t+1$ |
| $G_{2t-1}$ | $Q^{(i)}_{2t-1}$ | $t$ | $i+(2t-2)t+1$ |

The computation can be done as follows:

---

**Algorithm 4.7.** *Multiplying with the Left and Right Aggregates from Bulge Introduction*

**Input:** *Aggregations of the $(2t-1)$ groups of the left and right Householders.*
**Output:** *Updated matrices.*

> **for** $(z = x_i;\ z \geq 1;\ z--)$ **do**
>     $q = x_i - z + 1;$
>     $y = t + \lfloor (q-1)/2 \rfloor;$
>     *Multiply $A[i + (z-1)t + 1 : i + (z-1)t + y;\ r(i) + 1 : N]$*
>         *from the left with $G^T_z$*
>     *Multiply $A[1 : a(i) - 1;\ i + (z-1)t + 1 : i + (z-1)t + y]$*
>         *from the right with $G_z$*
> **endfor**

---

#### 4.3.2.2 Aggregation: the bulge chasing phase

In the bulge chasing phase of the algorithm, repeatedly, $t$ bulges that are nailed at the top left corner of a large box in a tightly packed sequence are chased one by one over a distance of $k$. This would leave them at the bottom right corner of the large box, which is also the top left corner of the next large box. Each bulge is, thus, chased $k/t$ times from its current position. See Figure 4.4.

During the chasing steps, the Householders found are applied on the fly only to the relevant portions of the box. Afterwards, we group the Householders, aggregate the groups and apply the aggregates on portions outside the box.

Consider the processing of the slab beginning at column $i$. Suppose we are at the $h$-th large box of this slab. Let $\alpha = x_i + (h-1)k/t + 1$.

The Householders are constructed and applied inside the box in the following order: $Q_\alpha^{(i)} \ldots Q_{\alpha+k/t-1}^{(i)} Q_{\alpha-2}^{(i+1)} \ldots Q_{\alpha+k/t-3}^{(i+1)} \ldots Q_{\alpha-2t+2}^{(i+t-1)} \ldots Q_{\alpha-2t+k/t+1}^{(i+t-1)}$. But while updating the relevant portions outside the box they need not be applied in the same order. The observations made during the bulge introduction phase hold for these Householders too. Hence these too can be grouped and aggregated the same way. That is, all $Q$'s with the same subscript can be grouped together, and aggregated in the order in which they are constructed. Note that the smallest and largest subscripts are $\alpha - 2t + 2$ and $\alpha + k/t - 1$ respectively. See Figure 4.5.

For a fixed $h$, as $j$ varies over $[i, i+t-1]$, and as $p$ varies over $[1, k/t]$,

$$z = \left( \frac{(h-1)k}{t} + 2t \right) + p - 2j + 2i - 1 = \alpha + D(p, j)$$

where $D(p, j) = p - 2j + 2i - 1$. Note that each $z$ corresponds to a group. When $z$ is fixed, $D(p, j)$ is also fixed, as $\alpha$ is independent of $p$ and $j$.

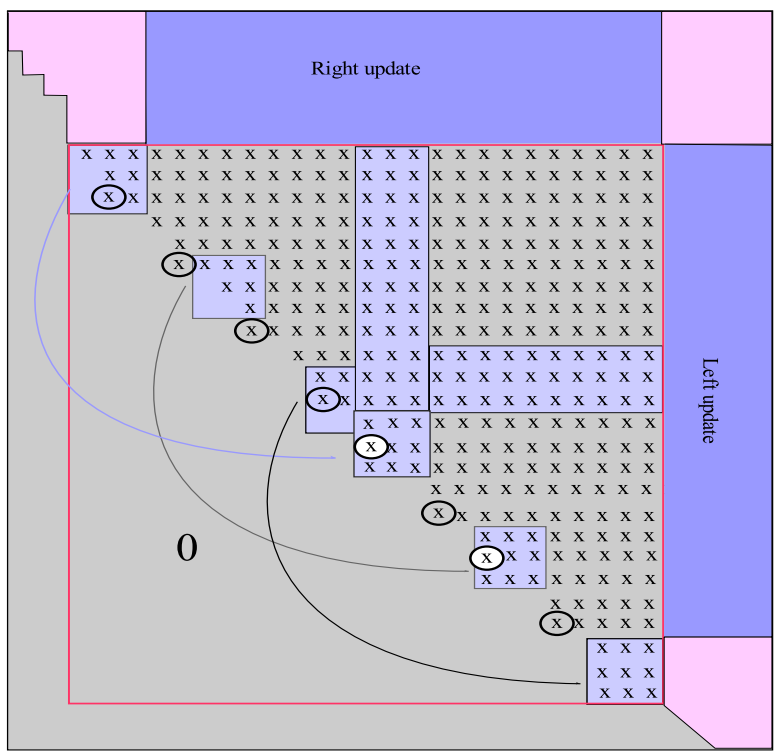

Figure 4.4: Chasing of 3 tightly packed bulges.

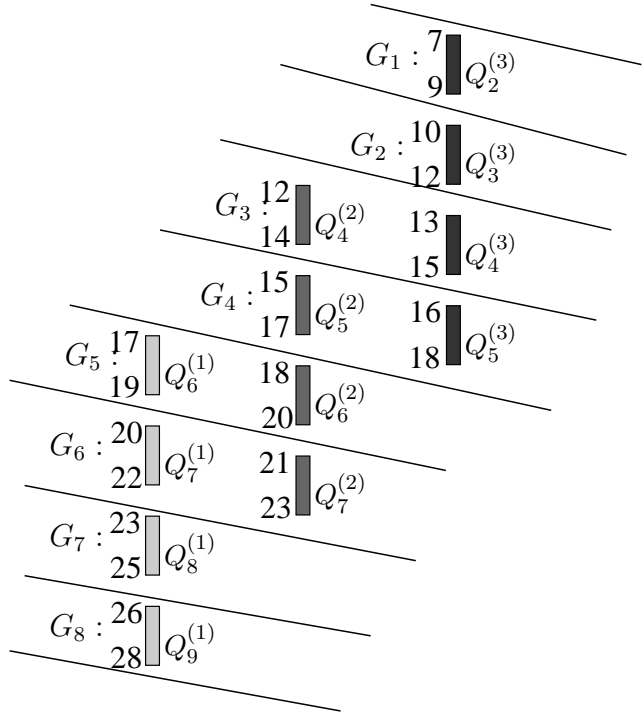

Figure 4.5: Householders constructed to chase 3 tightly packed bulges.

| $D(p,j)$ | $p=1$ | $p=2$ | $p=3$ | $\cdots$ | $p=k/t$ |
|---|---|---|---|---|---|
| $j=i$ | 0 | 1 | 2 | $\cdots$ | $k/t-1$ |
| $j=i+1$ | $-2$ | $-1$ | 0 | $\cdots$ | $k/t-3$ |
| $j=i+2$ | $-4$ | $-3$ | $-2$ | $\cdots$ | $k/t-5$ |
| $\vdots$ | | | | | |
| $j=i+t-1$ | $-(2t-2)$ | $-(2t-3)$ | $-(2t-4)$ | $\cdots$ | $k/t-2t+1$ |

Consult the above table. For $-(2t-2) \le D \le k/t-1$, set $\{(j,p)|D(p,j) = D\}$ corresponds to $G_{\alpha+D}$. If $D \ge 0$, then the leftmost cell $(j,p)$ with value $D$ is $(i, D+1)$. Otherwise, the leftmost cell $(j,p)$ is $(i + \lceil (-D)/2 \rceil, b)$, where $b = 1$ if $D$ is even, 2 if $D$ is odd. If $(j,p)$ is of value $D$, then the next cell of value $D$ to the right is $(j+1, p+2)$.

$Q_z^{(j)}$ operates on range $[j + (z-1)t + 1 : j + zt] = \alpha t + (j + Dt) + [-t+1 : 0]$.

For a given $z$, and $D = z - \alpha$, let $E(D)$ be the number of Householders aggregated into $G_z$. The range of $G_z$ then spans $E(D) - 1 + t$ dimensions starting at $\alpha t + Dt + (-t+1) + j^*(D)$, where $j^*(D)$ is the smallest row $j$ that contains $D$.

We assume that $2t - 2 \leq k/t$. If $D < 0$, then $E(D) = \lfloor (D + 2t)/2 \rfloor$. If $D \geq 0$ and $2t - 1 + D \leq k/t$, then $E(D) = t$. If $D \geq 0$ and $2t - 1 + D > k/t$, then $E(D) = \lceil \frac{1}{2}(\frac{k}{t} - D) \rceil$.

Therefore the updations can be done using the following algorithm:

---

**Algorithm 4.8.** *Multiplying with the Aggregates from Bulge Chasing*

---

**Input:** *The $k/t + (2t - 2)$ aggregations of the left and right Householders.*
**Output:** *Updated matrices.*

---

**for** $(z = hk/t + 2t - 1; \;\; z \geq (h-1)k/t + 2; \;\; z--)$ **do**
    $D = z - \frac{(h-1)k}{t} - 2t;$
    *If $D \geq 0$, then $j^* = i$;    else $j^* = (i + \lceil (-D)/2 \rceil)$.*
    *If $D < 0$, then $E = \lfloor (D + 2t)/2 \rfloor$;*
    *else if $2t - 1 + D \leq k/t$, then $E = t$;*
    *else $E = \lceil \frac{1}{2}(\frac{k}{t} - D) \rceil$.*
    $L = zt + j^* - t + 1; \qquad U = L + t + E - 2;$
    *Multiply $A[L : U; \; r(i) + 1 : N]$ from the left using $G_z$;*
    *Multiply $A[1 : a(i, h) - 1; \; L : U]$ from the right using $G_z$;*
**endfor**

---

### 4.3.2.3   I/O complexity

We first calculate the number of blocks of data moved by the algorithm.

Without loss of generality, we assume that $t \leq \sqrt{M}$. In the previous chapter, we found that the I/O complexity of banded Hessenberg reduction is $O(N^3/B\sqrt{M})$ if $t \geq \sqrt{M}$. Therefore, there is no advantage to choosing $t > \sqrt{M}$ there.

Each bulge introduction phase processes a small box of size $(w + t) \times (w + 1) = \Theta(t^4)$. To process a column of the slab, on an average, half of the small box needs to be updated. This amounts to a total of $\Theta(t^5/B)$ blocks of data movement. At the end of a bulge introduction phase for the slab beginning at the $i$-th column of the matrix, a $w \times (N - i - w)$ submatrix is to be updated from the left and a $i \times (w + 1)$ submatrix is to be updated from the right. These involve the application of the group aggregates one by one. Each group aggregate applies on at most $2t - 1$ dimensions. For no $z$ do the ranges $G_z$ and $G_{z+2}$ overlap. Therefore, these updations require each updated element to be handled at most a constant number of times, and so would incur $O((w(N - i) + wi)/B) = O(Nt^2/B)$ blocks of data movement. There are $\Theta(N/t)$ bulge introduction phases. Over all of them, the algorithm needs to move $\Theta\left(\frac{t^4 N + t N^2}{B}\right)$ blocks.

Now consider bulge chasing phases.

Each bulge chasing phase processes a large box of size $(k + w - t + 1) \times (k + w - t + 1) = \Theta((k + w)^2/B)$. To process a column of the slab, more than half of the large box needs to be updated. This amounts to a total of $\Theta((k + w)^2 t/B)$ blocks of data movement. At the end of the $h$-th bulge chasing phase for the slab beginning at the $i$-th column of the matrix, a $(w - 2t + k + 1) \times (N - i - w - hk)$ submatrix is to be updated from the left and an $(i + 2t + hk - k - 1) \times (w - t + k)$ submatrix is to be updated from the right. These involve the application of the group aggregates one by one. Each group aggregate applies on at most $2t - 1$ dimensions. For each $z$, the ranges of $G_z$ and $G_{z+2}$ begin at least $2t - 1$ apart. $G_z$ aggregates at most $t$ Householders ($E(D) \leq t$). $G_z$ spans at most $E(D) + t - 1 \leq 2t - 1$ dimensions. That is, the ranges of $G_z$ and $G_{z+2}$ cannot overlap. Therefore, these updations require each updated element to be handled at most a constant number of times. That means the processing of each large box would incur a total of $O\left(\frac{(k + w)^2 t + N(w + k)}{B}\right)$ blocks of data movement. This is continued for a $\Theta(N/k)$ times to reduce a slab of $t$ columns. There are $N/t$ slabs. When $k = \Theta(w) = \Theta(t^2)$, the data movement is minimised, and is $O(\frac{t^2 N^2}{B} + \frac{N^3}{tB})$. The cost of chasing the bulges out of the matrix from the last large box cannot be larger than the above. So the overall data movement cost of the algorithm is $O\left(\frac{t^2 N^2}{B} + \frac{N^3}{tB}\right)$

Now we calculate the number of seeks needed. Assume without loss of generality that the matrix is kept in column major order. With $k = \Theta(t^2)$ the large and small boxes are all of $\Theta(t^4)$ size. Therefore, we analyse only the large boxes. There are $O(N^2/t^3)$ large boxes overall.

There are $\Theta(t^2)$ inside-the-box right updates of $t^2 \times t$ vertical slabs. Each takes $t$ seeks if $t^3 < M$ and $t^4/M$ otherwise; in the latter case, we read an $(M/t) \times t$ submatrix in $t$ seeks; there $t^3/M$ such submatrices in a vertical slab. The total number of seeks incurred by these right updates is $O(N^2)$ if $t^3 < M$ and $O(N^2t/M)$ otherwise.

There are $\Theta(t^2)$ inside-the-box left updates of $t \times t^2$ horizontal slabs. These can be clubbed with the inside-the-box right updates. Since $t^2 < M$, we can keep all the Householder vectors generated to chase a particular column in-core. When a vertical panel is accessed for right updates, first apply all the pending left updates, then the right update.

For each large box, the algorithm needs to perform a left update of the right side of the box. This is performed as a series of left updates with Householder aggregates of horizontal panels of size $t \times N$. Each update of the series takes $N$ seeks, for a total of $O(Nt)$ seeks per large box. However, if $t^3 < M$, then all the aggregated Householder groups of a box can be kept in-core. All the left updates of a large box can therefore be performed $O(N)$ seeks: read the columns of the slab one by one, update and write back. Thus, the total seeks over all large boxes is $O(N^3/t^3)$ if $t^3 < M$, $O(N^3/t^2)$ otherwise.

Analogously, a right update of the portion above the box is also performed. The application of a Householder aggregate to an $N \times t$ vertical slab requires one seek if $Nt < M$ ($t$ consecutive columns of the matrix can be accessed in one seek), and $Nt^2/M$ otherwise ($t$ seeks to access an $(M/t) \times t$ segment of the panel; $Nt/M$ such segments). Thus, the total seeks over all large boxes is $O(N^2/t^2)$ if $Nt < M$, and $O(N^3/M)$, otherwise.

Combining all of the above, we find that the seek complexity of the algorithm is: $O(N^2 + N^3/t^3)$ if $t^3 < M$, and $O(N^3/t^2)$ otherwise.

In the above analysis we assumed that the boxes do not fit in the main memory. But if $t = \Theta(M^{1/4})$, then a $\Theta(t^2 \times t^2)$ box can be kept in-core. If we repeat the analysis with this assumption, we find that the total I/O complexity of the Algorithm is $O(\frac{tN^2}{B} + \frac{N^3}{tB} + \frac{N^3}{t^2})$.

### 4.3.2.4   The two step Hessenberg reduction

In Section 4.2.1, we saw that both the unblocked and blocked reductions of a full non-symmetric matrix to Hessenberg form cost $O(N^3/B)$ I/Os. The I/O performance can be improved if the reduction is split into two steps:

1. Reduce the matrix to banded Hessenberg form

2. Reduce the banded Hessenberg matrix to Hessenberg form

The first step, we saw in Chapter 3, takes $\Theta\left(\frac{N^3}{tB}\right)$ I/Os if $t < \sqrt{M}$ and $\Theta\left(\frac{N^3}{\sqrt{M}B}\right)$ if $t > \sqrt{M}$, where $t$ is the band width. We analysed the second step above. The second step dominates the I/O complexity.

We conclude the following:

- If $N \geq M\sqrt{M}$, choose $t = \Theta(\sqrt{M})$. The total I/O cost will be $O\left(\frac{N^3}{\sqrt{M}B}\right)$. That is, for large values of $N$, our algorithm matches the lower bound on number of data movement for the problem [13].

- If $M^{3/4} \leq N < M\sqrt{M}$, then choose $t = \Theta(N^{1/3})$, the total I/O complexity will be $O\left(\frac{N^3}{N^{1/3}B}\right)$. (We assume that $M^{1/4} < B$.)

- If $\sqrt{M} < N < M^{3/4}$, then choose $t = \Theta(M^{1/4})$. The total I/O complexity will be $O\left(\frac{N^3}{\sqrt{M}}\right)$.

Thus, our two stage reduction process improves the asymptotic I/O complexity of Hessenberg reduction by the following factors, over the direct single step unblocked and blocked reductions.

- a factor of $\sqrt{M}$, if $N \geq M\sqrt{M}$

- a factor of $N^{\frac{1}{3}}$, if $M^{3/4} \leq N < M\sqrt{M}$

- a factor of $\sqrt{M}/B$, if $\sqrt{M} < N < M^{3/4}$

## 4.4   The Reduction of a Symmetric Banded Matrix to Tridiagonal Form

Direct reduction of a full symmetric matrix to tridiagonal form was discussed in Subsection 4.2.2. If the matrix is a symmetric banded one then these reduction algorithms are not the methods of choice, as they don't take advantage of the sparsity of the banded matrix. Furthermore, after a few reduction steps of these algorithms, the matrix becomes full with the bulges created, resulting $O(N^3)$ flops and $O(N^3/B)$ I/Os [21, 77, 78].

The banded structure of the matrix makes it possible to employ the bulge chasing technique [21, 55, 72, 77, 95, 97]. We briefly discuss a few of these algorithms.

The algorithm of [97] chases a bulge as soon as it forms using Jacobi rotations. On an $N \times N$ matrix of semi-bandwidth $t$ the algorithm proceeds as follows: Annihilate element $a_{1\,t+1}$ using a Jacobi rotation $U_{t\,t+1}$ along dimensions $t$ and $t+1$. Because of symmetry, it is enough to consider the elements above the diagonal. The rotation introduces a bulge at $a_{t\,2t+1}$. Annihilate this bulge using a rotation $U_{2t\,2t+1}$. A bulge forms at $a_{2t\,3t+1}$. Repeat this process until the bulge is chased out of the matrix. Now start the next iteration with the annihilation of $a_{1t}$. Once all the unwanted elements of the first row are reduced, move on to the next row. The algorithm returns a tridiagonal matrix when all the $N-2$ rows are reduced. To reduce any row, the rest of the matrix has to be read and written twice. So total I/O complexity of the algorithm is $O(N^2t/B)$.

To access each row costs $O(t)$ seeks, so the seek complexity of the algorithm is $O(N^2t)$.

This algorithm was improved in [72] by incorporating vector operations on vector machines. The bulges are chased in a delayed fashion. Several bulges are introduced before any is chased out of the matrix. The bulge created by an element is chased and placed at a safe distance before the next element of the band is annihilated. Once a number of elements have been annihilated and bulges formed along the diagonals at a distance of $t$ from each other, vector operations are invoked to chase several bulges simultaneously.

Though this algorithm uses vector operations to exploit the locality, its asymptotic I/O complexity is the same as that of the previous algorithm, namely $O(N^2t/B)$. Similarly the asymptotic seek complexity remains same as that of the previous algorithm.

Algorithms have been designed using Householders instead of rotations too [77, 95]. These impart more data locality to the problem. We discuss the algorithm of [95].

This algorithm processes the columns of the input matrix $A$ (symmetric, banded with a semi-bandwidth of $t$) one by one. When the $j$-th column is taken up for processing, the first $(j - 1)$ columns (and rows) would have already been reduced. So the leading $j \times j$ submatrix of $A$ would be in tridiagonal form. The unwanted elements of the $j$-th column (the band elements $a_{j+2\,j}, \ldots, a_{j+t\,j}$) are annihilated at one go by applying a Householder on rows and columns numbered from $j + 1$ to $j + t$, from the left and right respectively. This will introduce a triangular bulge in columns $j + 1$ to $j + t$ at rows $j + t + 1$ to $j + 2t$. (We talk only of the lower triangular part.) If this submatrix is QR-decomposed and the resultant $Q$ matrix is applied to $A$ from the left and right, the whole of the bulge would vanish, but a new bulge would appear $t$ positions below to the right. This step is continued till the bulge is chased out of the matrix. The algorithm moves on to the next column.

The reduction of the $j$-th column involves $O\left(\frac{N-j}{t}\right)$ bulge chases, each of which causes a $QR$ update that costs $\Theta(t^2/B)$ I/Os. So the I/O complexity of the algorithm is $O(N^2t/B)$.

The problem with the above algorithm is that it chases the whole of a $t \times t$ bulge away as soon as it forms. But it is sufficient to chase the first column of the bulge away, for the next column to be ready for reduction. The remaining part of the bulge will be subsumed by bulge created by the next column, and therefore is harmless for the correctness of the algorithm. This observation led to the parallel algorithm of [77].

This algorithm is similar in structure to the above one. Hence the details are omitted here. Each chase here involves a single Householder, rather than an aggregate of $t$ Householders. This reduces computation. However, the asymptotic I/O complexity remains $O(N^2t/B)$.

To reduce each column, the leading submatrix is partitioned into $t \times t$ blocks. So the seek complexity of the algorithm is $O((tN/\sqrt{M} + N)N) = O(tN^2/\sqrt{M} + N^2)$.

### 4.4.1   The two step tridiagonal reduction

In Subsection 4.2.2, we saw that both the unblocked and blocked reductions of a full symmetric matrix to tridiagonal form takes $O(N^3/B)$ I/Os. The I/O performance can be improved if the reduction is split into two steps:

1. Reduce the matrix to symmetric banded form

2. Reduce the symmetric banded matrix to tridiagonal form

The first step, we saw in Chapter 3, takes $\Theta\left(\frac{N^3}{tB}\right)$ I/Os if $t < \sqrt{M}$ and $\Theta\left(\frac{N^3}{\sqrt{M}B}\right)$ if $t > \sqrt{M}$, where $t$ is the semi-bandwidth of the symmetric banded form. We analysed the second step above.

We calculate the combined I/O complexity of the two steps, and pick the $t$ that optimises this quantity. We conclude the following:

- If $N \geq M$, choose $t = \Theta(\sqrt{M})$. The total I/O cost will be $O\left(\frac{N^3}{\sqrt{M}B}\right)$ I/Os. That is, for large values of $N$, our algorithm matches the lower bound on the data movement for the problem [13].

- If $B^2 \leq N < M$, choose $t = \sqrt{N}$. The total I/O cost will be $O\left(\frac{N^3}{\sqrt{N}B}\right)$.

- If $N < B^2$, choose $t = B$. The total I/O cost will be will be $O(N^2)$.

Thus, our two stage reduction process improves the asymptotic I/O complexity of tridiagonal reduction by the following factors, over the direct single step unblocked and blocked reductions.

- a factor of $\sqrt{M}$, if $N \geq M$

- a factor of $N^{\frac{1}{2}}$, if $B^2 \leq N < M$

Note that the seek complexity of the first step is $O(N^3/kt^2)$ if only one tile fits in the main memory and $O(N^3/t^3)$ otherwise. (See Chapter 3.) This is inconsequential.

## 4.5 The Reduction of a Banded Matrix to Bidiagonal Form

Let $A \in \mathbb{R}^{N \times P}$ be a banded matrix with a lower band width of $t$ and an upper band width of $m$, i.e., $A(i,j) = 0$ for $i > j + t$ or $i < j - m$. Without loss of generality, we assume that $N \geq P$, and that $A$ is to be reduced to upper bidiagonal form; the case of $N \leq P$ is analogous, and only a minor modification is necessary to reduce $A$ to lower bidiagonal form.

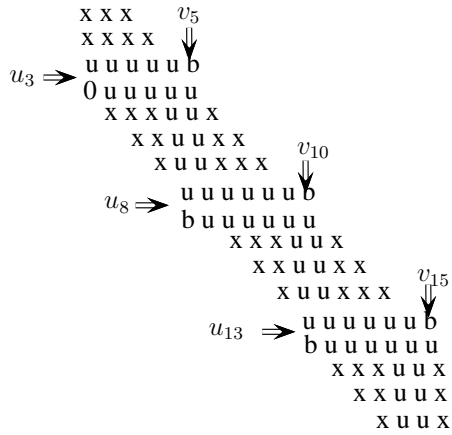

Figure 4.6: The first sweep in the reduction of a banded matrix $A \in \mathbb{R}^{17 \times 17}$ with lower bandwidth $t = 3$ and upper bandwidth $m = 2$ to upper bidiagonal form using rotators.

The direct bidiagonal reductions discussed in Section 4.2.2 are not the methods of choice when the input is a banded matrix. Those algorithms, when applied on a banded matrix, fill the matrix with unwanted entries in a few steps, and therefore fail to take advantage of its sparsity [21, 77, 78]. The LAPACK library [7] contains a rotation based bidiagonal reduction algorithm that chases a bulge along the band and out of the matrix as soon as it is formed. One sweep of this algorithm is demonstrated in Figure 4.6. In the Figure, 'x' denotes an unchanged element, 'u' denotes an element that is modified during the current sweep, 'b' denotes an element of a bulge and '0' denotes an element that is zeroed during the current sweep. This algorithm starts by annihilating $a_{t+1\,1}$ using a rotator $u_{t\,t+1}$ along rows $t$ and $t + 1$. Applying of this rotator generates a bulge at position $(t\,t + m + 1)$ above the band. This bulge is chased immediately using a right rotation $v_{t+m\,t+m+1}$ on columns $t + m$ and $t + m + 1$. A new bulge forms at position $(2t + m + 1\,t + m)$ below the band. Chase it too, and continue like this until a bulge is chased out of the matrix. The cost of this is $O(\min\{N, P\})$ flops and $O(\min\{N, P\}/B)$ I/Os. The next sweeps annihilate $a_{t\,1}, \ldots, a_{2\,1}$ in that order. After this $a_{1\,m+1}, \ldots, a_{1\,3}$ are eliminated in that order. Now the first row and the first column are done. Continue for $O(\min\{N, P\})$ times to reduce the

---

**Algorithm 4.9.** *Reduction of a banded matrix to bidiagonal form using Householders [79]*

**Input:** *A banded matrix $A \in \mathbb{R}^{N \times P}$, $N > P$ of bandwidth $t$*

**Output:** *A is overwritten by the upper bidiagonal matrix.*

---

**for** $j = 1, \ldots, P$ **do**

     *Let $X = A[j : N; \; j : P]$; $x = P - j + 1$;*

     *Partition $X$ as shown above;*

     *Find a Householder $U_1^{(j)}$ for reducing $A_{11}$;*

     $A_{11} := U_1^{(j)}.A_{11}$;

     **for** $k = 2, \ldots, k_{max}$ **do**

         $A_{k-1\,k} := U_{k-1}^{(j)}.A_{k-1\,k}$;

         *Find a Householder $V_k^{(j)}$ for reducing $A_{k-1\,k}$'s first row*

         $A_{k-1\,k} := A_{k-1\,k}.V_k^{(j)}$;    $A_{k\,k} := A_{k\,k}.V_k^{(j)}$;

         *Find a Householder $U_k^{(j)}$ for reducing $A_{k\,k}$'s first column*

         $A_{k\,k} := U_k^{(j)}.A_{k\,k}$;

     **endfor**

**endfor**

---

matrix to upper bidiagonal form. The reduction takes $O(\min\{N, P\}^2 (t + m))$ flops and $O(\min\{N, P\}^2 (t + m)/B)$ I/Os.

The seek complexity of the algorithm is $O(\min\{N, P\}^2 (t + m))$.

In [61, 78, 79], parallel and sequential algorithms are proposed for the bidiagonalisation of a banded matrix using Householder transformations. We discuss an adaptation of an algorithm from [79]. This algorithm is similar to the Householders based tridiagonalisation, we saw earlier and is described in Algorithm 4.9.

Given a banded matrix $A \in \mathbb{R}^{N \times P}$ of bandwidth $t$ with $N \geq P$, this algorithm repeatedly performs the following partition of a matrix $X \in \mathbb{R}^{(N-P+x) \times x}$ where $x = P - j + 1$:

• the first block column consists of the first column of $X$

• for $2 \leq k < k_{max} = \lceil (x-1)/(t+m) \rceil + 1$, the $k$-th block column consists of columns $(k-2)(t+m) + 2$ to $(k-1)(t+m) + 1$ of $X$.

• the $k_{max}$-th block column consists of columns $(k_{max} - 2)(t + m) + 2$ to $x$ of $X$.

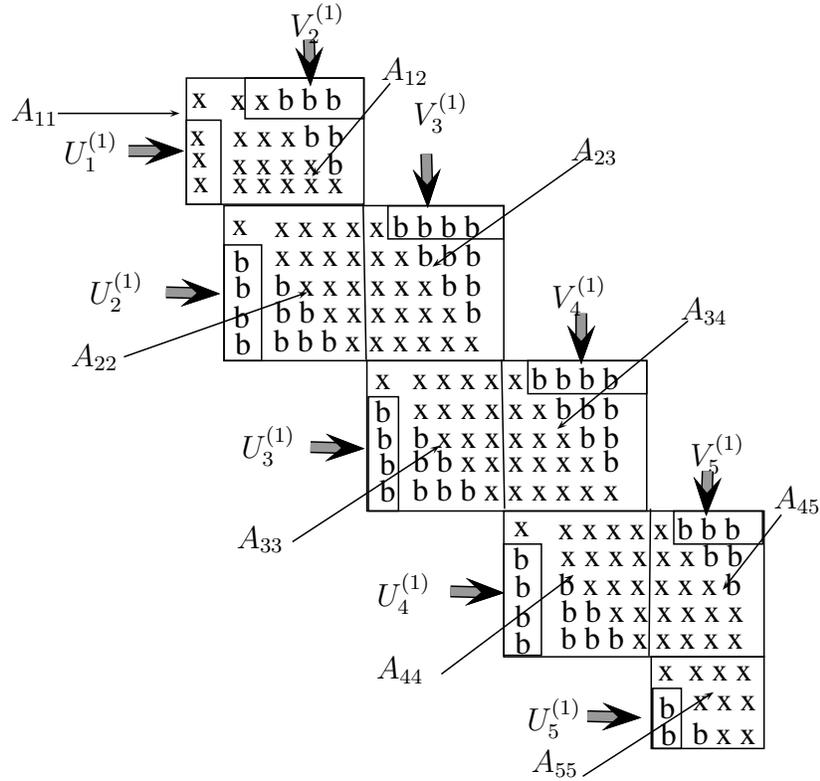

Figure 4.7: The first sweep in the reduction of a banded matrix $A \in \mathbb{R}^{22 \times 20}$ with lower bandwidth $t = 3$ and upper bandwidth $m = 2$ to upper bidiagonal form using Householder transformations.

- the first block row consists of rows 1 to $t + 1$ of $X$

- for $2 \leq k < k_{max} = \lceil (x - 1)/(t + m) \rceil + 1$, the $k$-th block row consists of rows $(k - 2)(t + m) + t + 2$ to $(k - 1)(t + m) + t + 1$ of $X$.

- the $k_{max}$-th block row consists of rows $(k_{max} - 2)(t + m) + t + 2$ to $x$ of $X$.

Let $X_{i,j}$ denote the tile that is the intersection of the $i$-th block row and $j$-th block column of $X$.

The first sweep of the algorithm is demonstrated in Figure 4.7. In the Figure 'x' denotes an element that was nonzero before the sweep, 'b' denotes an element of a bulge, and the small boxes inside tile denote the elements that are to be zeroed during the sweep.

This algorithm takes $O(P^2(t+m))$ flops and $O(P^2(t+m)/B)$ I/Os. In general, bidiagonal reduction using this algorithm incurs $O(\min\{N, P\}^2(t+m)/B)$ I/Os.

To reduce each column, the leading banded matrix is partitioned into blocks which takes $O(\min\{N, P\}(t+m)/\sqrt{M} + \min\{N, P\})$ seeks. So the seek complexity of the reduction algorithm is $O(\min\{N, P\}^2(t+m)/\sqrt{M} + \min\{N, P\}^2)$.

## 4.5.1　The two step bidiagonal reduction

Without loss of generality, we assume that the banded matrix is upper banded. Let $t$ be its bandwidth. The Algorithm 4.9 which takes a banded matrix of nonzero lower and upper bandwidths can easily be modified to handle upper banded matrices; the asymptotic complexity remains the same.

In Subsection 4.2.3, we saw that both the unblocked and blocked reductions of a full matrix to bidiagonal form takes $O(N^3/B)$ I/Os. The I/O performance can be improved if the reduction is split into two steps [61, 78, 79]:

1. Reduce the matrix to banded form

2. Reduce the banded matrix to bidiagonal form

The first step, we saw in Chapter 3, takes $O\left(\frac{NP\min\{N,P\}}{tB}\right)$ I/Os if $t < \sqrt{M}$ and $O\left(\frac{NP\min\{N,P\}}{\sqrt{M}B}\right)$ I/Os if $t > \sqrt{M}$, where $t$ is the bandwidth of the banded matrix. We analysed the second step above.

We calculate the combined I/O complexity of the two steps, and pick the $t$ that optimises this quantity. We conclude the following:

- If $\min\{N, P\} > \sqrt{M}$ and $\max\{N, P\} > M$, choose $t = \Theta(\sqrt{M})$. The total I/O cost will be $O\left(\frac{\min\{N,P\}.NP}{\sqrt{M}B}\right)$. That is, for large matrices, our algorithm matches the lower bound on data movement for the problem [13].

- If $\sqrt{M} < P, N < M$, choose $t = \Theta(\sqrt{\max\{N, P\}})$. The total I/O cost will be $O\left(\frac{\min\{N,P\}NP}{\sqrt{\max\{N,P\}}B}\right)$.

Thus, our two stage reduction process improves the asymptotic I/O complexity of bidiagonal reduction by the following factors, over the direct single step unblocked and blocked reductions.

- a factor of $\sqrt{M}$, if $\min\{N, P\} > \sqrt{M}$ and $\max\{N, P\} \geq M$

- a factor of $\sqrt{\max\{N, P\}}$, otherwise if $\sqrt{M} < P, N < M$

The seek complexity of the two stage reduction process is $O(\min\{N, P\}^2 + NP\min\{N, P\}/(M\sqrt{M}))$ with $t \approx \sqrt{M}$, if $\min\{N, P\} > \sqrt{M}$ and $\max\{N, P\} > M$, and $O(\min\{N, P\}^2\sqrt{\max\{N, P\}}/\sqrt{M})$ with $t \approx \sqrt{\max\{N, P\}}$, if $\sqrt{M} < P, N < M$.

# Chapter 5

# The Generalised Eigenvalue Problem

## 5.1 Introduction

Many scientific and engineering applications [98, 99, 104, 105] instantiate the Generalised Eigenvalue Problem (GEVP), where we seek to find, for two matrices $A, B \in \mathbb{C}^{N \times N}$, every $\lambda$ that for some nonzero $v$ satisfies

$$Av = \lambda Bv \tag{5.1}$$

The GEVP becomes the standard eigenvalue problem (EVP) when $B = I$. A nonzero $v$ that satisfies Equation 5.1 for some value of $\lambda$ is called an eigenvector of the ordered pair $(A, B)$, and $\lambda$ is called the associated eigenvalue; $\lambda$ and $v$ together form an eigenpair. $Av = \lambda Bv$ implies that $(\lambda B - A)v = 0$, which in turn implies that $\lambda$ is an eigenvalue of $(A, B)$ if and only if the characteristic equation

$$\det(\lambda B - A) = 0$$

is satisfied. $\det(\lambda B - A)$ is a polynomial of degree $N$ or less and is called the characteristic polynomial of the pair $(A, B)$. The form $\lambda B - A$ is called a *matrix pencil*.

The pair $(A, B)$ has $N$ finite nonzero eigenvalues, counting multiplicity, iff $rank(B) = N$. If $B$ is rank deficient, then the $N$ eigenvalues of $(A, B)$ include 0's and $\infty$'s [115]. The matrix pair $(A, B)$ is said to be *singular* if $A$ and $B$ are singular. The pair is called *regular* otherwise.

Two pairs of matrices $(A, B)$ and $(\tilde{A}, \tilde{B})$ are said to be *equivalent* if there exist nonsingular matrices $U$ and $V$ such that $\tilde{A} = UAV$ and $\tilde{B} = UBV$, which implies $\tilde{A} - \lambda\tilde{B} = U(A - \lambda B)V$. Equivalent pairs have the same characteristic equation, and hence the same eigenvalues. An equivalence transformation converts a pair into an equivalent one.

If matrices $A$ and $B$ are symmetric, $B$ is positive definite, and for an upper triangular $R$ the Cholesky decomposition of $B$ is $RR^T$, then $(\lambda, R^T v)$ is an eigenpair of $R^{-1}AR^{-T}$ iff $(\lambda, v)$ is an eigenpair of $(A, B)$. If $A$ and $B$ are symmetric, $A$ is positive definite, and for an upper triangular $R$ the Cholesky decomposition of $A$ is $RR^T$, then $(1/\lambda, R^T v)$ is an eigenpair of $R^{-1}BR^{-T}$ iff $(\lambda, v)$ is an eigenpair of $(A, B)$. That is, in either case, an equivalence transformation with $U = R^{-1}$ and $V = R^{-T}$ reduces the GEVP instance into an EVP instance with a symmetric coefficient matrix, which can then be solved using either the symmetric QR-algorithm, Jacobi method or the Bisection method for the symmetric EVP [58, 87, 115]. Efficient out-of-core Cholesky factorisation is discussed in [16].

If $B$ is a nonsingular matrix, then $(\lambda, v)$ is an eigenpair of $(A, B)$ iff $(\lambda, v)$ is an eigenpair of $B^{-1}A$ iff $(\lambda, Bv)$ is an eigenpair of $AB^{-1}$. If $A$ is a nonsingular matrix, then $(\lambda, v)$ is an eigenpair of $(A, B)$ iff $(1/\lambda, v)$ is an eigenpair of $A^{-1}B$ iff $(1/\lambda, Av)$ is an eigenpair of $BA^{-1}$. Therefore, most instances of the GEVP can be reduced to the EVP. But, for a number of reasons, this may not be the best way to solve the GEVP [115]. Suppose $B$ is nonsingular, and consider the reduction to the EVP on $AB^{-1}$. If $B$ is ill conditioned, then the eigenvalues of the computed $AB^{-1}$, even if they are well conditioned, can be poor approximations to the eigenvalues of the pair $(A, B)$. Even if $A$ and $B$ are symmetric, $AB^{-1}$ may not be symmetric and it might be desirable to preserve symmetry. Even if $A$ and $B$ are sparse, $AB^{-1}$ may not be, and a computation with it could be therefore expensive. Thus it is useful to develop direct algorithms to solve the GEVP.

Schur theorem for the EVP guarantees that for every matrix $A \in \mathbb{C}^{N \times N}$ there exists a unitary matrix $U \in \mathbb{C}^{N \times N}$ such that $T = U^* A U$, where $T$ is upper triangular. That means $A$ is unitarily similar to $T$. The analogous result for the GEVP is called "the Generalized Schur Theorem" and is stated as follows [115]: "Let $A, B \in \mathbb{C}^{N \times N}$. Then there exists unitary $Q, Z \in \mathbb{C}^{N \times N}$ and upper triangular $T, S \in \mathbb{C}^{N \times N}$ such that $Q^* A Z = T$ and $Q^* B Z = S$. Thus $Q^*(A - \lambda B)Z = T - \lambda S$". That means $(A, B)$ is unitarily equivalent to a triangular-triangular pair $(T, S)$. If $(T, S)$ could be found, then the eigenvalues could be read off as $T[k, k]/S[k, k]$, for $1 \le k \le N$.

A matrix $T \in \mathbb{R}^{n \times n}$ is called quasi-triangular if it has the block upper triangular form

$$T = \begin{bmatrix} T_{11} & T_{12} & \cdots & T_{1m} \\ 0 & T_{22} & & T_{2m} \\ \vdots & \ddots & \ddots & \vdots \\ 0 & \cdots & 0 & T_{mm} \end{bmatrix},$$

where each main diagonal block is either $1 \times 1$ or $2 \times 2$, and each $2 \times 2$ block has complex eigenvalues. If $A$ and $B$ are real matrices then the real counterpart of the Wintner-Murnaghan Theorem applies. It is stated as follows: "If $A$ and $B$ are in $\mathbb{R}^{N \times N}$, then there exist orthogonal matrices $Q$ and $Z$ such that $Q^T A Z$ is upper quasi-triangular and $Q^T B Z$ is upper triangular." That means $(A, B)$ is orthogonally equivalent to a triangular–quasi-triangular pair.

Abel's Impossibility Theorem shows that there is no general formula for the roots of a polynomial equation of degree $N$, if $N > 4$. Consequently, there is no general formula for the eigenvalues of an $N \times N$ matrix, if $N > 4$ [115]. No algorithm can find the equivalence transformations guaranteed by Schur Theorem in a finite number of steps. Therefore, as in the case of EVP, we need iterative algorithms to solve GEVP.

Such algorithms for the GEVP are known, and typically proceed in two stages. In the first stage, the input pair $(A, B)$, both matrices of which could be full in general, is reduced first to an equivalent full-triangular (F-T) pair, and then to an equivalent Hessenberg-Triangular (H-T) pair. In the second stage, an iterative method called QZ algorithm [58, 86, 115, 117] is applied to the reduced matrix pair. The first stage takes $O(N^3)$ flops in these algorithms. But these algorithms are rich in vector-vector (V-V) and vector-matrix (V-M) operations, and not in matrix-matrix (M-M) operations.

To improve the performance, it has been proposed that the reduction in the first stage be split into two steps [1, 37, 70], the first step reducing a full-full (F-F) pair to an F-T pair and then to a banded Hessenberg-Triangular pair (BH-T) [1, 35–37, 70], and the second step further reducing it to an H-T pair [1, 37, 70]. All reductions use equivalence transformations. Though the modified first stage takes more flops, its richness in M-M operations, makes it more efficient on machines with multiple levels of memory.

For reducing an F-T pair to a BH-T pair, tile based algorithms are known [1, 35, 36].

In this chapter we focus on the first stage. We analyse some known algorithms for their I/O complexity. We present a new algorithm too.

In all the algorithms we discuss, it is assumed that we begin with an F-T pair, as the transformation from an F-F pair to an F-T pair can be achieved through a QR-decomposition.

## 5.1.1   Organisation of this Chapter

In Section 2, we study some of the unblocked direct H-T reductions, and their I/O performances. Reduction to H-T pair using the slab approach is discussed in Section 3. We discuss the tile based reduction algorithms for F-T pair to BH-T pair in Section 4. Reduction of BH-T pair to H-T pair is discussed in Section 5.

## 5.2    Unblocked Reduction from Full-Triangular Form to Hessenberg-Triangular Form

See Algorithm 5.1. This algorithm was first presented in [86]. It uses Givens rotations. As can be readily seen, it requires $O(N^3)$ flops. Use of fast Givens rotations here can approximately halve the computational costs [84].

---

**Algorithm 5.1.** *Reduction from F-T form to H-T form [86]*

**Input:** *Matrices $A, B \in \mathbb{R}^{N \times N}$, where $B$ is upper triangular*

**Output:** *Orthogonal matrices $Q, Z \in \mathbb{R}^{N \times N}$ such that $(H, T) = (Q^T A Z, Q^T B Z)$ is an H-T pair. $H$ and $T$ overwrite $A$ and $B$, respectively.*

**Remark:** *$G_i \in \mathbb{R}^{N \times N}$ denotes a Givens rotation acting on rows/columns $i - 1$ and $i$. $I_N$ stands for the identity matrix of order $N$.*

---

*Set $Q \leftarrow I_N$; $Z \leftarrow I_N$;*
**for** $j = 1$ **to** $N - 2$ **do**
  **for** $i = N$ **to** $j + 2$ **step** $-1$ **do**
    *Construct $G_i$ such that the $(i, j)$-th entry of $G_i^T A$ is zero;*
    *Update $A \leftarrow G_i^T A$, $B \leftarrow G_i^T B$, $Q \leftarrow Q G_i$;*
    *Construct $G_i$ such that the $(i, i - 1)$-th entry of $B G_i$ is zero;*
    *Update $A \leftarrow A G_i$, $B \leftarrow B G_i$, $Z \leftarrow Z G_i$;*
  **end for**
**end for**

---

The total amount of data the algorithm needs to move, measured in blocks, assuming that $Q$ and $Z$ need not be explicitly constructed, is $O(N^3/B)$. Explicit construction of $Q$ and $Z$ matrices also requires to move $O(N^3/B)$ data blocks.

Assume that the matrices are stored in column major order and that $N > B$. Then accessing a row will require a seek per element. Therefore, the seek complexity of the algorithm is $O\left(\sum_{j=1}^{N-2} \sum_{i=j+2}^{N} N - j\right) = O(N^3)$, for an overall I/O complexity of $O(N^3)$.

### 5.2.1   An $O(1)$ improvement

A drawback of the above algorithm is that the Givens rotations are applied as soon as they are constructed. The performance could be improved if a number of rotations are accumulated and applied together. The algorithm of [70] does that; the rotators that reduce $A[j + 2 : N; \; j]$ (respectively, annihilate the subdiagonal bulges in $B$ created by them) are not applied on $A$ and $B$ as soon as they are formed, but are accumulated then applied together from the left (resp., right). See Algorithm 5.2.

---

**Algorithm 5.2.** *Reduction from F-T form to H-T form [70]*

---

**Input:** *Matrices $A, B \in \mathbb{R}^{N \times N}$, where $B$ is upper triangular*

**Output:** *Orthogonal matrices $Q, Z \in \mathbb{R}^{N \times N}$ such that $(H, T) = (Q^T AZ, Q^T BZ)$ is an*
    *H-T pair. $H$ and $T$ overwrite $A$ and $B$, respectively.*

**Remark:** *row-givens and column-givens apply a sequence of Givens rotations to the rows*
    *and columns of a matrix respectively.*

---

*Set $Q \leftarrow I_N$, $Z \leftarrow I_N$*
**for** *$j \leftarrow 1, 2, \ldots, N - 2$* **do**
    *Construct a sequence of Given rotations $G_N, \ldots, G_{j+2}$ to reduce $A(j + 2 : N; \; j)$;*
    *row-givens($G_N, \ldots, G_{j+2}$, A);*
    *row-givens($G_N, \ldots, G_{j+2}$, B);*
    *column-givens($G_N, \ldots, G_{j+2}$, Q);*
    *Construct a sequence of Given rotations $G_N, \ldots, G_{j+2}$ to annihilate the subdiagonal*
      *bulges in $B$;*
    *column-givens($G_N, \ldots, G_{j+2}$, B);*
    *column-givens($G_N, \ldots, G_{j+2}$, A);*
    *column-givens($G_N, \ldots, G_{j+2}$, Z);*
**end for**

---

The total amount of data moved remains $O(N^3/B)$, as the computation involved is the same. But the seek complexity is now improved. This can be seen as follows:

Consider the case of $N \leq M/4$. Without loss of generality, assume that the matrices are in column major order. Maintain the *sines* and *cosines* of the accumulated rotators in the main memory. Load a slab of $\Theta(M/N)$ columns of the matrix in the main memory, update the relevant portions of this slab and write back. All this takes $O(1)$ seeks. $\Theta(N^2/M)$ (resp., $\Theta(LN/M)$) slabs have to be handled for a left (resp., right) multiplication, where $L$ is the number of rotators accumulated. Note that the average value of $L$ is $\Theta(N)$ over the iterations. Therefore, the number of seeks is at most $N * (N^2/M) = O(N^3/M)$. The I/O complexity is, thus, $O(N^3/B)$.

The above will not work, if $N > M/4$. In that case, we can still transpose the matrix before and after every left application so that it is done on a row major form. The transpositions, over the iterations, cause a net data movement of $O(N^3/B)$ blocks and seeks that number $O(N^3/\sqrt{M})$. The rest of the algorithm will now incur $O(N^3/M)$ seeks. The I/O complexity of the algorithm is, thus, $O(N^3/B + N^3/\sqrt{M}) = O(N^3/B)$, if $B^2 < M$.

## 5.3    Slab Based Reduction from Full-Triangular Form to Hessenberg-Triangular Form

All the operations of the above algorithms are of V-M. Data locality can be further improved [70] by incorporating M-M operations using the techniques proposed in [80], which we now describe.

Let $G_i^{(j)}$ denote the Givens rotation that annihilates the $(i, j)$-th element of $A$. Note that for $k \geq 1$, $G_{i_1}^{(j)}$ and $G_{i_2}^{(j+k)}$ can commute if $i_2 > i_1 + k$. This observation allows us to regroup the rotators as follows: Divide $A$ into slabs of $b$ columns each. Consider the slab that spans columns $j + 1, \ldots, j + b$. The rotators formed for the slab are $G_N^{(j+1)}, \ldots, G_{j+3}^{(j+1)}, \ldots, G_N^{(j+b)}, \ldots, G_{j+b+2}^{(j+b)}$. (See Figure 5.1. Here $N = 13$ and $b = 4$. The figure shows the rotators generated for the first four columns; $j = 0$.) If the rotators are written as in the figure, they form of a trapezoid of height $b$ and parallel sides of length $N - j - 2$ and $N - j - b - 1$. Form groups by repeatedly cutting off rhombuses of side length $b$ from the right end of the trapezoid. The rhombuses will be limited on the left

$$
\begin{array}{l}
G_{13}^{(1)} \quad G_{12}^{(1)} \quad G_{11}^{(1)} \mid G_{10}^{(1)} \quad G_{9}^{(1)} \quad G_{8}^{(1)} \quad G_{7}^{(1)} \mid G_{6}^{(1)} \quad G_{5}^{(1)} \quad G_{4}^{(1)} \quad G_{3}^{(1)} \\[4pt]
G_{13}^{(2)} \quad G_{12}^{(2)} \mid G_{11}^{(2)} \quad G_{10}^{(2)} \quad G_{9}^{(2)} \quad G_{8}^{(2)} \mid G_{7}^{(2)} \quad G_{6}^{(2)} \quad G_{5}^{(2)} \quad G_{4}^{(2)} \\[4pt]
G_{13}^{(3)} \mid G_{12}^{(3)} \quad G_{11}^{(3)} \quad G_{10}^{(3)} \quad G_{9}^{(3)} \mid G_{8}^{(3)} \quad G_{7}^{(3)} \quad G_{6}^{(3)} \quad G_{5}^{(3)} \\[4pt]
\mid G_{13}^{(4)} \quad G_{12}^{(4)} \quad G_{11}^{(4)} \quad G_{10}^{(4)} \mid G_{9}^{(4)} \quad G_{8}^{(4)} \quad G_{7}^{(4)} \quad G_{6}^{(4)}
\end{array}
$$

Figure 5.1: Sequence of Givens rotations used to reduce the first 4 columns of a $13 \times 13$ matrix $A$.

by the left boundary of the trapezoid. Each rhombus corresponds to a group. The groups of a slab can be applied in the left-to-right order. The aggregate of each group will be a matrix of dimensions $2b \times 2b$ at the most.

The right side rotators can be grouped the same way.

The algorithm of [70] uses the above grouping, and proceeds as follows: Assume that $j$ columns of $A$ and $B$ have been reduced. Partition the matrix pair like this:

$$
A = \begin{array}{c} j+1 \\ N-j-1 \end{array}\begin{array}{c} \overset{j}{\phantom{A}} \overset{b}{\phantom{A}} \overset{N-j-b}{\phantom{A}} \\ \left[\begin{array}{ccc} A_{11} & A_{12} & A_{13} \\ 0 & A_{22} & A_{23} \end{array}\right], \end{array} \quad B = \begin{array}{c} \overset{j+1}{\phantom{B}} \overset{N-j-1}{\phantom{B}} \\ \left[\begin{array}{cc} B_{11} & B_{12} \\ 0 & B_{22} \end{array}\right] \begin{array}{c} j+1 \\ N-j-1 \end{array} \end{array}
$$

Consider the next $b$ columns (that is, $A_{22}$). Reduce the columns one by one. When the rotators for each column are found, apply them cumulatively, as in Algorithm 5.2, to $A_{22}$ and $B_{22}$ but not to $A_{23}$. After this, accumulate the rotators into groups as described above. Next find the right rotators that would reduce the bulges in $B$ just formed. Apply them cumulatively to $A_{22}, A_{23}$ and $B_{22}$, but not to $A_{12}, A_{13}$ and $B_{12}$. After this, accumulate these rotators into groups. When all the $b$ columns are reduced, aggregate each of the left side and right side groups into a $2b \times 2b$ matrix. Apply the left side aggregates in the correct order to $A_{23}$ from the left. Apply the right side aggregates similarly to $A_{21}, A_{31}$ and $B_{12}$ from the right using level-3 operations. Now $j + b$ columns of $A$ and $B$ are reduced. Proceed with the next slab of $b$ columns.

A rotator can be applied on a $2b \times 2b$ matrix kept in row major order in $O(b/B)$ I/Os. A group has at most $b^2$ rotators. Therefore a group can be aggregated in $O(b^3/B)$ I/Os. An aggregate can be applied on the pairs in $O(bN/B)$ I/Os, if $b = O(\sqrt{M})$.

For $1 \leq k \leq \lceil (N-2)/b \rceil$, the $k$-th slab produces $\lceil (N-(k-1)b)/b \rceil$ aggregates. Therefore, the total number of I/Os spent in aggregation is $O(N^2b/B)$ and the total number of I/Os spent in applying them is $O(N^3/bB)$. This is an improvement, but it is the rest of the algorithm that dominates the asymptotic I/O complexity. For example, the total cost of all left multiplications on $B$ is $O(N^3/B)$, as in Algorithms 5.1 and 5.2.

## 5.4   Tile Based Reduction from Full-Triangular Form to Banded Hessenberg-Triangular Form

The above algorithms are rich in V-V and V-M operations. To improve the performance, it has been proposed that the reduction from F-T form to H-T from be split into two steps [1, 2, 35–38, 70], the first step reducing an F-T pair to a banded Hessenberg-Triangular pair (BH-T), and the second step further reducing it to an H-T pair. In this section, we discuss a tile based algorithm for the first step. This algorithm is similar to Algorithm 3.2.

### 5.4.1   An algorithm for $t \leq \sqrt{M}$

Without loss of generality, assume that $N$ is an integer multiple of $t$. Partition the matrix pair $(A, B)$ into $N/t \times N/t$ tiles $A_{ij}$ and $B_{ij}$ of size $t \times t$ each. Assume that a tile-transposition has been performed, and the tiles are available in row major order. See Algorithm 5.3. This is an adaptation from [70].

As shown in [35], annihilations of the subdiagonal blocks and the fill-in elements can be done using various ways, i.e., using Givens rotations or Householder transformations or combination of Householders and Givens rotations. We use Householders to reduce subdiagonal blocks.

The functions "QR-decompose", "QR-Update", "QR-Multiply-from-left-1", and "QR-Multiply-from-right-2" were described in Chapters 2 and 3. We now describe the other functions.

**Function RQ-decompose**$(B_{ij})$**:**

Assume that the tile $B_{ij}$ is divided into horizontal slabs of width $k$ each. Let $r = t/k$. The invocation RQ-decompose finds matrices $Y_1, \ldots, Y_r, T_1, \ldots, T_r$ such that $R_{ij} = B_{ij}(I_t + Y_1 T_1^T Y_1^T) \ldots (I_t + Y_r T_r^T Y_r^T)$ is upper triangular. This is done as follows:

Read $B_{ij}$ into the main memory. Proceed in $r$ iterations. At the beginning of the $s$-th iteration, the subdiagonal elements of the first $(s-1)k$ rows of $B_{ij}$ are all zero. In the $s$-th iteration, calculate $Y_s$ and $T_s$ such that the subdiagonal elements of $B_{ij}(I_t + Y_s T_s^T Y_s^T)$ in the first $sk$ rows are all zero. Update $B_{ij}$ to $B_{ij}(I_t + Y_s T_s^T Y_s^T)$. Store $Y_s$ in the subdiagonal part of the $s$-th horizontal slab of $B_{ij}$.

When all slabs are processed, arrange the upper triangular part of $B_{ij}$ in column major order and lower triangular part in row major order, and write $B_{ij}$ back to the disk.

The number of I/Os needed is $2t^2/B$; $B_{ij}$ is read and written; the $B_{ij}$ and $T$ matrices are kept in the main memory till the right updates of the corresponding block columns of $A$ and $B$ are finished.

The number of seeks is $O(1)$ even for the one tile case.

**Function RQ-Update**$(B_{jj-1}, B_{jj})$

Here $B_{jj-1}$ and $B_{jj}$ are upper triangular, the invocation RQ-Update$(B_{jj-1}, B_{jj})$, finds a matrix $Q$ such that in $BQ^T$, every element of the $(j, j-1)$-th tile is zero, and the

---

**Algorithm 5.3.** *A tile based reduction from F-T form to BH-T form*

---

**Input:** *Matrices $A, B \in \mathbb{R}^{N \times N}$, where $B$ is upper triangular, and $t$, the desired bandwidth*

**Output:** *Orthogonal matrices $Q, Z \in \mathbb{R}^{N \times N}$ such that $(H, T) = (Q^T A Z, Q^T B Z)$ is a BH-T pair. $H$ and $T$ overwrite $A$ and $B$, respectively.*

---

**for** $i = 1$ **to** $N/t - 1$ **do**
  **for** $j = i + 1$ **to** $N/t$ **do**
    *QR-decompose*$(A_{ji})$;
        **for** $k = i + 1$ **to** $N/t$ **do**, *QR-Multiply-from-left-1*$(A_{ji}, A_{jk})$, **endfor**;
        **for** $k = j$ **to** $N/t$ **do**, *QR-Multiply-from-left-1*$(A_{ji}, B_{jk})$, **endfor**;
        /* $B_{jj}$ fills out forming a bulge */
  **endfor**;
  **for** $j = N/t$ **to** $i + 2$ **step** $-1$ **do**
    *QR-Update*$(A_{j-1\,i}, A_{ji})$;
        **for** $k = i + 1$ **to** $N/t$ **do**, *QR-Multiply-from-left-2*$(A_{ji}, A_{jk}, A_{j-1\,k})$, **endfor**;
        **for** $k = j - 1$ **to** $N/t$ **do**, *QR-Multiply-from-left-2*$(A_{ji}, B_{jk}, B_{j-1\,k})$, **endfor**;
        /* $B_{j\,j-1}$ fills out forming a bulge */
    *RQ-decompose*$(B_{j\,j-1})$;
        /* $B_{j\,j-1}$ is now upper triangular */
        **for** $k = 1$ **to** $N/t$ **do**, *RQ-Multiply-from-right-1*$(B_{j\,j-1}, A_{l\,j-1})$, **endfor**;
        **for** $k = 1$ **to** $j - 1$ **do**, *RQ-Multiply-from-right-1*$(B_{j\,j-1}, B_{l\,j-1})$ **endfor**;
    *RQ-decompose*$(B_{jj})$;
        /* $B_{jj}$ is now upper triangular */
        **for** $k = 1$ **to** $N/t$ **do**, *RQ-Multiply-from-right-1*$(B_{jj}, A_{lj})$ **endfor**;
        **for** $k = 1$ **to** $j - 1$ **do**, *RQ-Multiply-from-right-1*$(B_{jj}, B_{lj})$ **endfor**;
    *RQ-Update*$(B_{j\,j-1}, B_{jj})$;
        /* Now $B_{j\,j-1}$ is zero and $B_{jj}$ is upper triangular */
        **for** $k = 1$ **to** $N/t$ **do**, *RQ-Multiply-from-right-2*$(B_{j\,j-1}, A_{l\,j-1}, A_{lj})$ **endfor**;
        **for** $k = 1$ **to** $j - 1$ **do**, *RQ-Multiply-from-right-2*$(B_{j\,j-1}, B_{l\,j-1}, B_{lj})$ **endfor**;
  **endfor**;
  /* A bulge remains in $B_{i+1\,i+1}$ */
  *RQ-decompose*$(B_{i+1\,i+1})$;
    **for** $k = 1$ **to** $N/t$ **do**, *RQ-Multiply-from-right-1*$(B_{i+1\,i+1}, A_{l\,i+1})$ **endfor**;
    **for** $k = 1$ **to** $i$ **do**, *RQ-Multiply-from-right-1*$(B_{i+1\,i+1}, B_{l\,i+1})$ **endfor**;
  /* Now $B$ is upper triangular, again */
**endfor**;

---

$(j, j)$-th tile is upper triangular. This is done over $r = t/k$ iterations. The $s$-th iteration processes $B_{jj}^{(s)}$, the $s$-th rightmost vertical slab of width $k$ of $B_{jj}$. Consider the matrix $C_s = \left( \begin{array}{cc} B_{jj-1} & B_{jj}^{(s)} \end{array} \right)$. Let $\left( \begin{array}{c} Y_s \\ I_k \end{array} \right)$ and $T_s$ be matrices such that $C_s \left( I_{k+t} + \left( \begin{array}{c} Y_s \\ I_k \end{array} \right) T_s^T \left( \begin{array}{cc} Y_s^T & I_k \end{array} \right) \right)$ has zeroes in the leftmost $t$ columns of its $s$-th bottom-most horizontal slab. Then,

$$\left( \begin{array}{cc} B_{jj-1} & B_{jj} \end{array} \right) \left( I_{2t} + \left( \begin{array}{c} Y_s \\ 0_{(t-sk)} \\ I_k \\ 0_{(s-1)k} \end{array} \right) T_s^T \left( \begin{array}{cccc} Y_s^T & 0_{(t-sk)} & I_k & 0_{(s-1)k} \end{array} \right)^T \right)$$

too zeroes in the leftmost $2t - sk$ columns of its $s$-th bottom-most horizontal slab. Set $\left( \begin{array}{cc} B_{jj-1} & B_{jj} \end{array} \right)$ to this matrix. The $s$-th slab of $B_{jj-1}$ is now all zero.

The above can be implemented as follows: Read $B_{jj-1}$. Read $B_{jj}$ one vertical slab at a time. For each slab, update the slab, write it back and overwrite the corresponding horizontal slab of $B_{jj-1}$ with the $Y$ matrix. The number of I/Os is at most $3t^2/B$, irrespective of whether one tile or two tiles fit in the main memory: one tile and a half are both read and written.

Since both the tiles are upper triangular, with their nonzero elements stored contiguously during the RQ-decomposition, the number of I/Os is $2t^2/B$: two half tiles are both read and written. The $Y$ matrices is kept in-core for the future right updates.

The number of seeks is $O(t/k)$.

**Function RQ-Multiply-from-right-1**$(B_{ij}, D_{kj})$**:**

Invocation of this function after an RQ-decomposition multiplies tile $D_{kj}$ ($D$ here can be $A$ or $B$) from the right with $Q^T$.

Implementation: Read horizontal slabs of $D_{kj}$ into the main memory, update using the in-core $Y$ and $T$ matrices and write back. The number of I/Os is $2t^2/B$: read $D_{kj}$ and write back.

The number of seeks is $O(t/k)$.

**Function RQ-Multiply-from-right-2**$(B_{j\,j-1}, D_{l\,j-1}, D_{l\,j})$**:**

Invocation of this function after an RQ-Update over $B_{j\,j-1}$ and $B_{j\,j}$ updates $D_{l\,j-1}$ and $D_{l\,j}$ ($D$ can be $A$ or $B$) from the right using $Q^T$. The computation proceeds as follows:

Read $D_{l\,j-1}$. Read a vertical slab from $D_{l\,j}$ and the corresponding horizontal slab from $B_{j\,j-1}$. Update the vertical slab and $D_{l\,j-1}$ using the horizontal slab, and write the vertical slab back. When all slabs have been processed, write $D_{l\,j-1}$ back. If only one tile fits in the main memory, then the number of I/Os is $5t^2/B$: $D_{l\,j-1}$ and $D_{l\,j}$ are read and written; $B_{j\,j-1}$ is read. If two tiles fit in the main memory, then $B_{j\,j-1}$ can be kept in the main memory, as $l$ varies, and so the number of I/Os is $4t^2/B$.

The number of seeks is $O(t/k)$ if one or two tiles fit in the main memory; $O(1)$ if three tiles fit in the main memory.

### 5.4.1.1   Analysis of I/O and seek complexities

The I/O complexity of the one tile implementation is $\frac{11.625N^3}{tB} + O\left(\frac{N^2}{tB}\right)$. The I/O complexities of the two and three tile implementations are $\frac{9.833N^3}{tB} + O\left(\frac{N^2}{tB}\right)$.

If four or more tiles fit in the main memory, then the two single tile RQ-decompositions can be avoided. Instead of invoking RQ-decompose on $B_{jj-1}$ and $B_{jj}$ separately and then applying RQ-Update on them together, we can RQ-decompose ( $B_{j-1}\ \ B_{jj}$ ) directly. The resultant $Y$ and $T$ matrices can be kept in the main memory and while updating the $(j-1)$-th and $j$-th tile columns. This saves a little on the cost. Note that in each $i$-loop, there will be one single tile RQ-decompose; that of $B_{i+1\,i+1}$.

If one or two tiles fit in the main memory, the seek complexity is $O\left(\frac{N^3}{kt^2}\right)$. If three or more tiles fit, the seek complexity is $O\left(\frac{N^3}{t^3}\right)$.

As discussed in earlier chapters, the single, two and three tile implementations assume that $t^2 + 4kt \leq M$, $2t^2 + 3kt \leq M$, and $3t^2 + 2kt \leq M$ respectively. The parameter $k$ is usually kept small.

The number of I/Os required to handle the $T$ matrices is $O\left(\frac{N(N-t)k}{tB}\right)$, as in the other tile based algorithms.

## 5.4.2   Reduction to the Banded Hessenberg-Triangular form when $p \geq 2$ tiles fit in the main memory

We now briefly describe an algorithm that is applicable when $p \geq 2$ tiles fit in the main memory [37, 70], and then analyse it for its I/O and seek complexities.

Let $(A, B)$ be an F-T matrix pair. Let $A_{(i,p)}$ denote the submatrix of $A$ formed by the $p$ bottom-most tiles of its $i$-th tile column. QR-decompose $A_{(1,p)}$ into $Q_1 R_1$. Multiply the bottom-most $p$ tile rows of $A$ and $B$ with $Q_1^T$ from the left. A $pt \times pt$ bulge forms in the bottom-right corner of $B$. Let $\hat{B}$ denote the submatrix of the bulge formed by excluding its topmost tile row.

Suppose $R_2 Q_2$ is an RQ-decomposition of $\hat{B}$. Then $\hat{B} Q_2^T = R_2$. Let $Q_2^T = (Q_3^T\ Q_4^T)$, where $Q_3$ comprises of the first tile row of $Q_2$. QR-factorise $Q_3^T$ into, say, $Q_5(D^T\ 0\ \dots\ 0)^T$, where $D \in \mathbb{R}^{t \times t}$ is a diagonal matrix in which each diagonal element is $\pm 1$. Note that here $Q_5$ is a product of $t$ Householder transformations, and hence can be represented as $Q_5 = I + Y_5 T_5 T_5^T$, where $T_5 \in \mathbb{R}^{t \times t}$ and $Y_5 \in \mathbb{R}^{pt \times t}$.

$$\hat{B} Q_5 \begin{pmatrix} I_t \\ 0 \\ \vdots \\ 0 \end{pmatrix} = \hat{B} Q_5 \begin{pmatrix} D \\ 0 \\ \vdots \\ 0 \end{pmatrix} D^{-1} = \hat{B} Q_3^T D^{-1} = \hat{B} Q_2^T \begin{pmatrix} D^{-1} \\ 0 \\ \vdots \\ 0 \end{pmatrix}$$

This implies that the leftmost tile column of $\hat{B}$ can be annihilated by applying $Q_5$ to the last $p$ tile columns of $B$. The rest of the bulge is not annihilated; it stays on as a harmless bulges. Now the bottom most $p-1$ tiles of the first tile column of $A$ are done. Reduce the next set of $p-1$ tiles. This would first introduce a $pt \times pt$ bulge adjacent along the diagonal to the harmless one left behind earlier; the first column of this bulge again can be zeroed. Like this when the whole of the first column of $A$ is reduced a sequence of harmless bulge would have appeared along the diagonal of $B$. These are called harmless because when the next tile column of $A$ is reduced, the bulges created will subsume these. When all the tile columns of $A$ are reduced and $A$ becomes banded Hessenberg, $B$ would have become triangular again.

### 5.4.2.1    Analysis of I/O and seek complexities

The number of flops required by the algorithm is $O(N^3)$ [70].

Let us assume that the main memory can accommodate two panels of size $pt \times t$ together with one and a half tiles (to store $T$ and temporary values); i.e. $M \geq (2p + 3/2)t^2$.

To RQ factorise a $(p-1)t \times pt$ matrix, for each tile row, starting from the bottom-most, reduce it and, keeping the $Y$ and $T$ matrices in the main memory, update the other tile rows, until the submatrix that remains to be reduced fits in the main memory. Then it is read, RQ factorised in-core and written back; let this happen at the $l$-th tile row; then $M = t^2((p-l)(p-l+1) + (p-l+1) + 3/2)$ or $l = (p+1) \pm \sqrt{M/t^2 - 3/2}$. Once the $Y$ and $T$ matrices are computed, the first $t$ rows of the $Q$ matrix can be computed easily without constructing the $Q$ matrix explicitly. The number of I/Os needed by all this is:

$$x(p) = \sum_{i=1}^{l-1} 2\left(\frac{(p-i)t \times (p-(i-1))t}{B} + \frac{t \times (p-(i-1))t}{B}\right) + 2\left(\frac{M + t \times (p-(l-1))t}{B}\right),$$

where $l = (p+1) \pm \sqrt{M/t^2 - 3/2}$.

Similarly, the RQ factorisation of a $pt \times pt$ matrix and construction first $t$ rows of the $Q$ matrix require $y(p)$ I/Os;

$$y(p) = \sum_{i=1}^{l-1} 2\left(\frac{(p-(i-1))t \times (p-(i-1))t}{B} + 2\frac{t \times (p-(i-1))t}{B}\right) + 2\left(\frac{M + t \times (p-(l-1))t}{B}\right),$$

where $l = (2p+3) \pm \sqrt{4M/t^2 - 5}$.

The QR factorisations and the subsequent left updates require

$$\frac{t^2}{B}\sum_{i=1}^{N/t-1}\left(\sum_{j=1}^{Z-1}\left(2p + \sum_{k=i+1}^{N/t} 2p + \sum_{l=a}^{N/t} 2p\right)\right) + \frac{t^2}{B}\sum_{i=1}^{N/t-1}\sum_{j=b}^{N/t}\left(2 + \sum_{k=i+1}^{N/t} 2 + \sum_{l=b}^{N/t} 2\right)$$

I/Os, where $Z = \lceil\frac{N/t-i-1}{p-1}\rceil$, $a = i+1+(j-1)(p-1)$ and $b = i+1+(Z-1)(p-1)$. Here the first term corresponds the QR factorisations of $pt \times t$ submatrices and the subsequent left updates. If the number of tiles $N/t - i - 1$ to be reduced in $i$-th tile column of $A$ is not an integer multiple of $p-1$, then there would a QR factorisation involving $p-1$ or fewer tiles in that column. The second term stands for those QR factorisations and the left updates.

The I/O cost of RQ decompositions and the subsequent QR decompositions and right updates is:

$$\sum_{i=1}^{N/t-p}\left(y(p) + \sum_{j=1}^{Z-2} x(p)\right) + \sum_{i=1}^{N/t-p}\sum_{k=1}^{N/t-b} x(N/t - b - (k-2)) + \sum_{i=N/t-p+1}^{N/t-1} y(N/t - i)$$

$$+ \frac{t^2}{B}\sum_{i=1}^{N/t-1}\left(\sum_{m=1}^{i} 2p + \sum_{j=1}^{Z-2}\left(\sum_{k=1}^{N/t} 2p + \sum_{l=1}^{a} 2p\right)\right) + \frac{t^2}{B}\sum_{i=1}^{N/t-1}\sum_{j=b}^{N/t}\left(2 + \sum_{k=1}^{N/t} 2 + \sum_{l=1}^{j-1} 2\right)$$

where $Z = \lceil\frac{N/t-i-1}{p-1}\rceil$, $a = i+1+(j-1)(p-1)$ and $b = i+1+(Z-1)(p-1)$.

Thus the overall I/O complexity is $O(N^3/tB)$. The seek complexity can be similarly shown to be $O\left(\frac{N^3}{t^3} + \frac{p^2 N^2}{t}\right)$.

If $pt \times pt + t \times pt + 3t^2/2 \leq M$, then the RQ factorisations can be performed in-core. The I/O complexity of the QR factorisations and the left updates remain same as the above. But the RQ factorisations and the corresponding right updates cost

$$\frac{1}{B}\sum_{i=1}^{N/t-p}\left(((pt)^2 + pt^2) + \sum_{j=1}^{Z-2} 2((p-1)pt^2 + t \times pt)\right) + \frac{1}{B}\sum_{i=1}^{N/t-p}\sum_{k=1}^{N/t-b} 2\left((c-1)ct^2 + ct^2\right)$$

$$+ \sum_{i=N/t-p+1}^{N/t-1} 2\frac{(t(N/t-i))^2 + t^2(N/t-i)}{B}$$

$$+\frac{t^2}{B}\sum_{i=1}^{N/t-1}\left(\sum_{m=1}^{i}2p+\sum_{j=1}^{Z-2}\left(\sum_{k=1}^{N/t}2p+\sum_{l=1}^{a}2p\right)\right)+\frac{t^2}{B}\sum_{i=1}^{N/t-1}\sum_{j=b}^{N/t}\left(2+\sum_{k=1}^{N/t}2+\sum_{l=1}^{j-1}2\right)$$

I/Os where $Z = \lceil\frac{N/t-i-1}{p-1}\rceil$, $a = i + 1 + (j-1)(p-1)$, $c = N/t - b - (k-2)$ and $b = i + 1 + (Z-1)(p-1)$.

The asymptotic I/O and seek complexities remain the same as before.

Clearly it is better to do the RQ factorisations in-core. So the parameter $p$ should be chosen accordingly.

### 5.4.3   An improvement

We suggest the following improvement over the above. Partition the matrix pair into $(p-1)t \times t$ sized rectangular tiles instead of $t \times t$ square tiles. Now run the above algorithm. Assume that each bulges formed fits in the main memory, and leaves room for the $Y$ and $T$ matrices of the QR factorisation and for temporary data. That is, $M \geq p(p-1)t^2 + 2(p-1)t^2 + t^2/2$. Reading or writing of $pt^2$ elements here does not take $p$ seeks as before, but only one seek.

The I/O complexity of the algorithm is same as the previous algorithm. But the seek time complexity improves to $O\left(\frac{N^3}{pt^2} + \frac{N^2}{t^2}\right)$.

### 5.4.4   The case of $t > \sqrt{M}$

Clearly, the I/Os required would be $O\left(\frac{N^3}{\sqrt{M}B}\right)$; run the above algorithm as it is.

## 5.5   Reduction from Banded Hessenberg-Triangular Form to Hessenberg-Triangular Form

Unblocked and blocked algorithms using Givens rotations are discussed in [37, 70]. We study these first.

### 5.5.1   Unblocked algorithms using Givens rotations

Let $(A, B)$ be the input BH-T pair; let $t$ be the band width of $A$. First eliminate element $(t+1, 1)$ of $A$ by applying a Givens rotation from the left on rows $t$ and $t+1$ of $A$ and $B$. This creates a bulge at position $(t+1, t)$ of $B$. Eliminate this bulge by applying a Givens rotation from the right on columns $t$ and $t+1$ of $A$ and $B$. This forms a bulge at position $(2t+1, t)$ of $A$. Continue to chase this bulge out along the $(t+1)$-st subdiagonal of $A$ and the subdiagonal of $B$. Now we are done with the processing of $A[t+1, 1]$. Repeat this with $A[t, 1], \ldots, A[3, 1]$, and the first column would be done. Repeat for the other columns.

If the matrix pair is stored in row major order and $B < N$, then accessing any $k$ consecutive elements of a row would require $k$ I/Os. So the I/O complexity of the algorithm is, approximately $O(N^3)$.

The above algorithm accesses each row and column of $A$ and $B$ twice. This is because, each rotation is applied as soon as it is formed. [70] presents an algorithm that delays the applying of the rotators. In the first iteration of this algorithm, the unwanted elements of the first column are eliminated by applying a sequence $G_{t\,t+1}, \ldots, G_{23}$ of Givens rotations from the left, where $G_{i\,i+1}$ applies over rows $i$ and $i+1$. This introduces nonzero subdiagonal elements at positions $(t+1, t), \ldots, (3, 2)$ of $B$.

Suppose the sequence $G_{t+1\,t}, G_{t\,t-1}, \ldots, G_{32}$ of Givens rotations eliminate the unwanted elements of $B$ when applied from the right. If these rotations were applied together to $A$, then a dense $t \times t$ bulge would form below the $t$-th subdiagonal of $A$. So, instead, apply the rotations one-by-one; the bulge caused by one is to be chased a distance of $t$ before applying the next. For example, first apply $G_{t+1\,t}$ to $A$ creating bulge at position $(2t+1, t)$. Chase this bulge by applying only on $A$ a rotation $G_{2t\,2t+1}$ from the left. Then apply the next right rotation $G_{t\,t-1}$ creating a bulge at position $(2t, t-1)$. Chase this using a rotation $G_{2t-1\,2t}$ from left. Apply the remaining right rotations in this manner, forming a sequence $G_{2t\,2t+1}, G_{2t-1\,2t}, \ldots, G_{t+2\,t+3}$ of left rotations; apply this sequence together on $B$ now. This shifts the subdiagonal bulges of $B$ over $t$ columns and rows towards the bottom right corner. Repeat this process till the bulges are all chased out the matrices. (These steps are illustrated in Figure 5.2.) Now the first column is reduced. Repeat this process for the other columns.

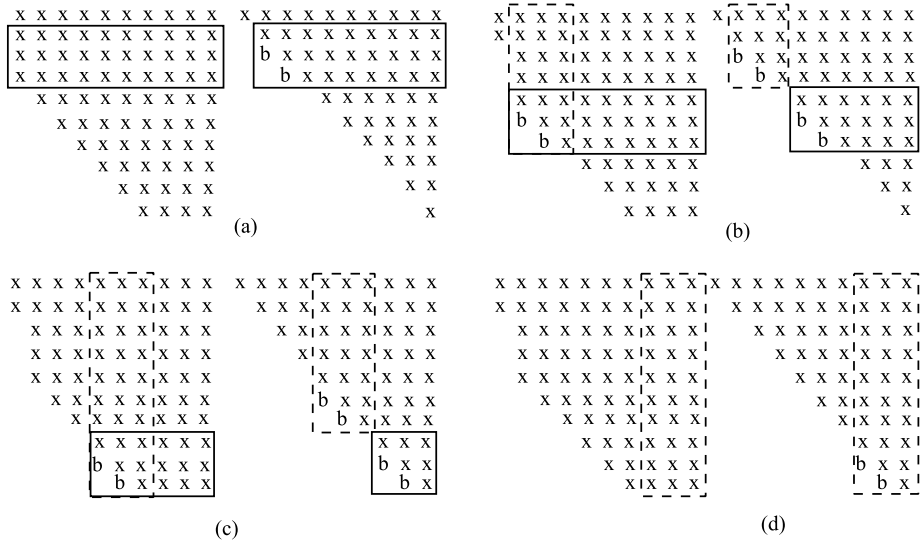

Figure 5.2: Reduction of first column of BH-T pair to H-T pair

If the matrices are in column major order, then an accumulated set of $t$ Givens rotations can be applied on them from the left in $O(Nt/B)$ I/Os and $O(N)$ seeks. Forming of the right rotations for annihilating the consequent bulges takes $O(t^2/B)$ I/Os and $O(t)$ seeks. Applying of these rotations from the right and forming the left rotations for annihilating the consequent bulges takes $O(Nt/B)$ I/Os and $O(t)$ seeks. Therefore, algorithm runs in $O(N^3/B)$ I/Os and $O(N^3/t)$ seeks.

### 5.5.2 Block algorithms using Givens rotations

A blocked algorithm for BH-T pair to H-T pair reduction is presented in [37], which we now briefly describe and analyse.

To reduce the $i$-th column of $A$, the submatrix pair $(A[1 : N; \ i + 1 : N], \ B[1 : N; \ i + 1, N])$ is partitioned into $s = \lceil (N - i)/t \rceil$ block columns consisting of $t$ columns each. (The last block column may have fewer columns.) Let $(A_k, B_k)$ denote the $k$-th block column pair. $A_k$ is further divided into $A_k^{(r)}$ made up of rows 1 to $kt + i$ of $A_k$ and $A_k^{(t)}$ made up of rows $kt + i + 1$ to $(k + 1)t + i$ of $A_k$. $B_k$ is further divided into $B_k^{(r)}$ made up of rows 1 to $(k - 1)t + i$ of $B_k$ and $B_k^{(t)}$ made up of rows $(k - 1)t + i + 1$ to $kt + i$ of $B_k$. The block partitioning and subsequent reference pattern to reduce the first column of $A$ is illustrated in Figure 5.3.

On the first column of $A$, the algorithm proceeds as follows. Find a set $row_1$ of Givens rotators to annihilate the unwanted elements of the first column of $A$ using the block marked (1) in the figure. Apply $row_1$ on $B_1^{(t)}$ ((2) in the figure). This introduces $t - 1$ single element bulges in $B_1^{(t)}$. Generate a set $col_1$ of rotators to annihilate those bulges. Apply $col_1$ to $B_1^{(r)}$ ((3) in the figure). Apply both $col_1$ and $row_1$ to $A_1^{(r)}$ ((4) in the figure). Apply $col_1$ to $A_1^{(t)}$ ((5) in the figure) introducing $t - 1$ single element bulges in it. Generate a set $row_2$ of rotators to annihilate these bulges. Apply $row_2$ to $B_2^{(t)}$ ((6) in the figure), introducing bulges in it and generating a rotator set $col_2$ for them. Apply $col_2$ to $B_2^{(r)}$ ((7) in the figure), $row_1$ and $row_2$ together with $col_1$ to $A_2^{(r)}$ ((8) in the figure), and $col_2$ to $A_2^{(t)}$ ((9) in the figure) introducing bulges and generating a rotator set $row_3$ for them. Continue like this until the bulges generated by the first column are chased out of the matrices, which completes the processing of the first column of $A$. Repeat for the remaining columns.

Assume that the matrix pair is available in column major order. To reduce any column $i$, we need to access the column panels to its right and update those using sets of rotations, both from the left and right, introducing bulges and chasing them. So the number of blocks of data moved in the processing of the $i$-th column is $O(N(N-i)/B)$. Therefore, the total number of blocks moved in the algorithm is $O(N^3/B)$. Similarly, the total number of seeks is $O(N^3/t)$.

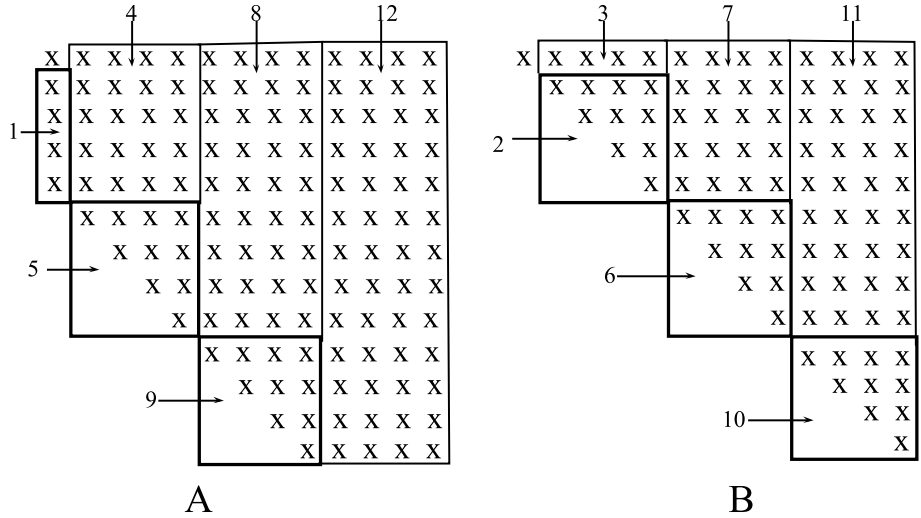

Figure 5.3: Block partitioning and reference pattern to reduce 1st column of BH-T pair to H-T pair using Givens rotations

This procedure can be extended so that $p$ columns are reduced and the resultant bulges are chased together in a *super-sweep* [37]. Let $row_i^j$ and $col_i^j$ respectively denote the rotator sets for the $i$-th block row and column obtained while reducing column $j$. The algorithm proceeds in two parts. In the first part (called the reduce-part of the super sweep), the row and column sets $row_{1:p-j+1}^j$ and $col_{1:p-j}^j$ for $j = 1, \ldots, p$ are generated. After reducing $p$ columns, in the second part (called the chase-part of the super-sweep), the necessary block column updates are performed. The chasing of the bulges and the subsequent updates of block columns are performed iteratively one block column ahead in a pipelined fashion starting with the leading block column. Asymptotic I/O and seek complexities remain same as above.

### 5.5.3    A new Algorithm

The above algorithms do asymptotically no better than the direct Hessenberg-Triangular reduction. The main drawback of the above algorithm is that to reduce a single column of $A$, all larger numbered columns in $A$ and $B$ are read in and written back.

We propose a Householder based algorithm that mitigates this drawback. Our algorithm reduces the columns using Householders. Each left application of a Householder forms a $(t \times t)$ bulge below the subdiagonal of $B$. The first column of the bulge is chased using an opposite Householder, which when applied from the right, forms $(t \times t)$ bulge below the band of $A$. A bulge in $B$ and the subsequent bulge in $A$ together form, what we call, a bulge pair.

We make the observation that it is enough to chase the bulge pairs a short distance to make the next few columns ready for processing. We maintain a box pair in $A$ and $B$ about the column that is being reduced, and chase the bulge pair to the bottom right corner of the box pair. When a number of columns are reduced we would have lined up an equal number of bulge pairs at the bottom right corner of the box pair in a tightly packed sequence. While doing this, we need to update with the left and right Householders only the portion inside the box pairs. After lining up a number of bulges in $A$, we use the aggregate of the left and right Householders to update the portions outside the box pair at one go.

On a banded upper Hessenberg-Triangular matrix pair $A$ and $B$ in $\mathbb{R}^{N \times N}$, where $A$ is of bandwidth $t$, the algorithm proceeds as follows:

Divide the matrices into slabs of $t$ columns each. Process the slabs one by one. Begin with the leftmost slab.

Let $w$ be $2t^2 - t$. Define the "small box" pair as the intersections of columns $l = 1$ to $r = 1 + w$ (left and right) and rows $a = 2$ to $b = 1 + w + t$ (above and below) of $A$ and $B$.

A small box has $w + 1$ columns and $w + t$ rows. For $h \geq 1$, let the $h$-th large box pair be made up of the intersections of columns $l(h) = 1 + t + (h-1)k$ to $r(h) = 1 + w + hk$ and rows $a(h) = i + 2t + (h-1)k$ to $b(h) = 1 + w + t + hk$ of $A$ and $B$. Thus, a large box has $k + w - t + 1$ columns and rows. ($k$ is a parameter to be chosen later.)

Process the columns of the first slab of $A$ one by one. Begin with the first column. Find a Householder $Q_1^{(1)}$ s.t. $Q_1^{(1)} * A[2 : 1 + t; 1]$ is of the form $(* \ 0 \cdots 0)^T$. Multiply $A[2 : 1 + t, 1 : r]$ and $B[2 : 1 + t, 2 : r]$ from the left with $Q_1^{(1)}$. A $(t \times t)$ bulge forms in columns $2 : 1 + t$ at rows $2 : 1 + t$ of $B$. Chase the leftmost column of this bulge in $B$, i.e., $B[2 : t + 1, 2]$: Construct a Householder $V_1^{(1)}$ using the opposite Householder technique [37, 69, 74, 117] so that $B[2 : t + 1, 2] * V_1^{(1)}$ is of the form $(* \ 0 \cdots \ 0)^T$.

(The opposite Householder technique: Given $X \in \mathbb{R}^{N \times N}$, find an RQ decomposition $X = RQ$ of $X$. Then $XQ^T = R$. Let $q^T$ be the first row of $Q$. Find a Householder $\tilde{Q}$ such that $\tilde{Q}q = se_1$, where $s = \pm 1$ and $e_1 = (1 \ 0 \ \ldots \ 0)^T$. Then $(X\tilde{Q})e_1 = s(X\tilde{Q})(se_1) = sX(s\tilde{Q}e_1) = sXq = sX(Q^Te_1) = s(XQ^T)e_1 = sRe_1 = \gamma e_1$, for a scalar $\gamma$. That is, the first column of $X\tilde{Q}$ is of the form $\gamma e_1$.)

Multiply $B[a : b; 2 : 1+t]$ and $A[a : b; 2 : 1+t]$ from the right with $V_1^{(1)}$. Note that only the portion inside the small box pair is involved in these multiplications. A bulge forms in columns $2 : 1 + t$ at rows $2 + t : 1 + 2t$ of $A$.

Chase the leftmost column of this bulge: Find a Householder $Q_2^{(1)}$ s.t. $Q_2^{(1)} * A[2 + t : 1+2t, 2]$ is of the form $(* \ 0 \cdots \ 0)^T$. Multiply $A[2+t : 1+2t; 2 : r]$ and $A[2+t : 1+2t; 2+t : r]$ from the left with $Q_2^{(1)}$. A bulge forms in columns $2 + t : 1 + 2t$ at rows $2 + t : 1 + 2t$ of $B$.

Chase the leftmost column of this new bulge similarly, and continue like that till the right boundaries of the latest bulge in $A$ and the small box coincide. This happens after the $(2t - 1)$-th Householder. Nail the bulge in its present position.

The nailed bulge in $A$ is unreduced and hence spans $t$ columns, whereas each of the other (harmless) bulges has lost one column and so spans $t - 1$ columns.

Now process the second column of $A$, and chase the bulge pairs similarly, until the latest bulge in $A$ is $t - 1$ columns away from the nailed bulge of the first column. This involves finding of $(2t - 3)$ left and right Householders. Note that the bulges created in $A$ and $B$ while processing the second column subsume, when they are as yet unreduced, the harmless bulges left behind by the processing of the first column.

Process the $t$ columns of the first slab of $A$ like this. When all are done, we will be left with $t$ nailed bulges spanning columns $t + 1$ to $2t^2 - t + 1$ in $A$, and any two consecutive nailed bulges will be separated by a harmless bulge that spans $t - 1$ columns. $B$ has only harmless bulges, and these span columns $t + 2$ to $2t^2 - 2t$.

Note that the slab and the interspersed sequence of the nailed and harmless bulges are tightly packed inside the small box pair of $A$ and $B$.

The left and right Householders found till now have been applied only to the relevant portions of the small box pair. Now we form groups of these left and right Householders, aggregate each group into an $(I + YTY^T)$ representation and apply the left Householder groups from the left on the portion to the right of the small box pair, and apply the right Householder groups from the right on the portion above the small box pair. (The details of the aggregation are given later.)

Next consider the first large box pair. The nailed bulges of the small box of $A$ now occupy the top left corner of this box. Recall that the small box of $B$ has only harmless bulges. Move the nailed bulges of $A$ into the bottom right corner of the first large box of $A$, one by one, starting with the rightmost bulge, by chasing them with left and right Householders as before and arrange them there in a similar tightly packed manner. This process would leave harmless bulges each spanning $t - 1$ columns in the small boxes of $A$ and $B$. In particular, columns $t + 2$ to $2t$ of $A$ and $B$ hold the harmless remnant of the bulge introduced by the reduction of column $t$. But column $t + 1$ of $A$ is now without a bulge, and so is ready for processing. The left and right Householders found during this process are applied on the fly only to the relevant portions of the first large box pair. Now group the left Householders, aggregate the groups and apply the aggregates on the portions of $A$ and $B$ to the right of the first large boxes from the left. Apply the right Householders on the portions of $A$ and $B$ above the first large boxes likewise. (The details of the aggregation are given later.)

We repeat the process for the remaining large boxes in a left to right order, until the nailed bulges in $A$ are all within the last $k + w$ columns of the matrix.

Next we chase the bulges from the bottom right corner of the last large box, and out of the matrix. We employ the same technique that shifts the tightly packed bulge pairs from the bottom right corner of one large box to the bottom right corner of the next. The only difference here is, instead of chasing each bulge pair exactly $k/t$ times, each bulge pair is chased till it is out of the matrix pair. For example, if $x$ chases are required for the rightmost bulge pair, chase the second bulge pair $(x + 2)$ times, the third bulge pair $(x + 4)$ times, and so on. Householders into groups and apply them to the top of the last box pair.

Now the first slab of $A$ has been processed. Continue with remaining slabs, until at most $k + w - t + 1$ rightmost columns of the matrix $A$ remain to be reduced. Then reduce the remaining columns one by one, completely chasing, for each, the bulges out of the matrix, and perform the necessary updates.

Now the matrix pair will be in Hessenberg-Triangular form.

A more formal description of the algorithm follows:

---

**Algorithm 5.4.** *BH-T pair to H-T pair using tightly coupled bulge chasing*

**Input:** *BH-T pair* $A, B \in \mathbb{R}^{N \times N}$.

**Output:** *A is overwritten with Hessenberg matrix and B is overwritten with triangular*

   *matrix.*

---

*Let* $w = 2t^2 - t$

**for** $(i = 1; i \leq \lceil (N - k - w - t)/t \rceil; i = i + t)$ **do**

   *Let the small box be the intersection of columns* $l(i) = i$ *to* $r(i) = i + w$ *and*

   *rows* $a(i) = i + 1$ *to* $b(i) = i + w + t$.

   *Thus, it has* $w + 1$ *columns and* $w + t$ *rows.*

   **for** $(j = i; j \leq i + t - 1; j ++)$ **do** /* *process the* $j^{th}$ *column* */

      *Initialise* $G_1 \ldots G_{2(t-1)+1}$ *and* $F_1 \ldots F_{2(t-1)+1}$ *to identity matrices of appropriate*
         *dimensions.*

      *Find Householder* $Q_1^{(j)}$ *s.t.* $Q_1^{(j)} * A[j + 1 : j + t; j]$ *has form* $(* \, 0 \cdots 0)^T$.

      *Aggregate* $Q_1^{(j)}$ *into* $G_1$.

      *Multiply* $A[j + 1 : \ j + t, \ j : r(i)]$ *with* $Q_1^{(j)}$ *from the left.*

      *Multiply* $B[j + 1 : \ j + t, \ j + 1 : r(i)]$ *with* $Q_1^{(j)}$ *from the left.*

      *A bulge forms in columns* $j + 1 : \ j + t$ *and rows* $j + 1 : \ j + t$ *of matrix* $B$

      *Find an opposite Householder* $V_1^{(j)}$ *such that the first column of*

         $B[j + 1 : j + t; \ j + 1 : j + t] * V_1^{(j)}$ *has form* $(* \, 0 \cdots 0)^T$.

      *Aggregate* $V_1^{(j)}$ *into* $F_1$.

      *Multiply* $B[a(i) : \ j + t, \ j + 1 : \ j + t]$ *with* $V_1^{(j)}$ *from the right.*

      *Multiply* $A[a(i) : \ j + 2t, \ j + 1 : \ j + t]$ *with* $V_1^{(j)}$ *from the right.*

      *A bulge forms in columns* $j + 1 : \ j + t$ *and rows* $j + t + 1 : \ j + 2t$ *of matrix* $A$

      *Let* $x_j = 2 * (t - (j - i + 1)) + 1$; *Then* $x_j - 1$ *is the number of times the bulge pair*
         *will be chased before it's fixed in a small box.*

**for** $(z = 2; z \leq x_j; z + +)$ **do**

    *Find $Q_z^{(j)}$ s.t. $Q_z^{(j)} * A[j + (z-1)t + 1 : j + zt, \; j + (z-2)t + 1]$ has form $(* \, 0 \cdots \, 0)^T$.*

      *Aggregate $Q_z^{(j)}$ into $G_z$.*

    *Multiply $A[j + (z-1)t + 1 : j + zt; \; j + (z-2)t + 1 : r(i)]$ from the left with $Q_z^{(j)}$*

    *Multiply $B[j + (z-1)t + 1 : j + zt; \; j + (z-2)t + 1 : r(i)]$ from the left with $Q_z^{(j)}$*

    *A bulge forms in columns $j + (z-1)t + 1 : j + zt$ and*

      *rows $j + (z-1)t + 1 : j + zt$ of B. See remark below*

  *Find an opposite Householder $V_z^{(j)}$ such that the first column of*

    *$B[j + (z-1)t + 1 : \; j + zt; \; j + (z-1)t + 1 : j + zt] * V_1^{(j)}$ has form $(* \, 0 \cdots 0)^T$.*

    *Aggregate $V_z^{(j)}$ into $F_z$.*

    *Multiply $B[a(i) : \; j + zt; \; j + (z-1)t + 1 : j + zt]$ from the right with $V_z^{(j)}$*

    *Multiply $A[a(i) : \; j + (z+1)t; \; j + (z-1)t + 1 : j + zt]$ from the right with $V_z^{(j)}$*

    *A bulge forms in columns $j + (z-1)t + 1 : j + zt$ and*

      *rows $j + zt + 1 : j + (z+1)t$ of A. See remark below*

  **endfor** */* Nail the bulge at the present position */*

**endfor**

*Multiply $A[a(i) : b(i) - t, \; r(i) + 1 : \; N]$ and $B[a(i) : b(i) - t, \; r(i) + 1 : \; N]$*

  *(right of the small box) from the left with $(G_1 \cdots G_{2(t-1)+1})^T$*

*Multiply $B[1 : a(i) - 1, \; l(i) : \; r(i)]$ and $A[1 : a(i) - 1, \; l(i) : \; r(i)]$*

  *(above the small box) from the right with $(F_{2(t-1)+1} \ldots F_1)$*

*See Algorithm 5.5*

*Let the h-th large box be the intersection of*

*columns $l(i, h) = i + t + (h-1)k$ to $r(i, h) = i + w + hk$ and*

*rows $a(i, h) = i + 2t + (h-1)k$ to $b(i, h) = i + w + t + hk$.*

*Thus, it has $k + w - t + 1$ columns and rows.*

*$k$ is a parameter to be chosen later.*

**for** $(h = 1; r(i, h) \leq N - t; h++)$ **do** /* move the bulges to the h-th large box */

  **for** $(j = i; j \leq i + t - 1; j++)$ **do**

    *Initialise* $G_{hk/t+2t-1} \ldots G_{(h-1)k/t+2}$ *and* $F_{hk/t+2t-1} \ldots F_{(h-1)k/t+2}$ *to identity matrices*

      *of appropriate dimensions.*

    $x_j = 2 * (t - (j - i + 1)) + 1;$

    **for** $(p = 1; p \leq k/t; p++)$ **do**

      *Let* $z = p + (h-1)k/t + x_j.$    */* for each column j we pick up the*

      *z-count from where the previous box (large or small) left off */*

      *Find* $Q_z^{(j)}$ *s.t.* $Q_z^{(j)} * A[j + (z-1)t + 1 : j + zt; j + (z-2)t + 1]$

      *has form* $(* \, 0 \ldots 0)^T.$ *Aggregate* $Q_z^{(j)}$ *with* $G_z.$

      *Multiply* $A[j + (z-1)t + 1 : j + zt; j + (z-2)t + 1 : r(i,h)]$ *and*

      $B[j + (z-1)t + 1 : j + zt; j + (z-1)t + 1 : r(i,h)]$ *from the left with* $Q_z^{(j)}.$

      *A bulge forms in columns* $j + (z-1)t + 1 : j + zt$ *and*

      *rows* $j + (z-1)t + 1 : j + zt$ *in B.*

      *Find an opposite Householder* $V_z^{(j)}$ *such that the first column of*

        $B[j + (z-1)t + 1 : \; j + zt; j + (z-1)t + 1 : j + zt] * V_1^{(j)}$ *has form* $(* \, 0 \cdots 0)^T.$

      *Aggregate* $V_z^{(j)}$ *into* $F_z.$

      *Multiply* $B[a(i,h) : \; j + zt, \, j + (z-1)t + 1 : \; j + zt]$ *and*

      $A[a(i,h) : j + (z+1)t; \, j + (z-1)t + 1 : \; j + zt]$ *from the right with* $V_z^{(j)}.$

      *A bulge forms in columns* $j + (z-1)t + 1 : j + zt$ *and*

      *rows* $j + zt + 1 : j + (z+1)t$ *in A.*

  **endfor**

**endfor**

*The smallest and largest values assigned to $z$ in the above are*

*$(h-1)k/t + 2$ and $hk/t + 2t - 1$, when $(p = 1, j = i + t - 1)$ and*

*$(p = k/t, j = i)$ respectively. See Algorithm 5.6.*

*Multiply $A[a(i,h) : b(i,h) - t; \ r(i,h) + 1 : \ N]$ and $B[a(i,h) : b(i,h) - t; \ r(i,h) + 1 : \ N]$*

*(right of the $h$-th large box) from the left with $(G_{(h-1)k/t+2} \ldots G_{hk/t+2t-1})^T$*

*Multiply $A[1 : a(i,h) - 1; l(i,h) : \ r(i,h)]$ and $B[1 : a(i,h) - 1; l(i,h) : \ r(i,h)]$*

*(above the $h$-th large box) from the right with $(F_{hk/t+2t-1} \ldots F_{(h-1)k/t+2})$*

*See Algorithm 5.6.*

**endfor**

*Chase the tightly packed bulge pairs that are now at the bottom right corner of the*

*$(h-1)$-th large box, one after another, out of the matrix pair.*

*Aggregate the right Householders and apply to the corresponding columns of $A$ and $B$.*

**endfor**

*Reduce the submatrix pair $A[N + 1 - \lceil \frac{(N-k-w+t)}{t} \rceil t : N; \ N + 1 - \lceil \frac{(N-k-w+t)}{t} \rceil t : N]$ and.*

*$B[N + 1 - \lceil \frac{(N-k-w+t)}{t} \rceil t : N; \ N + 1 - \lceil \frac{(N-k-w+t)}{t} \rceil t : N];$*

*Aggregate the right Householders and apply to the the submatrix pair*

*$A[1 : N - \lceil \frac{(N-k-w+t)}{t} \rceil t; \ N + 2 - \lceil \frac{(N-k-w+t)}{t} \rceil t : N]$ and*

*$B[1 : N - \lceil \frac{(N-k-w+t)}{t} \rceil t : N; \ N + 2 - \lceil \frac{(N-k-w+t)}{t} \rceil t : N]$*

**Remark:** The bulge that forms in columns $j + (z-1)t + 1 : j + zt$ and rows $j + zt + 1 : j + (z+1)t$ of $A$ subsumes any harmless elements that may have been present there. Columns $j + (z-1)t + 1 : j + zt$ cannot have elements below this bulge; column $j + zt$ has been cleared of any bulge before proceeding to this step. Similarly, the bulge that forms in columns $j + (z-1)t + 1 : j + zt$ and rows $j + (z-1)t + 1 : j + zt$ of $B$ subsumes any harmless elements that may have been present there.

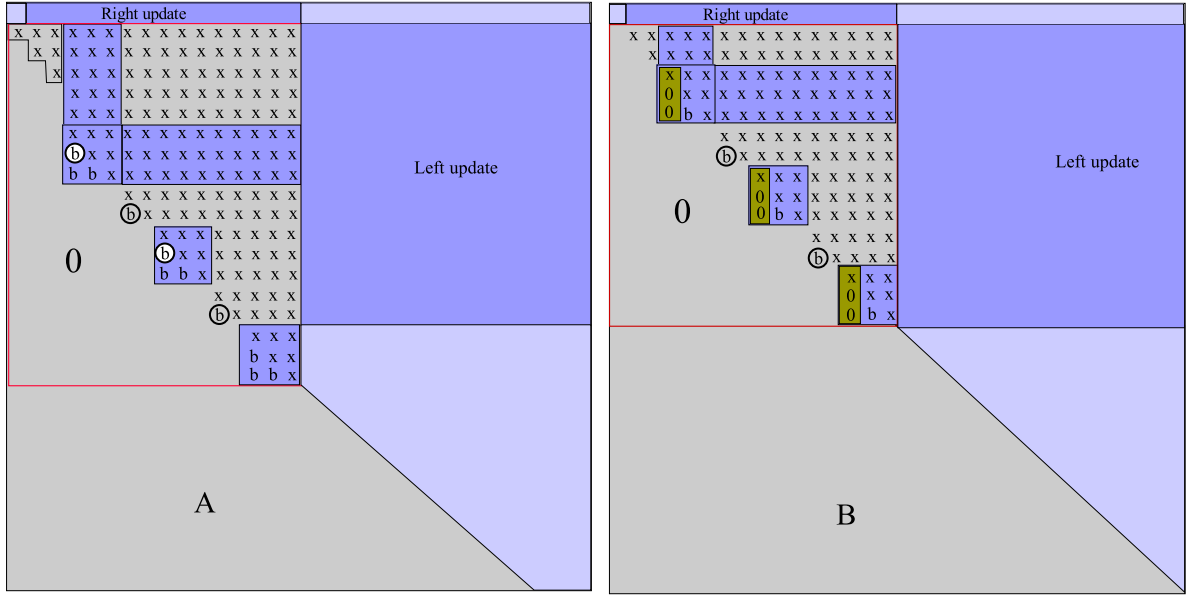

Figure 5.4: Introduction of $t = 3$ tightly packed bulge pairs after reducing first 3 columns of BH-T pair.

A snapshot of the matrix after applying the bulge introduction phase to the first slab ($t = 3$) is shown in Figure 5.4. In the figure, the submatrices $A[i+1 : i+2t^2, \; i : i+2t^2-t] = A[2 : 19, \; 1 : 16]$ and $B[i+1 : i+2t^2-t, \; i : i+2t^2-t] = B[2 : 16, \; 1 : 16]$ are updated from both the left and right, the submatrices $A[i+1 : i+2t^2-t, \; i+w+1 : N] = A[2 : 16, \; 17 : N]$ and $B[i+1 : i+2t^2-t, \; i+w+1 : N] = B[2 : 16, \; 17 : N]$ are updated from the left and the submatrices $A[1 : i, \; i+1 : i+w] = A[1 : 1, \; 2 : 16]$ and $B[1 : i, \; i+1 : i+w] = B[1 : 1, \; 2 : 16]$ are updated from the right.

### 5.5.3.1    Aggregation: the bulge introduction phase

Let $Q_1^{(j)}$ be the left Householder that reduces the $j$-th column of $A$, and for $z > 1$, let $Q_z^{(j)}$ be the left Householder that reduces the first column of the $z$-th bulge induced by $j$-th column in $A$. $Q_z^{(j)}$ acts along dimensions $j + 1 + (z - 1)t$ to $j + zt$.

Consider the processing of the slab beginning at column $i$ of $A$.

The Householders are constructed and applied inside the small box pair in the following order: $Q_1^{(i)} \ldots Q_{x_i}^{(i)} Q_1^{(i+1)} \ldots Q_{x_i-2}^{(i+1)} \ldots Q_1^{(i+t-1)}$ from the left. But while updating the portions to the right of the small boxes they need not be applied in the same order. We can aggregate them into different groups using a technique proposed in [78], because of the following observations: (See Figure 5.5 and Figure 5.4.)

1. $Q_*^{(j)}$'s, which reduce the $j$-th column and the bulges induced by it, act along disjoint dimensions. That means, if $Q_*^{(j)}$'s are to be applied on a submatrix only from one side (left or right), then they can be applied in any order.

2. The dimensions of $Q_z^{(j)}$ and $Q_y^{(j-1)}$ overlap iff $y = z$ or $y = z+1$. Therefore, $Q_z^{(j)}$ can be applied only after $Q_z^{(j-1)}$ and $Q_{z+1}^{(j-1)}$.

3. $Q_z^{(*)}$'s are to be applied in the order of their construction. Let $G_z$ be the aggregation of $Q_z^{(*)}$'s in the order of construction.

4. If the $G_z$'s are applied in the decreasing order of $z$, then the above two requirements will be met.

The computing of $(G_1 \ldots G_{2(t-1)+1})^T A[a(i) : b(i) - t, \ r(i) + 1 : \ N]$ and $(G_1 \ldots G_{2(t-1)+1})^T B[a(i) : b(i) - t, \ r(i) + 1 : \ N]$ can be done as follows. Note that $x_i = 2(t-1)+1$, $G_{x_i} = Q_{x_i}^{(1)}$ and $G_{x_i-1} = Q_{x_i-1}^{(1)}$. In a slab only the first column is subjected to more than $x_i - 2$ chases. $G_{x_i-2}$ is an aggregation of $Q_{x_i-2}^{(1)}$ and $Q_{x_i-2}^{(2)}$. $G_{x_i-3}$ is an aggregation of $Q_{x_i-3}^{(1)}$ and $Q_{x_i-3}^{(2)}$. In a slab only the first two columns are subjected to more than $x_i - 4$ chases. Continuing like this, we find that for $1 \leq z \leq x_i$, if $q = x_i - z + 1$, then $1 + \lfloor (q-1)/2 \rfloor$ is the number of Householders aggregated into $G_z$, which therefore operates on $y = t + \lfloor (q-1)/2 \rfloor$ rows.

| Group | Householders | $y$ | left end of the range |
|-------|--------------|-----|----------------------|
| $G_1$ | $Q_1^{(i)}, Q_1^{(i+1)}, \ldots, Q_1^{(i+t-2)}, Q_1^{(i+t-1)}$ | $2t-1$ | $i+1$ |
| $G_2$ | $Q_2^{(i)}, Q_2^{(i+1)}, \ldots, Q_2^{(i+t-2)}$ | $2t-2$ | $i+t+1$ |
| $G_3$ | $Q_3^{(i)}, Q_3^{(i+1)}, \ldots, Q_3^{(i+t-2)}$ | $2t-2$ | $i+2t+1$ |
| $\vdots$ | | | |
| $G_{2r}$ | $Q_{2r}^{(i)}, Q_{2r}^{(i+1)}, \ldots, Q_{2r}^{(i+t-r-1)}$ | $2t-r-1$ | $i+(2r-1)t+1$ |
| $G_{2r+1}$ | $Q_{2r+1}^{(i)}, Q_{2r+1}^{(i+1)}, \ldots, Q_{2r+1}^{(i+t-r-1)}$ | $2t-r-1$ | $i+2rt+1$ |
| $\vdots$ | | | |
| $G_{2t-2}$ | $Q_{2t-2}^{(i)}$ | $t$ | $i+(2t-3)t+1$ |
| $G_{2t-1}$ | $Q_{2t-1}^{(i)}$ | $t$ | $i+(2t-2)t+1$ |

The right Householders $V_*^{(j)}$s satisfy all the criteria discussed above in the context of the left Householders $Q_*^{(j)}$s. Therefore, aggregate them into groups $F_1 \ldots F_{2(t-1)+1}$, and compute $A[1:a(i)-1;\ l(i):\ r(i)](F_{2(t-1)+1} \ldots F_1)$ and $B[1:a(i)-1;\ l(i):\ r(i)](F_{2(t-1)+1} \ldots F_1)$ in the same manner.

The left and right applications of the aggregated Householders to submatrix pairs can be done as follows:

---

**Algorithm 5.5.** *Multiplying with the Left and Right Aggregates from Bulge Introduction*

---

**Input:** *Aggregations of the $(2t-1)$ groups of the left and right Householders.*
**Output:** *Updated matrices.*

    **for** $(z = x_i;\ z \geq 1;\ z--)$ **do**

        $q = x_i - z + 1;$

        $y = t + \lfloor (q-1)/2 \rfloor;$

        *Multiply $A[i+(z-1)t+1 : i+(z-1)t+y;\ r(i)+1 : N]$ and*

            $B[i+(z-1)t+1 : i+(z-1)t+y;\ r(i)+1 : N]$ *from the left with $G_z^T$*

        *Multiply $A[1:a(i)-1;\ i+(z-1)t+1 : i+(z-1)t+y]$ and*

            $B[1:a(i)-1;\ i+(z-1)t+1 : i+(z-1)t+y]$ *from right with $F_z$*

    **endfor**

---

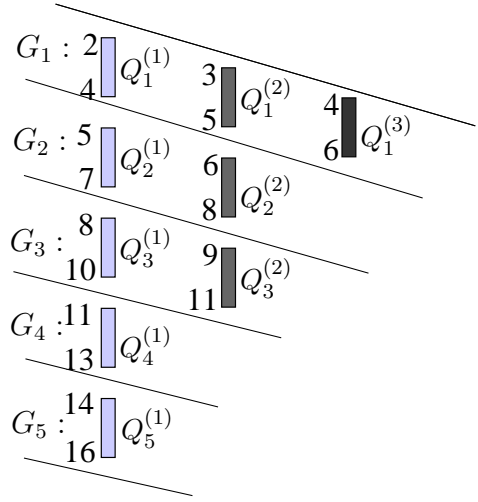

Figure 5.5: Left Householders constructed to introduce $t = 3$ tightly packed bulge pairs after reducing first block(first 3) column.

### 5.5.3.2    Aggregation: the bulge chasing phase

In the bulge chasing phase of the algorithm, repeatedly, $t$ bulges that are nailed at the top left corner of a large box in a tightly packed sequence in $A$, are chased one by one over a distance of $k$. This would leave them at the bottom right corner of the large box, which is also the top left corner of the next large box. In the corresponding large box of $B$ the harmless bulges will lie below the diagonal. Each bulge pair is, thus, chased $k/t$ times from its current position. See Figure 5.6.

During the chasing steps, the left and right Householders found are applied on the fly only to the relevant portions of the box pair. Afterwards, we group the Householders, aggregate the groups and apply the aggregates on portions outside the box pair.

Consider the processing of the slab beginning at column $i$ of $A$. Suppose we are at the $h$-th large box of this slab. Let $\alpha = x_i + (h-1)k/t + 1$.

The left Householders are constructed and applied inside the box pair in the following order: $Q_\alpha^{(i)} \dots Q_{\alpha+k/t-1}^{(i)} Q_{\alpha-2}^{(i+1)} \dots Q_{\alpha+k/t-3}^{(i+1)} \dots Q_{\alpha-2t+2}^{(i+t-1)} \dots Q_{\alpha-2t+k/t+1}^{(i+t-1)}$. Similarly the corre-sponding right Householders are constructed and applied inside the box pair in the follow-ing order: $V_\alpha^{(i)} \dots V_{\alpha+k/t-1}^{(i)} V_{\alpha-2}^{(i+1)} \dots V_{\alpha+k/t-3}^{(i+1)} \dots V_{\alpha-2t+2}^{(i+t-1)} \dots V_{\alpha-2t+k/t+1}^{(i+t-1)}$. But while updating the relevant portions outside the box pair they need not be applied in the same order. The observations made during the bulge introduction phase hold for these left and right House-holders too. Hence these left and right Householders too can be grouped separately and aggregated the same way. That is, all $Q$'s with the same subscript can be grouped to-gether, and aggregated in the order in which they are constructed. Similarly, all $V$'s with the same subscript can be grouped together, and aggregated in the order in which they are constructed. Note that the smallest and largest subscripts are $\alpha - 2t + 2$ and $\alpha + k/t - 1$ respectively. See Figure 5.7.

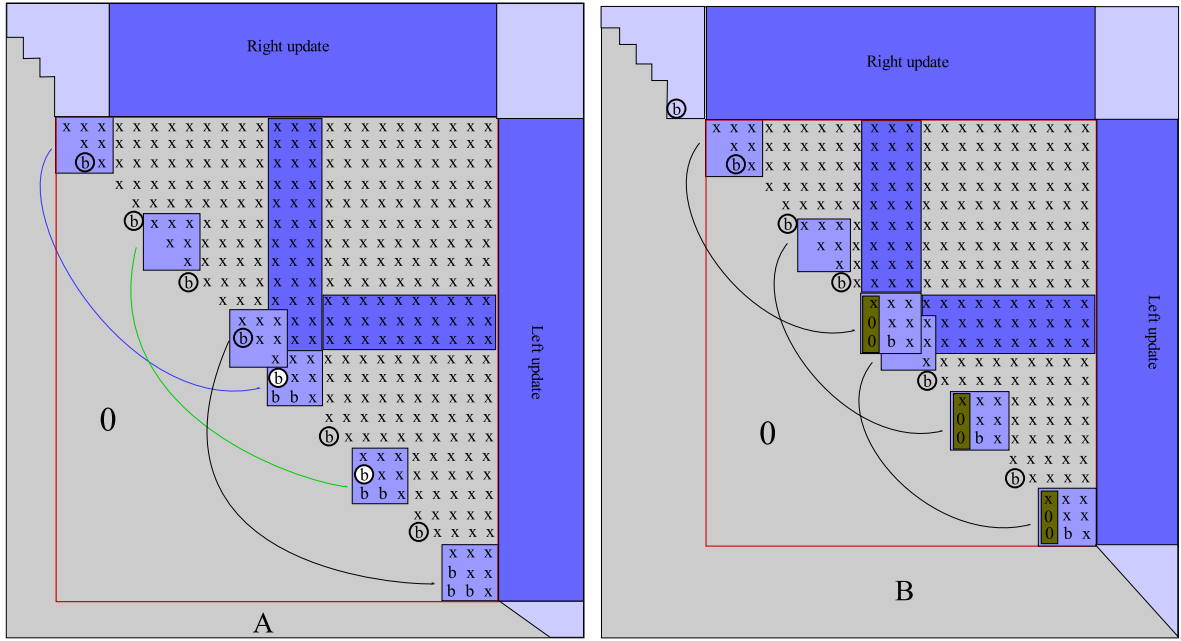

Figure 5.6: Chasing of 3 tightly packed bulge pairs during stage-2 of BH-T to H-T reduction

For a fixed $h$, as $j$ varies over $[i, i + t - 1]$, and as $p$ varies over $[1, k/t]$,

$$z = \left(\frac{(h-1)k}{t} + 2t\right) + p - 2j + 2i - 1 = \alpha + D(p,j)$$

where $D(p,j) = p - 2j + 2i - 1$. Note that each $z$ corresponds to a group. When $z$ is fixed,

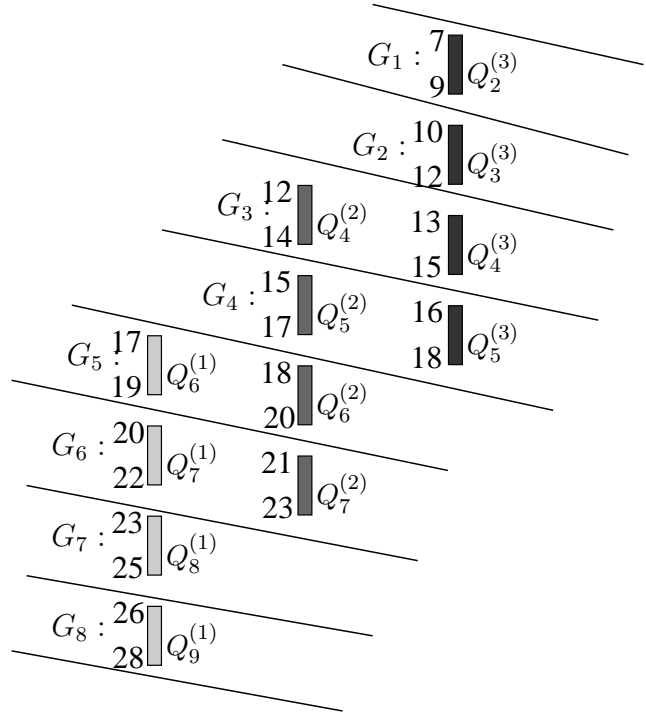

Figure 5.7: Left Householders constructed to chase 3 tightly packed bulge pairs.

$D(p, j)$ is also fixed, as $\alpha$ is independent of $p$ and $j$.

| $D(p, j)$ | $p = 1$ | $p = 2$ | $p = 3$ | $\cdots$ | $p = k/t$ |
|---|---|---|---|---|---|
| $j = i$ | 0 | 1 | 2 | $\cdots$ | $k/t - 1$ |
| $j = i + 1$ | $-2$ | $-1$ | 0 | $\cdots$ | $k/t - 3$ |
| $j = i + 2$ | $-4$ | $-3$ | $-2$ | $\cdots$ | $k/t - 5$ |
| $\vdots$ | | | | | |
| $j = i + t - 1$ | $-(2t - 2)$ | $-(2t - 3)$ | $-(2t - 4)$ | $\cdots$ | $k/t - 2t + 1$ |

Consult the above table. For $-(2t - 2) \leq D \leq k/t - 1$, set $\{(j, p) | D(p, j) = D\}$ corresponds to $G_{\alpha + D}$. If $D \geq 0$, then the leftmost cell $(j, p)$ with value $D$ is $(i, D + 1)$. Otherwise, the leftmost cell $(j, p)$ is $(i + \lceil (-D)/2 \rceil, b)$, where $b = 1$ if $D$ is even, 2 if $D$ is odd. If $(j, p)$ is of value $D$, then the next cell of value $D$ to the right is $(j + 1, p + 2)$.

$Q_z^{(j)}$ and $V_z^{(j)}$ operate on range $[j + (z - 1)t + 1 : j + zt] = \alpha t + (j + Dt) + [-t + 1 : 0]$.

For a given $z$, and $D = z - \alpha$, let $E(D)$ be the number of Householders aggregated into $G_z$(or $F_z$). The range of $G_z$(or $F_z$) then spans $E(D) - 1 + t$ starting at $\alpha t + Dt + (-t + 1) + j^*$, where $j^*$ is the smallest row $j$ that contains $D$.

We assume that $2t - 2 \leq k/t$. If $D < 0$, then $E(D) = \lfloor (D + 2t)/2 \rfloor$. If $D \geq 0$ and $2t - 1 + D \leq k/t$, then $E(D) = t$. If $D \geq 0$ and $2t - 1 + D > k/t$, then $E(D) = \lceil \frac{1}{2}(\frac{k}{t} - D) \rceil$.

Therefore the updations can be done using the Algorithm 5.6

---

**Algorithm 5.6.** *Multiplying with the Aggregates from Bulge Chasing*

**Input:** *The $k/t + (2t - 2)$ aggregations of the left and right Householders*
**Output:** *Updated matrices.*

---

**for** $(z = hk/t + 2t - 1; \quad z \geq (h - 1)k/t + 2; \quad z - -)$ **do**
    $D = z - \frac{(h-1)k}{t} - 2t;$
    *If $D \geq 0$, then $j^* = i;$ else $j^* = (i + \lceil (-D)/2 \rceil).$*
    *If $D < 0$, then $E = \lfloor (D + 2t)/2 \rfloor;$*
    *else if $2t - 1 + D \leq k/t$, then $E = t;$*
    *else $E = \lceil \frac{1}{2}(\frac{k}{t} - D) \rceil.$*
    *$L = zt + j^* - t + 1; \quad U = L + t + E - 2;$*
    *Multiply $A[L : U; \ r(i) + 1 : N]$ and $B[L : U; \ r(i) + 1 : N]$ from the left using $G_z;$*
    *Multiply $A[1 : a(i, h) - 1; \ L : U]$ and $B[1 : a(i, h) - 1; \ L : U]$ from right using $F_z;$*
**endfor**

---

### 5.5.3.3  I/O Complexity

The I/O complexity of this algorithm is at most twice that of Algorithm 4.6 of Chapter 4. The analysis is identical and is therefore omitted.

### 5.5.3.4  The two step reduction to Hessenberg-Triangular form

Analogous to our Hessenberg reduction of Chapter 4, the results to the last two sections can be combined to get an algorithm for Hessenberg-Triangular reduction with the same asymptotic I/O complexity.

### 5.5.4    A new algorithm using rotations

With a little modification, the algorithm similar to the Algorithm 5.4 can be devised using rotations instead of reflectors. The application of rotations from left would form $t - 1$ bulges along the subdiagonal of $B$. Chase all the $t - 1$ bulges using right rotations. In this case there would be no harmless bulges in $B$. Other steps are similar to the algorithm with the same asymptotic I/O and seek complexities. So we omit the detailed description and analysis here.

# Chapter 6

# The QR and QZ Algorithms

## 6.1 Introduction

The QR algorithm [51, 52, 58, 111, 115, 116] is popular for computing all eigenvalues of a matrix. More advantageous algorithms have been proposed for the symmetric eigenvalue problem [7, 34, 88], but the QR algorithm continues to be the method of choice for the nonsymmetric eigenvalue problem.

The QZ algorithm [58, 86, 115] and its variants [2, 37, 69, 71, 109, 110] are popular for computing all generalised eigenvalues of a matrix pair.

In this chapter, we study these algorithms, analyse them for their I/O and seek complexities, and propose their variants that use fewer seeks overall, and fewer I/Os in some cases.

### 6.1.1 The QR Algorithm

Given a matrix $A \in \mathbb{C}^{N \times N}$, the QR algorithm iteratively computes a Schur decomposition $T = Q^* A Q$ of $A$, where $Q \in \mathbb{C}^{N \times N}$ is unitary and $T \in \mathbb{C}^{N \times N}$ is upper triangular. The basic QR algorithm starts with $A_0 = A$ and iterates as follows:

$$A_{m-1} \to Q_m R_m, \ R_m Q_m \to A_m,$$

which generates a sequence of unitarily similar matrices $A_0, A_1, A_2 \ldots$, which under appropriate conditions, converges to an upper triangular matrix [58, 115, 118]. The eigenvalues of a triangular matrix are its diagonal elements. An iteration of the algorithm is called a *QR iteration* or *QR step*.

When the QR algorithm is applied on a full matrix $A$, each QR step takes $O(N^3)$ flops and $O\left(\frac{N^3}{\sqrt{M}B}\right)$ I/Os, as our discussion in Chapter 2 shows. If $A$ is first reduced to Hessenberg form using OSTs, and then the QR algorithm is applied to the reduced matrix, then a QR step would take only $O(N^2)$ flops; the Hessenberg reduction would need $O(N^3)$ flops, but that is done only once. A QR step applied on a Hessenberg matrix produces a Hessenberg matrix.

The explicit (single shift) QR algorithm introduces shifts in the basic QR algorithm to accelerate its convergence:

$$A \to A_0; \quad (A_{m-1} - \rho_{m-1}I) \to Q_m R_m; \quad R_m Q_m + \rho_{m-1}I \to A_m$$

Here the scalar $\rho_{m-1}$ is called a shift, and is typically chosen to approximate the eigenvalue which emerged at the lower right corner of the matrix. The explicit (double shift) QR algorithm is a further improvement, and avoids complex arithmetic of real matrices.

The explicit variants of the QR algorithm are not used in practice, some bulge chasing algorithms called the implicit (shifted) QR algorithms [51, 52, 58] are used. In these, an iteration starts with an upper Hessenberg matrix, introduces a $1 \times 1$ or $2 \times 2$ bulge in the upper left corner of the matrix, and chases it one column at a time along the subdiagonal to the bottom right corner. An iteration is completed when the bulge is chased out of the matrix restoring it to the upper Hessenberg form.

If after an iteration of the QR algorithm, a subdiagonal entry of the matrix is very near to zero, we could explicitly set it to zero. This would divide (or deflate) the EVP into smaller instances, which could then be conquered independently. The subdiagonal entries can be monitored either using this classical deflation technique [51, 52, 118] or the more advanced aggressive early deflation technique [26, 76]. The QR algorithm is said be converged when all the deflated diagonal blocks are either $1 \times 1$ or $2 \times 2$ for real or complex conjugate eigenvalues [51, 52, 58].

In the single or double shift algorithms the bulges are $1 \times 1$ or $2 \times 2$, and are typically chased using rotators. So these algorithms use only V-V operations. [12] proposes the multishift QR algorithm that incorporates V-M and M-M operations using $m > 2$ shifts. The bulges in this algorithm are $m \times m$. These bulges are chased using Householders, which can be aggregated before updations to make the algorithm rich in M-M operations.

However the performance of the multishift algorithm is not found to be good [75, 112–114]. This is due to a phenomenon called *shift blurring*: when the number of shifts is large, the eigenvalues of the bulge tend to get contaminated by roundoff errors during the bulge chasing process. A technique called the small bulge multishift method has been proposed [25, 67, 73, 80] to address shift blurring. This method, instead of introducing an $m \times m$ bulge at one go, constructs a chain of small bulges of size $2 \times 2$, where each bulge propagates information of only one pair of shifts. The bulges of the chain are chased one after another starting from the lowest.

In this chapter we briefly describe and analyse the small bulge multishift algorithm of [25]. We also propose a tile based variant of it that improves the seek complexity by a factor of $m$, and improves the I/O complexity if $m$ is very large.

## 6.1.2   The QZ Algorithm

Given a pair of matrices $A, B \in \mathbb{C}^{N \times N}$, the QZ algorithm iteratively computes a generalised Schur decomposition of $(A, B)$; that is, it computes unitary matrices $Q$ and $Z$ such that $S = Q^*AZ$ and $T = Q^*BZ$ are upper triangular. (If $A, B \in \mathbb{R}^{N \times N}$, then $Q$ and $Z$ would be orthogonal and $S$ would be quasi-upper triangular. The complex conjugate (resp., real) eigenvalues appear in $2 \times 2$ (resp., $1 \times 1$) diagonal blocks of the matrix pair $(S, T)$.) The basic (explicit) QZ algorithm (with a single shift) starts with $(A_0, B_0) = (A, B)$ and iterates as follows:

$$(A_{m-1}B_{m-1}^{-1} - \rho_{m-1}I) \rightarrow Q_m R_m, \ (B_{m-1}^{-1}A_{m-1} - \rho_{m-1}I) \rightarrow Z_m S_m,$$

$$Q_m^* A_{m-1} Z_m \rightarrow A_m, \ Q_m^* B_{m-1} Z_m \rightarrow B_m,$$

which generates a sequence of equivalent pairs $(A_0, B_0), (A_1, B_1), (A_2, B_2), \ldots$, which under appropriate conditions, converges to a triangular-triangular pair [115]. The eigenvalues of a triangular-triangular pair $(S, T)$ are $T[i, i]/S[i, i]$, for $1 \leq i \leq N$. An iteration of the algorithm is called a *QZ iteration* or *QZ step*. The shift is typically chosen to approximate the eigenvalue which emerged at the lower right corners of the matrices.

The algorithm converges in fewer and faster iterations when applied on Hessenberg-Triangular (H-T) pairs. A QZ iteration applied on an H-T pair $(H, T)$ brings $H$ closer to triangular form while preserving the triangularity of $T$. Therefore, a pair of matrices is usually reduced to an H-T pair before applying the QZ algorithm. Each iteration takes $O(N^2)$ flops then.

The explicit (double shift) QZ algorithm is a further improvement, and avoids complex arithmetic of real matrices.

The explicit variants of the QZ algorithm are not used in practice, some bulge chasing algorithms called the implicit (shifted) QZ algorithms are [58, 86]. In these, an iteration starts with an upper Hessenberg-Triangular pair $(H, T)$, introduces a $1 \times 1$ or $2 \times 2$ bulge in the upper left corner of $T$, and chases it back and forth between $H$ and $T$, one column at a time to the bottom right corners. An iteration is completed when the bulge is chased out of the matrix pair restoring it to the Hessenberg-Triangular form [117].

If after an iteration of the QZ algorithm, a subdiagonal entry of $H$ is very near to zero, we could explicitly set it to zero. This would divide (or deflate) the GEVP into smaller instances, which could then be conquered independently. The subdiagonal entries can be monitored either using this classical deflation technique [58, 86] or the more advanced aggressive early deflation technique [2, 3, 69].

In the single or double shift unblocked algorithms the bulges are $1 \times 1$ or $2 \times 2$, and are typically chased using rotators [37, 58, 86]. So these algorithms use only V-V operations. Blocked versions are also known, where the application of the left and right transformations are delayed till required [37]. V-V operations dominate this algorithm too. An implicit shifted QZ algorithm using $m > 2$ shifts is given in [69,117], in which $m \times m$ bulge pairs are chased using Householders and opposite Householders from the left and right respectively. The multishift QZ algorithms too suffer from shift blurring.

To mitigate shift blurring, the small bulge multishift variant of the QR algorithm can be extended into a QZ algorithm [69]. This method, again, instead of introducing an $m \times m$ bulge pair at one go, constructs a chain of small bulge pairs (of size $2 \times 2$ or $4 \times 4$ in [69]), where each bulge pair propagates information of only one pair of shifts. The bulge pairs of the chain are chased one after another starting from the lowest.

In this chapter we briefly describe and analyse the small bulge multishift algorithm of [69]. We also propose a tile based variant of it that improves the seek complexity by a factor of $m$, and improves the I/O complexity if $m$ is very large.

### 6.1.3  Organisation of this Chapter

The rest of the chapter is organised as follows: Section 2 discusses the QR algorithm. Section 3 is on the QZ algorithm.

## 6.2   The QR Algorithm

In this section we consider some variants of the QR algorithm.

### 6.2.1   The implicit multishift QR algorithm

The explicit or implicit QR algorithms using one or two shifts chase $1 \times 1$ or $2 \times 2$ bulges and therefore are rich in V-V operations. [12] proposes the multishift QR algorithm that incorporates more V-M and M-M operations using $m > 2$ shifts. (See Algorithm 6.1.)

---

**Algorithm 6.1.** *The Implicit Multishift QR Algorithm [12]*

**Input:** *An $N \times N$ properly Hessenberg matrix $H$ and the number of shifts $m \in [2, N]$.*

**Output:** *An orthogonal matrix $Q \in \mathbb{R}^{N \times N}$ such that $Q^T H Q$ orthogonally similar to $H$, which is overwritten by $Q^T H Q$.*

**Notation:** *Let $\mathfrak{H}_j(x)$ denote a Householder matrix that maps the last $N - j$ elements of vector $x \in \mathbb{R}^N$ to zero without modifying the leading $j - 1$ elements.*

---

*Determine $m$ shifts $\rho_1, \ldots, \rho_m$;*
*$x = (H - \rho_1 I)(H - \rho_2 I) \cdots (H - \rho_m I) e_1$;*
*/* Only the topmost $m + 1$ elements of $x$ can be nonzero */*
*$Q = \mathfrak{H}_1(x)$;   /* $Q$ operates on dimensions 1 to $m + 1$ */*
*$H = Q^T H Q$;*
**for** *$j = 1$ to $N - 2$* **do**
    *$\tilde{Q} = \mathfrak{H}_{j+1}(H e_j)$;   /* $\tilde{Q}$ operates on dimensions $j + 1$ to $j + m$ */*
    *$H = \tilde{Q}^T H \tilde{Q}$;*
    *$Q = Q \tilde{Q}$;*
**endfor**

---

Several ways in which the $m$ shifts can be chosen are reported in [12]. We do not discuss them further. Once the shifts are chosen, the first column of $\Pi = (H - \rho_1 I)(H - \rho_2 I) \cdots (H - \rho_m I)$ can be constructed in $O(m^3/B)$ I/Os, since $\Pi$ has exactly $m$ subdiagonals.

When $H$ is updated with the first Householder matrix, it becomes an upper Hessenberg matrix with an $m \times m$ bulge in the leftmost $m$ columns. In the for-loop of the algorithm, $H$ is further reduced to the Hessenberg form using an OST. Instead of using Givens rotations as in single and double shift algorithms, here Householders are used to chase the $m \times m$ bulge. The direct Hessenberg reductions discussed in Chapter 4 could be used here. In particular, if Algorithm 4.2 were to be used, the total amount of data moved would be $O(N^2(m+k)/B)$, where $k$ is the slab width.

If $(m+k)k \leq M$ then the number of seeks is $O\left(\frac{N^2(m+k)}{k\sqrt{M}} + \frac{N^2(m+k)}{M}\right)$.

## 6.2.2 A small-bulge variant of the implicit multishift QR algorithm

As was mentioned in Subsection 6.1.1, because of shift blurring, the performance of the multishift algorithm is disappointing when the number of shifts is more than ten [114], which is too few to make the algorithm rich in M-M operations. There are several small bulge variants of the QR algorithm [25, 67, 73, 80]. But here we discuss the small bulge variant of the multishift QR algorithm described in [25]. This algorithm lines up a chain of $m$ tightly packed bulges, where each bulge is $2 \times 2$ as only double implicit shifts are used in each bulge to avoid complex arithmetic; that is the number of shifts is $2m$.

Let $(s_i, t_i) \in \mathbb{C} \times \mathbb{C}$, for $1 \leq i \leq m$, be the $m$ pairs of shifts; either $\overline{s_i} = t_i$ or $(s_i, t_i) \in \mathbb{R} \times \mathbb{R}$.

The algorithm has three phases. In the first phase, a tightly packed chain of $2 \times 2$ bulges is introduced at the top left corner of the matrix. In the second phase, the bulge chain is chased down to the bottom right corner along the diagonal. In the third phase, the chain of bulges is chased out of the matrix, restoring it to Hessenberg form.

We now discuss each phase in detail.

The first phase deals with the submatrix $H[1:3m+1; \ 1:3m+1]$. A $2 \times 2$ bulge is introduced at its top left corner using $(s_1, t_1)$ and is then chased to its bottom right corner.

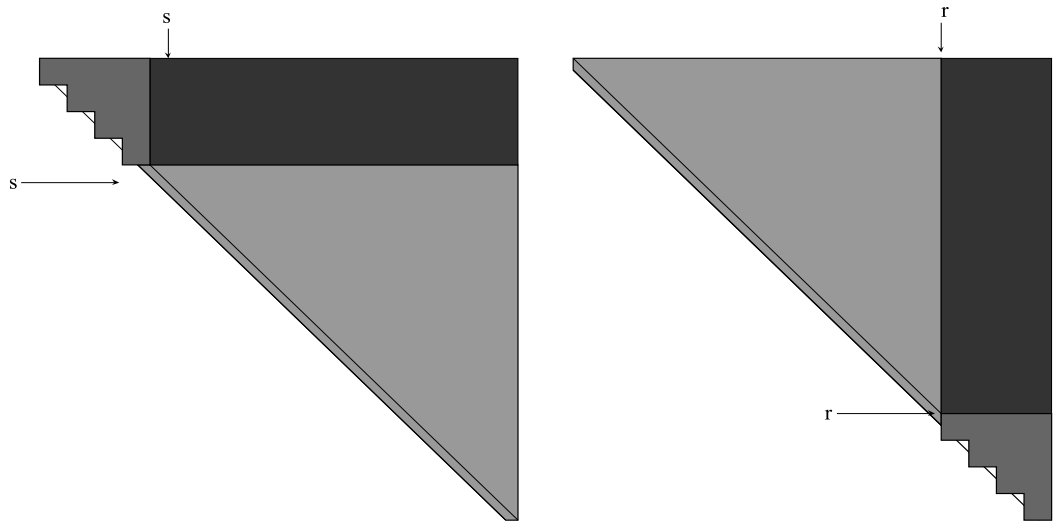

Figure 6.1: Left:Introdcucing bulge chain from row 1 through $s = 3m + 1$. Right:Bulge chain at the bottom right corner from row $r = N - (3m + 1)$ through $N$.

Another bulge is then introduced using $(s_2, t_2)$ and is then chased until it lies one column away from the first bulge. This is continued till $m$ bulges are lined up. This requires the application of $m(3m - 1)/2$ OSTs using $3 \times 3$ Householders on the submatrix. Next the Householders are accumulated into a thickly banded orthogonal matrix and the rest of $H$ is updated from both left and right. This is illustrated in the left side of Figure 6.1; the lightly shaded portion in the top left is the submatrix, and the darkly shaded panel to its right is the portion updated from the left.

The second phase has a number of iterations, each of which chases the bulges by $k$ rows towards the bottom right corner of $H$. Suppose at the beginning of an iteration, the bulges are in rows $r$ through $r + 3m$. At the end of the iteration, the bulge chain will span rows $r + k$ to $r + 3m + k$. The submatrix $H[r : r + 3m + k; \ r : r + 3m + k]$, dealt in this iteration, is updated using the $3 \times 3$ Householders as soon as they are formed. At the end of the iteration, the Householders are accumulated into an orthogonal matrix and the rest of the matrix is updated from the left and right. This is performed as follows:

If $U \in \mathbb{R}^{(3m+k-1) \times (3m+k-1)}$ is the accumulate of the $3 \times 3$ Householders generated in the iteration, then define $\tilde{U}$ as:

$$\tilde{U} = \begin{pmatrix} 1 & 0 & 0 \\ 0 & U & 0 \\ 0 & 0 & 1 \end{pmatrix}$$

Apply $\tilde{U}$ to the submatrix $H[r : 3m + k; \; r + 3m + k + 1 : N]$ from the left

$$H[r : r + 3m + k, \; r + 3m + k + 1 : N] \leftarrow \tilde{U}^T H[r : r + 3m + k, \; r + 3m + k + 1 : N]$$

Similarly apply $\tilde{U}$ to the submatrix $H[1 : r - 1; \; r : r + 3m + k]$ from the right

$$H[1 : r - 1; \; r : r + 3m + k] \leftarrow H[1 : r - 1; \; r : r + 3m + k]\tilde{U}$$

An iteration is illustrated in Figure 6.2; the darkly shaded portions are updated only from one side.

If $m << N$, and $k << N$, then the cost of the M-V operations (to update the submatrix $H[r : r + 3m + k, \; r : r + 3m + k]$ and to accumulate $U$) is small compared to the cost of the M-M operations.

In the third phase, the chain of bulges is chased off the matrix from their positions is $H[N - 3m : N; \; N - 3m : N]$. Then the accumulate of the Householders is applied on the the vertical panel $[1 : N - 3m - 1, \; N - 3m + 1 : N]$ from the right. See Figure 6.1; the darkly shaded vertical panel is updated from the right.

### 6.2.2.1 I/O and seek complexities

It has been observed that $k \approx 3m$ minimises the number of flops at $O(mN^2)$ [25]. If the bulge chasing window, i.e. $H[r : r + 3m + k, \; r : r + 3m + k]$, along with matrix $U$ fit in the main memory (i.e., $m = O(\sqrt{M})$), then the updates with the OSTs inside the window and update of $U$ can all be performed in-core. The portions outside the window that are to be updated with $U$ can be read in, updated in-core and written back. So the total I/O cost would be $O(N^2/B)$. The seek complexity of the algorithm would be $O(N^2/m)$. (We assume that the matrix is in column major order. To access an $O(m \times N)$ submatrix, $O(N)$ seeks are needed. To access an $O(N \times m)$ submatrix, $O(Nm^2/M) = O(N)$ seeks are needed. There are $N/m$ such submatrices to handle.)

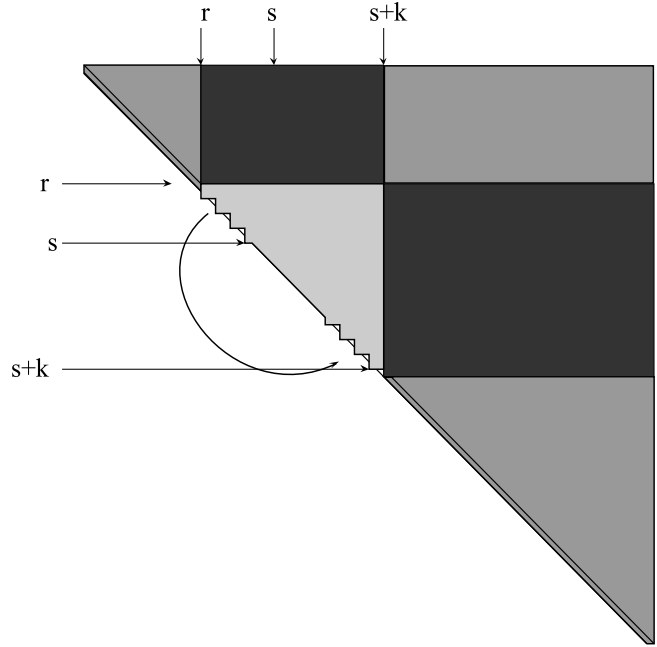

Figure 6.2: Bulge chasing phase: bulge chain is chased from rows $r$ through $s = r + 3m$ down to rows $r + k$ through $s + k$.

If the bulge chasing window does not fit in the memory, then all the operations, i.e., shifting the bulge chain inside the small window, constructing the $U$ matrix explicitly and updating the required portions from right and left are performed out-of-core. The V-V operations take $O(m^2N/B)$ I/Os and the out-of-core M-M operations take $O\left(\frac{mN^2}{\sqrt{M}B}\right)$. So the total I/Os required by the algorithm is $O\left(\frac{m^2N}{B} + \frac{mN^2}{\sqrt{M}B}\right)$. The seek complexity in this case would be $O(N^2m/M + Nm^2)$. ($\Theta(m^2)$ OSTs are applied inside the bulge window, each of which requires $m$ seeks for a total of $\Theta(m^3)$ seeks per window. The updating of $U$ requires fewer seeks. Updating the rest of the matrix costs a matrix multiplication of dimensions $\Theta(m)$, $\Theta(m)$, and $N$ that incurs $\Theta(m^2N/M)$ seeks. There are $\Theta(N/m)$ windows to process.)

### 6.2.3   An improvement using tiles

We now present a variant of the above algorithm of [25] that uses a tile partitioning of the matrix to improve the seek complexity.

Suppose the input Hessenberg matrix $H$ is tiled as in Algorithm 6.2.

---

**Algorithm 6.2.** *Tile partitioning for our small bulge multishift QR algorithm*

**Input:** *An $N \times N$ Hessenberg matrix $H$ in row/column major order.*
**Output:** *A tile partitioning of $H$.*

---

*Let $w = 3m + 1$ and $h = (N - w)/(w - 1)$;*
$H_{11} = H[1 : w, 1 : w]$;
**for** $j = 1$ *to $h$* **do**
    $H_{1\,j+1} = H[1 : w, (w + 1) + (j - 1)(w - 1) : (w + 1) + j(w - 1) - 1]$;
**endfor**;
**for** $i = 2$ *to $h + 1$* **do**
    *Let $x = (w + 1) + (i - 2)(w - 1)$ and $y = x + w - 2$;*
    $H_{ii} = H[x : y, x - 3 : y]$;
    **for** $j = i$ *to $h$* **do**
        $H_{i\,j+1} = H[x : y, y + (j - i)(w - 1) + 1 : y + (j - i + 1)(w - 1)]$;
    **endfor**;
**endfor**;

---

Here, without loss of generality, we assume that $N = (3m + 1) + h(3m)$; otherwise, the last diagonal tile would be of a smaller size. The partitioning has the following properties: The first block row is made up of $w = 3m + 1$ rows; each of the remaining block rows is made up of $w - 1$ rows. The diagonal tile of the first row is made up of the leftmost $w$ columns; each of the other tiles of the first row is made up of $w - 1$ columns. For $i > 1$, the diagonal tile of the $i$-th block row spans $w + 2$ columns starting at column $2 + (i - 1)(w - 1)$; each of the other tiles of the $i$-th block row spans $w - 1$ columns.

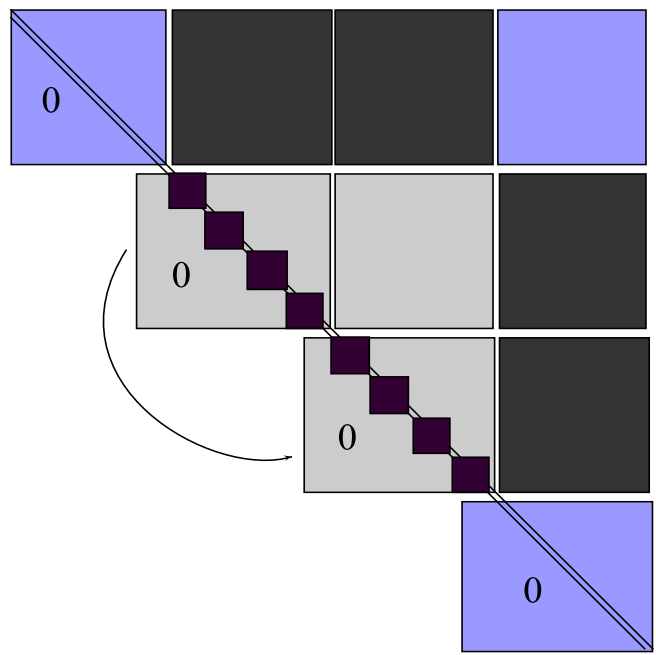

Figure 6.3: Bulge chasing phase: bulge chain is chased from one diagonal tile to the next diagonal tile.

We assume that the matrix has been " tile transposed" according to this partitioning; that is, the elements of each tile are stored contiguously.

The small bulge multishift QR algorithm of [25] is applied as follows: Read $H_{11}$ into the main memory; introduce $m$ bulges in it, and construct the corresponding $U$ all in-core. Then, read, update and write back $H_{1*}$, the first block row of tiles, one tile at a time. A bulge chasing iteration is depicted in Figure 6.3: read the lightly shaded tiles into the main memory, chase the bulge chain, and update the lightly shaded tiles using OSTs; also construct the corresponding $U$ in the main memory. Now multiply the darkly shaded block rows and block columns with $U$ from the left and right respectively. Continue this step until the bulge chain touches the bottom row. Then chase the bulge chain out of the matrix starting with the bottom-most bulge, construct the corresponding $U$ and multiply the last block column with $U$ from the right. This completes one QR iteration.

It is easy to see that the amount of data moved remains the same as in [25], namely, $O(N^2/B)$ blocks if a bulge chase window fits in the main memory, and $O\left(\frac{m^2 N}{B} + \frac{mN^2}{\sqrt{M}B}\right)$ blocks otherwise.

However, there is an improvement in the seek complexity. In the algorithm of [25], if the matrix is stored in column major order, accessing a horizontal slab during the bulge chasing phase stage incurs $O(N)$ seeks each time. In our tile based implementation, if we choose $m = O(\sqrt{M})$, a horizontal slab can be read in and written in $O(N/m)$ seeks (as one tile can be read in with a single seek), for a total of $O(N^2/m^2)$ seeks, which is asymptotically the same as the cost of reading, updating and writing $H$ when $m = \Theta(\sqrt{M})$

On the other hand, if $m > \sqrt{M}$, then instead of performing the computations out-of-core, we do the following rescheduling. Choose $\tilde{m} < m$, and introduce the shift pairs in batches of $\tilde{m}$. Chase the bulges of one batch out of the matrix before introducing the next batch. Since the bulge chasing window fits in the main memory now, all computations are done in-core. Repeat this step for $m/\tilde{m}$ times to complete one QR iteration of the algorithm. It can be easily seen that this modification generates the same bulges and the same sequence of $3 \times 3$ Householders as the algorithm of [25].

The I/O cost of processing one batch of shifts is $O(N^2/B)$. As there are $m/\tilde{m}$ batches, the total I/O cost is $O\left(\frac{N^2 m}{\tilde{m}B}\right)$, which is $O\left(\frac{N^2 m}{\sqrt{M}B}\right)$, when $\tilde{m} = \Theta(\sqrt{M})$. Similarly, the seek complexity is $O\left(\frac{N^2 m}{\tilde{m}^2}\right)$, which is $O\left(\frac{N^2 m}{M}\right)$, when $\tilde{m} = \Theta(\sqrt{M})$.

Note that here the tile partitioning must be done with $\tilde{m}$ instead of $m$.

## 6.3   The $QZ$ Algorithm

In this section we consider some variants of the QZ algorithm.

## 6.3.1 The Implicit Multishift QZ Algorithm

The explicit or implicit QZ algorithms [2, 37, 58, 69, 71, 86, 109, 110, 115] using one or two shifts chase $1 \times 1$ or $2 \times 2$ bulge pairs and therefore are rich in V-V operations. A multishift QZ algorithm that incorporates more V-M operations using $m > 2$ shifts is known [69]. (See Algorithm 6.3.)

The QZ algorithm is an extension of the QR algorithm; each QZ step effectively applies one QR step on $HT^{-1}$, where $(H, T)$ is the input pair. The implicit QZ algorithm avoids a costly, and possibly unstable, explicit computation of $HT^{-1}$.

---

**Algorithm 6.3.** *The implicit multishift QZ algorithm [69]*

**Input:** *An $N \times N$ properly-Hessenberg-Triangular matrix pair $(H, T)$ and the number of shifts $m \in [2, N]$.*

**Output:** *Orthogonal matrices $Q, Z \in \mathbb{R}^{N \times N}$ such that $(Q^T H Z, Q^T T Z)$ is equivalent to $(H, T)$, which is overwritten by $(Q^T H Z, Q^T T Z)$.*

**Notation:** *Let $\mathfrak{H}_j(x)$ denote a Householder matrix that maps the last $N - j$ elements of vector $x \in \mathbb{R}^N$ to zero without modifying the leading $j - 1$ elements.*

---

*Determine $m$ shifts $\rho_1, \ldots, \rho_m$;*
*$x = (HT^{-1} - \rho_1 I)(HT^{-1} - \rho_2 I) \cdots (HT^{-1} - \rho_m I)e_1$;*
*$Q = \mathfrak{H}_1(x);$   $(H, T) = (QH, QT);$*
*$Z = \mathfrak{H}_1(T^{-1}e_1);$   $(H, T) = (HZ, TZ);$*
**for** *$j = 1$* **to** *$N - 2$* **do**
   *$\tilde{Q} = \mathfrak{H}_{j+1}(He_j);$   $(H, T) = (\tilde{Q}H, \tilde{Q}T);$  $Q = Q\tilde{Q};$*
   *$\tilde{Z} = \mathfrak{H}_{j+1}(T^{-1}e_{j+1});$   $(H, T) = (H\tilde{Z}, T\tilde{Z});$  $Z = Z\tilde{Z};$*
**endfor**

---

There are several ways in which the $m$ shifts can be chosen. For example, they could be the generalised eigenvalues of the bottom right $m \times m$ matrix pair; these are called the *generalized Francis shifts*. Once the shifts are chosen, the first column $x$ of $\Pi = (HT^{-1} - \rho_1 I)(HT^{-1} - \rho_2 I) \cdots (HT^{-1} - \rho_m I)$ can be constructed in $O(m^3/B)$ I/Os, since $\Pi$ has exactly $m$ subdiagonals.

Apply $\mathcal{H}_1(x)$ to the matrix pair from the left. That creates an $m \times m$ bulge in $T$. The first column of the bulge is chased using an opposite Householder transformation [37,69,74,117]: *If $T$ is an invertible matrix, then the first column of $T\mathcal{H}_1(T^{-1}e_1)$ is a scalar multiple of $e_1$.*

Applying the right Householder creates bulge in matrix $H$ from column 1 to $m + 1$. Then left and right Householders are alternatively applied to chase the bulge pair down along the subdiagonal, out of the matrix pair, restoring the matrix pair $(H, T)$ to the Hessenberg-Triangular form.

If $m > \sqrt{M}$, then the matrix inversions have to be computed out-of-core; this would add extra I/Os to the computation. So let us assume that $m \leq \sqrt{M}$. The algorithm takes $O(mN^2)$ flops. The I/O complexity of the algorithm is $O(mN^2/B)$. The seek complexity is $O(N^2)$ since $m \leq \sqrt{M}$.

## 6.3.2 A small-bulge variant of the implicit multishift QZ algorithm

In [69], a small bulge variant of the multishift QZ algorithm is proposed, which is a generalisation of the small bulge variant of the multishift QR algorithm of [25]. This algorithm chooses $n_s = 2$ or $4$, and then lines up a chain of $m$ tightly packed bulges, where each bulge is $n_s \times n_s$; that is the number of shifts is $n_s m$.

This algorithm too has three phases that are analogous to the three phases of the multishift QR algorithm of [25]. However in the QZ algorithm, when the matrix pair $(H, T)$ is multiplied from the left with a Householder matrix meant to chase a bulge in $H$, a bulge appears in $T$, and when that is chased by multiplying the pair from the right, a bulge appears in $H$. Apart from this, the two algorithms are similar, with the same asymptotic I/O and seek complexities. So we omit a detailed description and analysis of the QZ algorithm.

---

**Algorithm 6.4.** *Tile partitioning for our small bulge multishift QZ algorithm*

---

**Input:** *A Hessenberg-Triangular matrix pair $(H, T)$ in row/column major order.*

**Output:** *A partitioning of $(H, T)$.*

---

*Let $w = m(n_s + 1) + 1$ and $h = (N - w)/(w - 1)$;*
*$H_{11} = H[1 : w, 1 : w]$, $T_{11} = T[1 : w, 1 : w]$;*
**for** $j = 1$ *to* $h$ **do**
    *$H_{1j+1} = H[1 : w, (w + 1) + (j - 1)(w - 1) : (w + 1) + j(w - 1) - 1]$;*
    *$T_{1j+1} = T[1 : w, (w + 1) + (j - 1)(w - 1) : (w + 1) + j(w - 1) - 1]$;*
**endfor**;
**for** $i = 2$ *to* $h + 1$ **do**
    *Let $x = (w + 1) + (i - 2)(w - 1)$ and $y = x + w - 2$;*
    *$H_{ii} = H[x : y, x - (n_s + 1) : y]$;*
    *$T_{ii} = T[x : y, x - (n_s + 1) : y]$;*
    **for** $j = i$ *to* $h$ **do**
        *$H_{ij+1} = H[x : y, y + (j - i)(w - 1) + 1 : y + (j - i + 1)(w - 1)]$;*
        *$T_{ij+1} = T[x : y, y + (j - i)(w - 1) + 1 : y + (j - i + 1)(w - 1)]$;*
    **endfor**;
**endfor**;

---

## 6.3.3   An improvement using tiles

The technique we used to design a tile based small bulge multishift QR algorithm can also be extended to this case. Without loss of generality, assume that $N = m(n_s + 1) + 1 + hm(n_s + 1)$; otherwise, the last diagonal tile pair would be of lesser dimension than the rest of the diagonal tile pairs. Partition the Hessenberg-Triangular matrix pair as described in Algorithm 6.4, which is a straight forward generalisation of the partitioning we did for the QR algorithm.

The small bulge multishift QZ algorithm of [69] is applied on the partitioned pair as follows: Read $(H_{11}, T_{11})$ into the main memory; introduce $m$ bulge pairs in them, and construct the corresponding $U$ and $V$ (resp., the left and right accumulates) all in-core. Then, read, update and write back $H_{1*}$ and $T_{1*}$, the first block rows of tiles, one tile at a time. A bulge chasing iteration is depicted in Figure 6.4: read the lightly shaded tile pairs into the main memory, chase the bulge chain, and update the lightly shaded tiles using OSTs; also construct the corresponding $U$ and $V$ in the main memory. Now multiply the darkly shaded block rows and block columns with $U$ and $V$ from the left and right respectively. Continue this step till the bulge chain touches the bottom rows of $(H, T)$. Then chase the bulge chain out of the matrix pair starting with the bottom-most bulge pair, construct the corresponding $V$ and update the last block column of $(H, T)$ with $V$ from the right. This completes one QZ iteration.

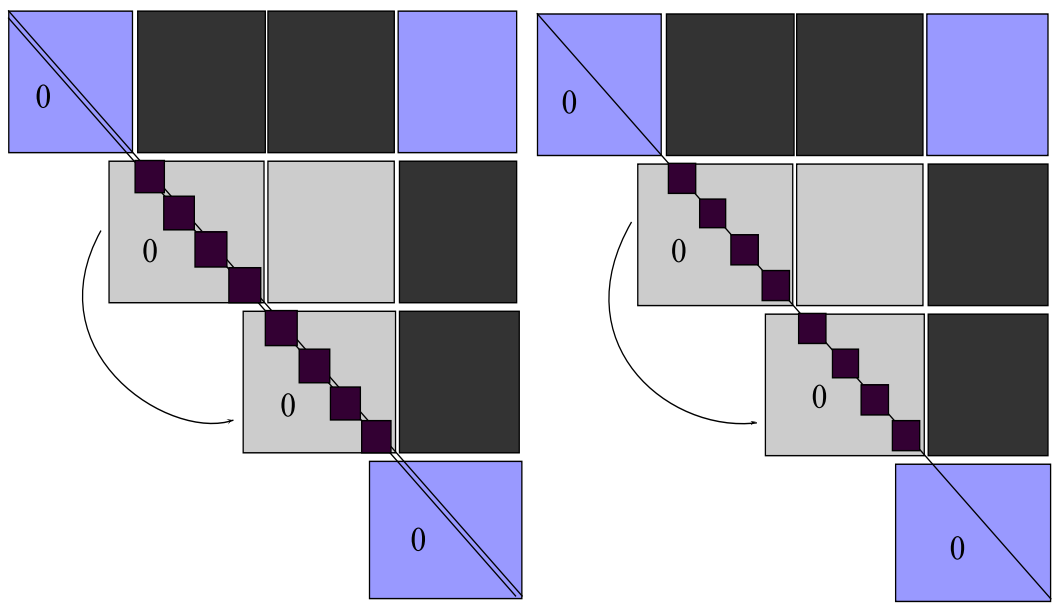

Figure 6.4: Bulge chasing phase: chain of bulge pair is chased from one tile to another tile towards bottom right.

I/O and seek complexities remain same as in the tile based small bulge multishift QR algorithm. We omit the details. Note that the case of, number of shift pairs, $mn_s > \sqrt{M}$ can also be handled similarly, with similar results.

# Conclusions

We have studied some problems of matrix computations. We analysed several known algorithms. We proposed some new algorithms too. The analyses were carried out on the external memory model of Aggarwal and Vitter with an intent of estimating the asymptotic I/O and seek complexities of the algorithms.

A seek is the task of positioning the read/write head of the disk on the data block needed next. In reality, the cost of seeking a block will depend on the location of the block on the disk relative to the block being read or written now. We make the simplistic assumption that every block can be sought at the same cost. We define the seek complexity of an algorithm as the number of times the algorithm chooses to read or write a block that is not physically contiguous with the block being read or written now, thereby necessitating a seek.

The I/O complexity of an algorithm is the sum of the number of blocks of data it moves between the two levels of memory, and its seek complexity. In all the algorithms we have seen, the former asymptotically dominates the latter for almost all choices of problem parameters. Therefore (and also because, new memory technologies (such as flash memory) tend to make switches in data localities cheap), in all the algorithms we designed, our attempt has been to optimise on the net data movement.

Note that while translating our complexity bounds into real time on a disk system of today, the data movement complexity is to be scaled using the disk access rate, while the seek complexity using the average of "seek time + latency"; the latter scaling factor is usually much larger. Also, reductions in the former are not always concomitant with reductions in the latter, as some examples in this thesis show.

For the problems we consider in this thesis a lower bound of $\Omega(N^3/\sqrt{M}B)$ on blocks of data to be moved is known [13]. We demonstrate matching upper bounds for most case. In particular, we present a new algorithm that matches the lower bound for Hessenberg reduction.

Two ratios of interest are the data utilisation ratio defined as the number of operations performed per element moved, and the data locality ratio defined as the number of blocks of data moved per seek, even though neither is a suitable performance metric.

A matrix multiplication on two $t \times t$ matrices, for $t = O(\sqrt{M})$, reads in two $t \times t$ matrices at the cost of $O(t^2/B)$ I/Os, and performs $O(t^3)$ operations on them. With $t = \Theta(\sqrt{M})$, the data utilisation ratio is maximised at $\Theta(\sqrt{M})$ operations per element moved. When $t = \omega(\sqrt{M})$, a matrix multiplication performs $O(t^3)$ operations and $O(t^3/B\sqrt{M})$ I/Os. The data utilisation ratio remains the same. The lower bound of on the amount of data to be moved for matrix problems [13], sets an upper bound of $O(\sqrt{M})$ on the data utilisation ratios of their algorithms.

The data locality ratios of matrix transposition and multiplication algorithms discussed in Chapter 1 are $O(\sqrt{M}/B)$. It is not known whether algorithms of larger ratios exist for these problems. We, however, demonstrate algorithms for other matrix problems (for eg., QR decomposition) with data locality ratio as high as $O(M/B)$.

The I/O performances of slab based algorithms, which are not quite state of the art anymore, depend on the the slab widths chosen. To get the best out of these algorithms, we find, surprisingly, that it is not always necessary that the slabs fit in the main memory.

(See Chapter 2 on QR-decomposition, for example.) Also, for sufficiently large matrices, with the matrix multiplications implemented using an optimal blocked OOC algorithm, we find, the slab based algorithm for QR decomposition runs with an I/O complexity of $O(N^3/B\sqrt{M})$. That means existing implementations of slab based algorithms may not have to be completely redesigned (into, for example, the tile based algorithms) for good OOC performance. If some of the elementary matrix operations in them (for example, matrix multiplication) are handled I/O efficiently, asymptotically optimal I/O performances might be achievable for most inputs. While this is true for QR-decomposition, it is not for Hessenberg reduction, as we have shown.